\documentclass{article}
\usepackage[utf8]{inputenc}

\usepackage{tikz}
\definecolor{colorGibbs}{HTML}{900C3F}
\definecolor{colorApprox}{HTML}{FFC300}
\definecolor{colorFGL}{HTML}{2471A3}
\definecolor{colorGLASSO}{HTML}{229954}

\usepackage{float}
\usepackage{threeparttable}
\usepackage{booktabs}

\usepackage{xcolor}

\usepackage[unicode,linktoc=all,hidelinks]{hyperref}
\usepackage[a4paper, margin=2cm]{geometry}
\usepackage{mathtools}

\usepackage{amsmath}
\usepackage{amssymb}
\usepackage{bm}
\DeclareMathOperator*{\argmax}{arg\,max}

\usepackage{algorithm}
\usepackage{algorithmic}

\usepackage[english]{babel}
\usepackage{natbib}

\usepackage{graphicx}

\usepackage[title]{appendix}

\begin{document}

\title{Copula Graphical Models for Heterogeneous Mixed Data}
\author{Sjoerd Hermes$^{1,2}$, Joost van Heerwaarden$^{1,2}$ and Pariya Behrouzi\thanks{Corresponding author.}$^{\text{ }1}$}
\date{%
    $^1$ Mathematical and Statistical Methods, Wageningen University\\%
    $^2$ Plant Production Systems, Wageningen University\\
}
\maketitle

\begin{abstract}
    \noindent This article proposes a graphical model that handles mixed-type, multi-group data. The motivation for such a model originates from real-world observational data, which often contain groups of samples obtained under heterogeneous conditions in space and time, potentially resulting in differences in network structure among groups. Therefore, the i.i.d.\ assumption is unrealistic, and fitting a single graphical model on all data results in a network that does not accurately represent the between group differences. In addition, real-world observational data is typically of mixed discrete-and-continuous type, violating the Gaussian assumption that is typical of graphical models, which leads to the model being unable to adequately recover the underlying graph structure. The proposed model takes into account these properties of data, by treating observed data as transformed latent Gaussian data, by means of the Gaussian copula, and thereby allowing for the attractive properties of the Gaussian distribution such as estimating the optimal number of model parameter using the inverse covariance matrix. The multi-group setting is addressed by jointly fitting a graphical model for each group, and applying the fused group penalty to fuse similar graphs together. In an extensive simulation study, the proposed model is evaluated against alternative models, where the proposed model is better able to recover the true underlying graph structure for different groups. Finally, the proposed model is applied on real production-ecological data pertaining to on-farm maize yield in order to showcase the added value of the proposed method in generating new hypotheses for production ecologists. 
\end{abstract}

\section{Introduction}
Gaussian graphical models are statistical learning techniques used to make inference on conditional independence relationships within a set of variables arising from a multivariate normal distribution \cite{lauritzen1996graphical}. These techniques have been successfully applied in a variety of fields, such as finance \citep{giudici2016graphical}, biology \citep{krumsiek2011gaussian}, healthcare (Gunathilake et al., \citeyear{gunathilake2020identification}) and others. Despite their wide applicability, the assumption of multivariate normality is often untenable. Therefore, a variety of alternative models have been proposed, in, for example, the case of Poisson or exponential data (Yang et al., \citeyear{yang2015graphical}), ordinal data (Guo et al., \citeyear{guo2015graphical}) and the Ising model in the case of binary data. More general, despite the availability of approaches that do not impose specific distributions on the data, they are limited by their inability to allow for non-binary discrete data (Liu et al., \citeyear{liu2012high}; Fan et al., \citeyear{fan2017high}) or contain a substantial number of parameters (Lee \& Hastie, \citeyear{lee2015learning}). Dobra and Lenkoski (\citeyear{dobra2011copula}) developed a type of Gaussian graphical model that allows for mixed-type data, by combining the theory of copulas, Gaussian graphical models and the rank likelihood method (Hoff, \citeyear{hoff2007extending}). Whereas this model consisted of a Bayesian framework, Abegaz and Wit (\citeyear{abegaz2015copula}) proposed a frequentist alternative, reasoning that the choice of priors for the inverse covariance matrix is nontrivial. Both the Bayesian and frequentist approaches have seen further development and application to real problems in the medical (Mohammadi et al., \citeyear{mohammadi2017bayesian}) and biomedical sciences (Behrouzi \& Wit, \citeyear{behrouzi2019detecting}).

Notwithstanding distributional assumptions, all aforementioned methods assume that the data is i.i.d.\ (independent and identically distributed).
However, real-world observational data often contain groups of samples obtained under heterogeneous conditions in space and time, potentially resulting in differences in network structure among groups. Therefore, the i.i.d.\ assumption is unrealistic, and fitting a single graphical model on all data results in a network that does not accurately represent the between group differences. Conversely, fitting each graph separately for each group fails to take advantage of underlying similarities that may exist between the groups, thereby possibly resulting in highly variable parameter estimates, especially if the sample size for each group is small (Guo et al., \citeyear{guo2011joint}). For these reasons, during the last decade, several researchers have developed graphical models for so-called heterogeneous data, that is, data consisting of various groups (Guo et al., \citeyear{guo2011joint}; Danaher et al., \citeyear{danaher2014joint}; Xie et al., \citeyear{xie2016joint}). Akin to graphical models for homogeneous data, research on heterogeneous graphical models has mainly pertained to the Gaussian setting, despite mixed-type heterogeneous data occurring in a wide variety of situations, such as multi-nation survey data, meteorological data measured at different locations, or medical data of different diseases. Consequently, the aim of this article is to fill the methodological gap that is graphical models for heterogeneous mixed data. 

Even though Jia and Liang (\citeyear{jia2020joint}) aimed to close this methodological gap using their joint mixed learning model, the effectiveness of said model has only been shown in the case where the data follow Gaussian or binomial distributions. This is not always the case in real-world applications. In addition, the model is unable to handle missing data, which tend to be the norm, rather than the exception in real-world data (Nakagawa \& Freckleton, \citeyear{nakagawa2008missing}). Despite Jia and Liang also including an R package with their method, it is currently depreciated and not usable for graph estimation. Motivated by an application of networks on disease status, Park and Won (\citeyear{park2022inference}) recently proposed the fused mixed graphical model: a method to infer graph structures of mixed-type (numerical and categorical) data for multiple groups. This approach is based on the mixed graphical model by Lee and Hastie (\citeyear{lee2013structure}), but extended to the multi-group setting. The proposed model assumes that the categorical variables given all other variables follow a multinomial distribution and all numeric variables follow a Gaussian distribution given all other variables, which is not realistic in the case of Poisson, or non-Gaussian continuous variables. Moreover, the imposed penalty function consists of 6 different penalty parameters to be estimated for 2 groups, which only grows further as the number of groups increases, resulting in the FMGM being prohibitively computationally expensive. Furthermore, no comparative analysis is done with existing methods, but only to a separate network estimation, giving no indication of comparative performance on different types of data. Finally, the FMGM is not accompanied by an R package that allows for such comparative analyses. There is a need for a method that can handle more general mixed-type data consisting of any combination of continuous and ordered discrete variables in a heterogeneous setting, which to the best of our knowledge does not exist at present. Borrowing from recent developments in copula graphical models, the proposed method can handle Gaussian, non-Gaussian continuous, Poisson, ordinal and binomial variables, thereby letting researchers model a wider variety of problems.  All code used in this article can be found at \url{https://github.com/sjoher/cgmhmd-analysis}, whilst the R package can be found at \url{https://github.com/sjoher/cgmhmd}.
 
\subsection{Application to production ecological data}
Interest in relationships between multiple variables based on samples obtained over different locations and time-points is particularly common in production-ecology, a science that aims to understand and predict the productivity of agricultural systems (e.g.\ yield) as a function of their genetic biological components (G), the production environment (E) and human management (M).  Production-ecological data typically consist of observations from different crops, seasons, environments, or management conditions and research is likely to benefit from the use of graphical models. Moreover, production ecological data tends to be of mixed-type, consisting of (commonly) Gaussian, non-Gaussian continuous and Poisson environmental data, but also ordinal and binomial management data. 

A typical challenge for production-ecological research lies in explaining variability in observed yields as a function of a wide set of potential enabling and constraining variables. This is typically done by employing linear models or basic machine learning methods such as random forest that model yield as a function of a set of covariates (Ronner et al., \citeyear{ronner2016}; Bielders \& Gérard, \citeyear{bielders2015millet}; Palmas \& Chamberlin, \citeyear{palmas2020fertilizer}). However, advanced statistical models such as graphical models have not yet been introduced to this field. As graphical models are used to represent the conditional dependencies underlying a set of variables, we expect that these models can greatly aid researchers' understanding  of G$\times$E$\times$M interactions by way of exposing new, fundamental relationships that affect plant production, which have not been captured by methods that are commonly used in the field of production ecology. Therefore, we use this field as a way to illustrate our proposed method and thereby introduce graphical models in general to production ecologists.
\\
\\
This article extends the Gaussian copula graphical model to allow for heterogeneous, mixed data, where we showcase the effectiveness of the novel approach on production-ecological data. To this end, in Section \ref{Methodology}, the proposed methodology behind the Gaussian copula graphical model for heterogeneous data is presented. Section \ref{Simulation study} presents an elaborate simulation study, where the performance of the newly proposed method compared to other types of graphical models is evaluated. An application of the new method on production-ecological data consisting of multiple seasons is given in Section \ref{Real world data example}. Finally, the conclusion can be found in Section \ref{Conclusion}.

\section{Methodology} \label{Methodology}
A Gaussian graphical model corresponds to a graph $G = (V,E)$ that represents the full conditional dependence structure between variables represented by a set of vertices $V = \{1,2,\ldots,p\}$ through the use of a set of undirected edges $E \subset V \times V$, and depends on a $n \times p$ data matrix $\bm{X} = (X_1, X_2,\ldots,X_p), X_j = (X_{1j}, X_{2j}, \ldots, X_{nj})^T, j = 1,\ldots,p$, where $X \sim N_p(0, \Sigma)$, with $\Sigma = \Theta^{-1}$. $\Theta$ is known as the precision matrix containing the scaled partial correlations: $\rho_{ij} = -\frac{\Theta_{ij}}{\sqrt{\Theta_{ii}\Theta_{jj}}}$. Thus, the partial correlation $\rho_{ij}$ represents the independence between $X_i$ and $X_j$ conditional on $X_{V\backslash ij}$. Therefore, $(i,j) \not\in E \Leftrightarrow \Theta_{ij} = 0$. 
\subsection{Copula graphical models for heterogeneous data}
\begin{figure}[H]
\centering
\includegraphics[width=0.5\textwidth]{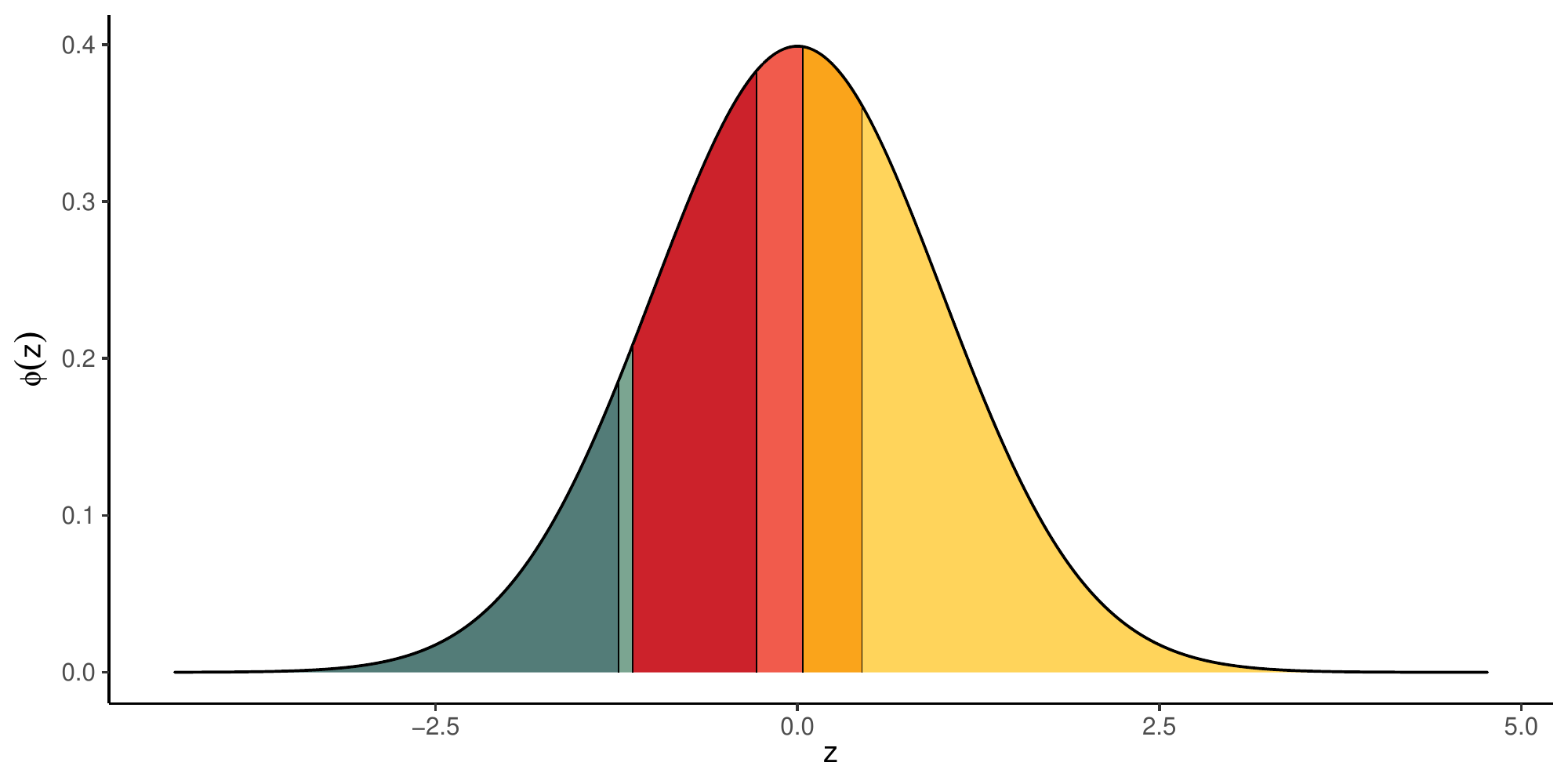}\hfill
\includegraphics[width=0.5\textwidth]{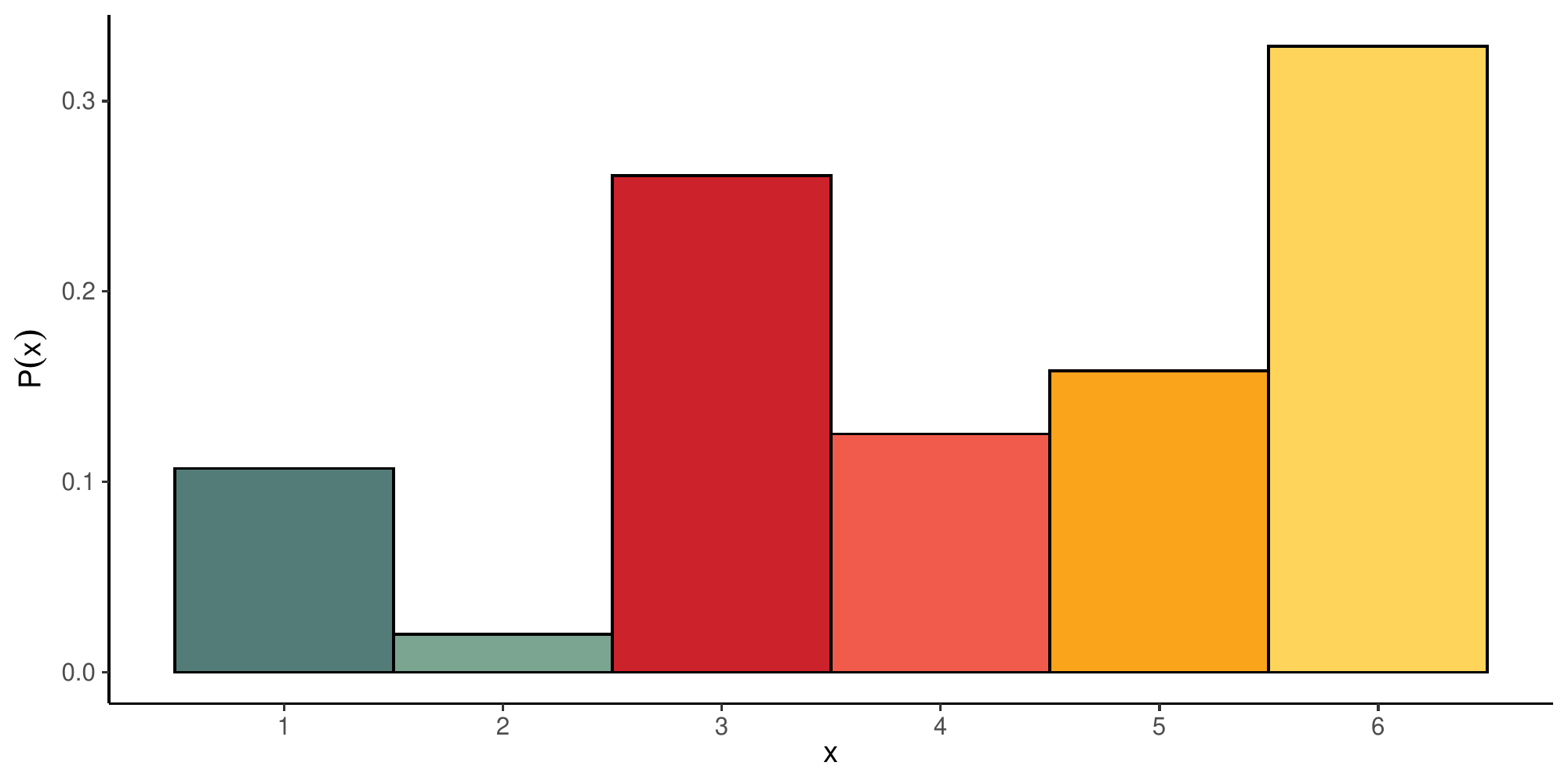}\hfill
\caption{Relationship between the latent and observed values for ordinal variable $X_j^{(k)}$.}
\label{fig:copulatransform}
\end{figure}

Let $X^{(k)} = X_1^{(k)}, X_2^{(k)}, \ldots, X_p^{(k)}$, where $k = 1,2,\ldots,K$ represents the group index, indicating differential genotypic, environmental or management situations, and $X_j^{(k)}$ is a column of length $n_{k}$, where $n_{k}$ is not necessarily equal to $n_{k'}$ for $k \neq k'$ and the data are of mixed-type, i.e. non-Gaussian, counts, ordinal or binomial data, as obtained from measurements on different genotypic, environmental, management and production variables. Moreover, the data across the different groups are not i.i.d.\ For group $k$, a general form of the joint cumulative density function is given by
\begin{equation}
\label{eq:jointdensity}
        F(x_1^{(k)},\ldots,x_p^{(k)}) = \mathbb{P}(X_1^{(k)} \leq x_1^{(k)},\ldots, X_p^{(k)} \leq x_p^{(k)}).
\end{equation}
As the Gaussian assumption is violated for the $X^{(k)}$, maximum likelihood estimation of $\Theta^{(k)}$ based on a Gaussian model will not suffice. For joint densities consisting of different marginals, as in (\ref{eq:jointdensity}), copulas can be applied to model the joint dependency structure between the variables (Nelsen \citeyear{nelsen2007introduction}). In the copula graphical model literature, each observed variable $X_j$ is assumed to have arisen by some perturbation of a latent variable $Z_j$, where $Z \sim N_p(0, \Sigma)$, with correlation matrix $\Sigma$. The choice for a Gaussian latent variable is motivated by the familiar closed-form of the density and the fact that the Gaussian copula correlation matrix enforces the same conditional (in)dependence relations as the precision matrix of graphical models (Dobra \& Lenkoski, \citeyear{dobra2011copula}; Behrouzi \& Wit, \citeyear{behrouzi2019detecting}; Abegaz \& Wit, \citeyear{abegaz2015copula}). This article also assumes a Gaussian distribution for the latent variables such that 
\begin{equation*}
    Z^{(k)} \sim N_p(0, \Sigma^{(k)}),
\end{equation*}
where $\Sigma^{(k)} \in \mathbb{R}^{p \times p}$ represents the correlation matrix for group $k$. The latent variables are linked to the observed data as
\begin{equation*}
    X_j^{(k)} = F_j^{(k)^{-1}}(\Phi(Z_j^{(k)})),
\end{equation*}
where the $F_j^{(k)}()$ are non-decreasing marginal distribution functions, the 
$F_j^{(k)^{-1}}()$ are quantile functions, $\Phi()$ the standard normal cdf and the $X_j^{(k)}, j = 1,2,\ldots, p$ are observed continuous and ordered discrete variables taking values (in the discrete case) in $\{0,1,\ldots, d_j^{(k)}-1\}, d_j^{(k)} \geq 2$, with $d_j^{(k)}$ being the number of categories of variable $j$ in group $k$. A visualization of the relationship between the latent and observed variables is given in Figure \ref{fig:copulatransform}.

\noindent The copula function joining the marginal distributions is denoted as
\begin{equation*}
        \mathbb{P}(X_1^{(k)} \leq x_1^{(k)},\ldots, X_p^{(k)} \leq x_p^{(k)}) = C(F_1^{(k)}(x_1^{(k)}),\ldots, F_p^{(k)}(x_p^{(k)})),
\end{equation*}
where $F_j^{(k)}(x_j^{(k)})$ is standard uniform (Casella \& Berger, \citeyear{casella2001statistical}) and, due to the Gaussian assumption of the $Z_j^{(k)}$, can be written as
\begin{equation*}
        \Phi_{\Sigma^{(k)}}(\Phi^{-1}(F_1^{(k)}(x_1^{(k)})),\ldots, \Phi^{-1}(F_p^{(k)}(x_p^{(k)}))) = \Phi_{\Sigma^{(k)}}(\Phi^{-1}(u_1^{(k)}),\ldots,\Phi^{-1}(u_p^{(k)})),
\end{equation*}
where $\Phi_{\Sigma^{(k)}}()$ is a cdf of a multivariate normal distribution with correlation matrix $\Sigma^{(k)} \in \mathbb{R}^{p \times p}$. As the $\Phi()$ is always nondecreasing and the $F_j^{(k)^{-1}}(t)$ are nondecreasing due to the ordered nature of the data, we have that $x_{ij}^{(k)} < x_{i'j}^{(k)}$ implies $z_{ij}^{(k)} < z_{i'j}^{(k)}$ and $z_{ij}^{(k)} < z_{i'j}^{(k)}$ implies $x_{ij}^{(k)} \leq x_{i'j}^{(k)}, 1 \leq i \neq i' \leq n_{k}$, see Hoff (\citeyear{hoff2007extending}). Thus, we have that $z_j^{(k)} \in D(x_j^{(k)}) = \{z_j^{(k)} \in \mathbb{R}^{n_{k}}: L_{ij}(x_{ij}^{(k)}) < z_{ij}^{(k)} < U_{ij}(x_{ij}^{(k)})\}$, where $L_{ij}(x_{ij}^{(k)}) = \max\{z_{i'j}^{(k)} : x_{i'j}^{(k)} < x_{ij}^{(k)}\}$ and $U_{ij}(x_{ij}^{(k)}) = \min\{z_{i'j}^{(k)} : x_{ij}^{(k)} < x_{i'j}^{(k)}\}$. From here on out, we refer to the set of intervals containing the latent data $D(x) = \{z \in \mathbb{R}^{\sum^K n_k \times p}:z_j^{(k)} \in D(x_j^{(k)})\}$ as $D$.

\noindent In order to facilitate the joint estimation of the different $\Theta^{(k)}$, the probability density function over all $K$ groups is given as
\begin{equation}
\label{eq:pdfgroups}
    \begin{gathered}
        f(x_1^{(1)},\ldots,x_p^{(1)},\ldots\ldots,x_1^{(K)},\ldots,x_p^{(K)})\\
        = \prod_{k = 1}^K \left[c\left(F_1^{(k)}(x_1^{(k)}),\ldots, F_p^{(k)}(x_p^{(k)})\right)\prod_{j = 1}^p f^{(k)}_j(x_j^{(k)})\right],
    \end{gathered}
\end{equation}
where $c(F_1^{(k)}(x_1^{(k)}), \ldots, F_p^{(k)}(x_p^{(k)}))$ is the copula density function and $f^{(k)}_j$ is the marginal density function for the $j$-th variable and the $k$-th group. This copula density is obtained by taking the derivative of the cdf with respect to the marginals. As the Gaussian copula is used, the copula density function can be rewritten as:
\begin{equation*}
    \begin{gathered}
        c(F_1^{(k)}(x_1^{(k)}), \ldots, F_p^{(k)}(x_p^{(k)}))\\
        = \frac{\partial^p C}{\partial F_1^{(k)}, \ldots, \partial F_p^{(k)}}\\
        = \frac{\Phi_{\Sigma^{(k)}}(\Phi^{-1}(u_1^{(k)}),\ldots,\Phi^{-1}(u_p^{(k)}))}{\prod_{i=1}^p\Phi(\Phi^{-1}(u_i^{(k)}))}\\
       = (2\pi)^{-\frac{p}{2}}\det(\Sigma^{(k)})^{-\frac{1}{2}}\frac{\exp(-\frac{1}{2}(\Phi^{-1}(u_1^{(k)}),\ldots,\Phi^{-1}(u_p^{(k)}))^T \Sigma^{(k)^{-1}}(\Phi^{-1}(u_1^{(k)}),\ldots,\Phi^{-1}(u_p^{(k)})))}{\prod_{i=1}^p (2\pi)^{-\frac{1}{2}}\exp(-\frac{1}{2}\Phi^{-1}(u_i^{(k)})\Phi^{-1}(u_i^{(k)}))}\\
       = (2\pi)^{-\frac{p}{2}}\det(\Sigma^{(k)})^{-\frac{1}{2}}\frac{\exp(-\frac{1}{2}Z^{(k)^T} \Sigma^{(k)^{-1}}Z^{(k)})}{(2\pi)^{-\frac{p}{2}}\exp(-\frac{1}{2}Z^{(k)^T} Z^{(k)})}\\
       = \det(\Theta^{(k)})^{\frac{1}{2}}\exp(-\frac{1}{2}Z^{(k)^T} (\Theta^{(k)}-I)Z^{(k)}),
    \end{gathered}
\end{equation*}
where $Z^{(k)}$ is used to shorten the $n_{k} \times p$ latent matrix $(\Phi^{-1}(u_1^{(k)}),\ldots,\Phi^{-1}(u_p^{(k)}))$ and $I$ is a $p \times p$ identity matrix. The full log-likelihood over $K$ groups is then given by 
\begin{gather}
    \ell(\{\Theta^{(k)}\}_{k = 1}^{K}|\bm{X}) = \log\left[\prod_{k = 1}^K \prod_{i = 1}^{n_{k}}\left( c(F_1^{(k)}(x_{i1}^{(k)}), \ldots, F_p^{(k)}(x_{ip}^{(k)}))\prod_{j = 1}^p f^{(k)}_j(x_{ij}^{(k)})\right)\right] \notag\\
    = \sum_{k = 1}^K \sum_{i = 1}^{n_{k}}\log(c(F_1^{(k)}(x_{i1}^{(k)}), \ldots, F_p^{(k)}(x_{ip}^{(k)}))) + \sum_{k = 1}^K \sum_{i = 1}^{n_{k}}\sum_{j = 1}^p \log(f^{(k)}_j(x_{ij}^{(k)})) \notag\\
    = \sum_{k = 1}^K \sum_{i = 1}^{n_{k}}\log\left(\det(\Theta^{(k)})^{\frac{1}{2}}\exp(-\frac{1}{2}Z_i^{(k)^T} (\Theta^{(k)}-I)Z_i^{(k)})\right)\notag\\
     + \sum_{k = 1}^K \sum_{i = 1}^{n_{k}}\sum_{j = 1}^p \log\left( \frac{1}{\sigma_j\sqrt{2\pi}}\exp\left(-\frac{1}{2}\frac{x_{ij}^{(k)^2}}{\sigma_j^2}\right)\right) \notag\\
    = \frac{1}{2}\sum_{k = 1}^K n_{k}\log(\det(\Theta^{(k)})) -\frac{1}{2}\sum_{k = 1}^K\sum_{i = 1}^{n_{k}}Z_i^{(k)^T} (\Theta^{(k)}-I)Z_i^{(k)}\notag\\
     - \frac{1}{2}\sum_{k = 1}^K n_{k}p\log(2\pi) - \frac{1}{2}\sum_{k = 1}^K \sum_{i = 1}^{n_{k}}\sum_{j = 1}^p  x_{ij}^{(k)^2} \notag\\
    \propto \frac{1}{2}\sum_{k = 1}^K n_{k}\log(\det(\Theta^{(k)})) -\frac{1}{2}\sum_{k = 1}^K\sum_{i = 1}^{n_{k}}Z_i^{(k)^T} (\Theta^{(k)}-I)Z_i^{(k)} 
    \label{eq:loglik}
\end{gather}

\noindent where $\bm{X} = (X^{(1)},\ldots, X^{(K)})^T$. We denote $\{\Theta^{(k)}\}_{k = 1}^{K}$ as $\bm{\Theta}$ for the purpose of simplicity. The two rightmost terms in the penultimate line of (\ref{eq:loglik}) were omitted, as they are constant with respect to $\Theta^{(k)}$ because of the standard normal marginals.

\subsection{Model estimation}
When estimating the marginals, a nonparametric approach is adhered to, as is common in the copula literature. This is due to the computational costs involved their estimating and because of the fact that we only care about the dependencies encoded in the $\Theta^{(k)}$. They are estimated as $\hat{F}^{(k)}_j(x) = \frac{1}{n_{k}+1}\sum_{i = 1}^{n_k}\mathbb{I}(X_{i j}^{(k)} \leq x)$. Whilst (\ref{eq:loglik}) allows for the joint estimation of the graphical models pertaining to the different groups, these models are not sparse and cannot enforce relations to be the same. Sparsity is a common assumption in biological networks and production ecology is not an exception. Consider for example the solubilization of fertiliser which is independent of root activity (de Wit, \citeyear{de1953physical}), the independence between nitrogen and yield for certain crops (Raun et al. \citeyear{raun2011independence}), or more general the independence between weather and various management techniques. Moreover, if certain groups are highly similar, for example different locations with similar climates, enforcing relations between those groups to be the same is both realistic and parsimonious. Therefore, a fused-type penalty is imposed upon the precision matrix, such that the penalised log-likelihood function has the following form
\begin{equation}
\label{eq:loglikpen}
\begin{gathered}
    \ell(\bm{\Theta}|\bm{X}) = \frac{1}{2}\sum_{k = 1}^K n_{k} \log(\det(\Theta^{(k)})) -\frac{1}{2}\sum_{k = 1}^K\sum_{i = 1}^{n_{k}}Z_i^{(k)^T} (\Theta^{(k)}-I)Z_i^{(k)}\\
     - \lambda_1\sum_{k=1}^K\sum_{j\neq j'}|\theta_{jj'}^{(k)}| - \lambda_2\sum_{k<k'}\sum_{j,j'}|\theta_{jj'}^{(k)} - \theta_{jj'}^{(k')}| 
\end{gathered}
\end{equation}
for $1 \leq k \neq k' \leq K$ and $1 \leq j \neq j' \leq p$. Here,  $\lambda_1$ controls the sparsity of the $K$ different graphs and $\lambda_2$ controls the edge-similarity between the $K$ different graphs. Higher values for $\lambda_1$ and $\lambda_2$ correspond to respectively more sparse and more similar graphs, where similarity is not only limited to similar sparsity patterns in the different $\Theta^{(k)}$, but also in terms of attaining the exact same coefficients across different $\Theta^{(k)}$. The fused-type penalty for heterogeneous data graphical models was originally proposed by Danaher et al. (\citeyear{danaher2014joint}). Whenever groups pertaining to seasons or environments share similar characteristics, production ecological research has hinted at similar edge values between groups (Hajjarpoor et al., \citeyear{hajjarpoor2021environmental}; Zhang et al., \citeyear{zhang1999water}; Richards \& Townley-Smith, \citeyear{richards1987variation}). Consider the case where groups represent different locations. If two groups have very similar environments, both weather patterns and soil properties, many conditional independence relations are expected to be similar between the groups, as the underlying production ecological relations are assumed to be invariant across (near) identical situations (Connor et al., \citeyear{connor2011crop}). Conversely, if the amount of shared characteristics is limited between the groups, the edge values between groups are expected to be different, resulting from the low value for $\lambda_2$, as obtained from a penalty parameter selection method. Moreover, this fused-type penalty has been shown to outperform other types of penalties (Danaher et al., \citeyear{danaher2014joint}), and, if the data contains only 2 groups, this type of penalty has a very light computational burden, due to the existence of a closed-form solution for (\ref{eq:loglikpen}) once the conditional expectations of the latent variables have been computed.
\\
\\
As direct maximization $\ell(\bm{\Theta}|\bm{X})$ is not feasible due to the nonexistence of an analytic expression of (\ref{eq:loglikpen}), an iterative method is needed to estimate the value of $\bm{\Theta}_{\lambda_1,\lambda_2}$. A common algorithm used in the presence of latent variables is the EM-algorithm (McLachlan \& Krishnan, \citeyear{mclachlan2007algorithm}). A benefit of this algorithm is that it can handle missing data, which is not uncommon in production ecology as plants can die mid-season due to external stresses such as droughts or pests. The EM algorithm alternates between an E-step and an M-step, where during the E-step the expectation of the (unpenalised) complete-data (both $\bm{X}$ and $\bm{Z}$) log-likelihood conditional on the event $D$ and the estimate $\bm{\hat{\Theta}}^{(m)}$ obtained during the previous M-step is computed

\begin{gather}
    Q(\bm{\Theta}|\bm{\hat{\Theta}}^{(m)}) = \mathbb{E}\left[\sum_{k = 1}^K \sum_{i = 1}^{n_{k}}\log(p(Z_i|\bm{\Theta}))|x_{i}^{(k)}, \bm{\hat{\Theta}}^{(m)}, D\right]\notag\\
    = \mathbb{E}\left[\frac{1}{2}\sum_{k = 1}^K n_{k}\log(\det(\Theta^{(k)})) -\frac{1}{2}\sum_{k = 1}^K\sum_{i = 1}^{n_{k}}Z_i^{(k)^T} (\Theta^{(k)}-I)Z_i^{(k)}\bigg| x_{i}^{(k)}, \bm{\hat{\Theta}}^{(m)}, D\right]\notag\\
    = \mathbb{E}\left[\frac{1}{2}\sum_{k = 1}^K n_{k}\log(\det(\Theta^{(k)})) -\frac{1}{2}\sum_{k = 1}^K\sum_{i = 1}^{n_{k}}Z_i^{(k)^T}\Theta^{(k)}Z_i^{(k)} + \frac{1}{2}\sum_{k = 1}^K\sum_{i = 1}^{n_{k}}Z_i^{(k)^T}Z_i^{(k)}\bigg| x_{i}^{(k)}, \bm{\hat{\Theta}}^{(m)}, D\right]\notag\\
    = \frac{1}{2}\sum_{k = 1}^K n_{k}\log(\det(\Theta^{(k)})) -\frac{1}{2}\sum_{k = 1}^K\sum_{i = 1}^{n_{k}}\mathbb{E}(Z_i^{(k)^T}\Theta^{(k)}Z_i^{(k)}|x_{i}^{(k)}, \bm{\hat{\Theta}}^{(m)}, D) \notag\\+ \frac{1}{2}\sum_{k = 1}^K\sum_{i = 1}^{n_{k}}\mathbb{E}(Z_i^{(k)^T}Z_i^{(k)}|x_{i}^{(k)}, \bm{\hat{\Theta}}^{(m)}, D)\notag\\
    = \frac{1}{2}\sum_{k = 1}^K n_{k}\log(\det(\Theta^{(k)})) -\frac{1}{2}\sum_{k = 1}^K\sum_{i = 1}^{n_{k}}\text{tr}(\Theta^{(k)}\mathbb{E}(Z_i^{(k)}Z_i^{(k)^T}|x_{i}^{(k)}, \bm{\hat{\Theta}}^{(m)}, D)) \notag\\
    + \frac{1}{2}\sum_{k = 1}^K\sum_{i = 1}^{n_{k}}\text{tr}(\mathbb{E}(Z_i^{(k)}Z_i^{(k)^T}|x_{i}^{(k)}, \bm{\hat{\Theta}}^{(m)}, D))\notag\\
    =  \frac{1}{2}\sum_{k = 1}^K n_{k}\left[\log(\det(\Theta^{(k)})) -\text{tr}(\Theta^{(k)}\bar{R}^{(k)}) + \text{tr}(\bar{R}^{(k)})\right]\notag\\
    \propto \frac{1}{2}\sum_{k = 1}^K n_{k}\left[\log(\det(\Theta^{(k)})) -\text{tr}(\Theta^{(k)}\bar{R}^{(k)})\right]
    \label{eq:e-step}
\end{gather}
where 
\begin{equation}
    \begin{gathered}
    \label{eq:test}
    \bar{R}^{(k)} = \frac{1}{n_{k}}\sum_{i = 1}^{n_{k}}\mathbb{E}(Z_i^{(k)}Z_i^{(k)^T}|x_{i}^{(k)}, \bm{\hat{\Theta}}^{(m)}, D)\\
    \mathbb{E}(Z_i^{(k)}Z_i^{(k)^T}|x_{i}^{(k)}, \bm{\hat{\Theta}}^{(m)}, D) = \mathbb{E}(Z_i^{(k)}|x_{i}^{(k)}, \bm{\hat{\Theta}}^{(m)}, D)\mathbb{E}(Z_i^{(k)}|x_{i}^{(k)}, \bm{\hat{\Theta}}^{(m)}, D)^T\\
     + \text{ cov}(Z_i^{(k)}|x_{i}^{(k)}, \bm{\hat{\Theta}}^{(m)}, D),
    \end{gathered}
\end{equation} 
which are the estimated correlation matrix and the first and second moments of a doubly truncated multivariate normal density, respectively. Expressions of these moments are given by Manjunath and Wilhelm (\citeyear{manjunath2021moments}). Note that $\text{tr}(\bar{R}^{(k)})$ does not depend on $\Theta^{(k)}$ and was therefore omitted in the last step of the derivation of $Q(\bm{\Theta}|\bm{\hat{\Theta}}^{(m)})$.
\\
\\
Despite the existence of a functional expression for $\mathbb{E}(Z_i^{(k)}Z_i^{(k)^T}|x_{i}^{(k)}, \bm{\hat{\Theta}}^{(m)}, D)$, Behrouzi and Wit (\citeyear{behrouzi2019detecting}) used two alternative methods to compute this quantity, as directly computing the moments is computationally expensive, even for moderate $p$ of 50. Accordingly, the faster alternatives proposed are a Gibbs sampling and an approximation-based approach. Whereas the former results in better estimates for the precision matrices, the latter is computationally more efficient. The Gibbs sampler is built on the Fortran-based truncated normal sampler in the \texttt{tmvtnorm} R package (Wilhelm \& Manjunath, \citeyear{tmvtnorm}).\\
\\
The Gibbs method consists of a limited number of steps, where the focus lies on drawing $N$ samples from the truncated normal distribution, where $t_{x_{i}^{(k)}}$ is a vector of length $p$ and contains the lower truncation points for observation $x_{i}^{(k)}$ and $t_{x_{i}^{(k)} + 1}$ is a vector of length $p$ and contains the upper truncation points for observation $x_{i}^{(k)}$. The method is summarised in Algorithm \ref{alg:gibbs}.

\begin{algorithm}[H]
\caption{Gibbs method}\label{alg:gibbs}
    \hspace*{\algorithmicindent} \textbf{Input:} $\bm{X}$ and $\{\Sigma^{(1)},\ldots,\Sigma^{(K)}\}$\\
    \hspace*{\algorithmicindent} \textbf{Output:} $\{\bar{R}^{(1)},\ldots,\bar{R}^{(K)}\}$
  \begin{algorithmic}[1]
    \FOR{$k = 1$ to $K$}
      \FOR{$i = 1$ to $n_k$}
        \STATE Compute $t_{x_{i}^{(k)}}^{(k)},t_{x_{i}^{(k)} + 1}^{(k)}$
        \STATE Generate $Z^{(k)}_{i*} = (Z^{(k)}_{i11}, \ldots, Z^{(k)}_{iN1}, \ldots, Z^{(k)}_{i1p}, \ldots, Z^{(k)}_{iNp}) \sim TN(0, \Sigma^{(k)}, t_{x_{i}^{(k)}}^{(k)}, t_{x_{i}^{(k)} + 1}^{(k)})$
      \ENDFOR
      \STATE Compute $\bar{R}^{(k)} = \frac{1}{n_{k}}\sum_{i = 1}^{n_{k}}\frac{1}{N}Z^{(k)^T}_{i*} Z^{(k)}_{i*}$
    \ENDFOR
  \end{algorithmic}
\end{algorithm}

\noindent When the Gibbs method is run for the first iteration ($m = 1$) of the EM algorithm, $\Sigma^{(k)} = I_p$, and otherwise $\Sigma^{(k)} = \Theta^{(m-1), (k)^{-1}}$. Computing the sample mean on simulated data from a truncated normal distribution leads to consistent estimates of $\bar{R}^{(k)}$ (Manjunath \& Wilhelm, \citeyear{manjunath2021moments}).
\\
\\
Guo et al.\ (\citeyear{guo2015graphical}) proposed an approximation method to estimate the conditional expectation of the covariance matrix. When $j = j'$, the variance elements of this matrix correspond to the second moment $\mathbb{E}(Z_i^{(k)^2}|x_{i}^{(k)}, \bm{\hat{\Theta}}^{(m)}, D)$ and the covariance elements, $j \neq j'$ are approximated by  $\mathbb{E}(Z_{ij}^{(k)}|x_{i}^{(k)}, \bm{\hat{\Theta}}^{(m)}, D)\mathbb{E}(Z_{ij'}^{(k)}|x_{i}^{(k)}, \bm{\hat{\Theta}}^{(m)}, D)$, which is simply a product of the first moment for variables $j$ and $j'$. The method is summarised in Algorithm \ref{alg:approx}, and details can be found in Appendix \ref{Approximate method details}.

\begin{algorithm}[H]
\caption{Approximate method}\label{alg:approx}
    \hspace*{\algorithmicindent} \textbf{Input:} $\bm{X}$\\
    \hspace*{\algorithmicindent} \textbf{Output:} $\{\tilde{R}^{(1)},\ldots,\tilde{R}^{(K)}\}$\\
    \hspace*{\algorithmicindent} \textbf{Initialise:} $\text{ }\mathbb{E}(Z_{ij}^{(k)}|x_{i}^{(k)}, \bm{\hat{\Theta}}^{(m)}, D) \approx \mathbb{E}(Z_{ij}^{(k)}|x_{ij}^{(k)}, \bm{\hat{\Theta}}^{(m)}, D),\\ \textcolor{white}{texdfdhfjsdfd}\text{ }\mathbb{E}(Z_{ij}^{(k)^2}|x_{i}^{(k)}, \bm{\hat{\Theta}}^{(m)}, D) \approx \mathbb{E}(Z_{ij}^{(k)^2}|x_{ij}^{(k)}, \bm{\hat{\Theta}}^{(m)}, D)$\\
    $\textcolor{white}{texdfdhfjsdfd}\text{ and }\mathbb{E}(Z_{ij}^{(k)}Z_{ij'}^{(k)}|x_{i}^{(k)}, \bm{\hat{\Theta}}^{(m)}, D) \approx \mathbb{E}(Z_{ij}^{(k)}|x_{ij}^{(k)}, \bm{\hat{\Theta}}^{(m)}, D)\mathbb{E}(Z_{ij'}^{(k)}|x_{ij'}^{(k)}, \bm{\hat{\Theta}}^{(m)}, D)$, \\
    $\textcolor{white}{texdfdhfjsdfd}\text{ }$for $i = 1,\ldots n_k, j,j'=1,\ldots,p$ and $k = 1,\ldots,K$
  \begin{algorithmic}[1]
    \FOR{$k = 1$ to $K$}
      \FOR{$i = 1$ to $n_k$}
        \IF{$j = j'$}
        \STATE Update $\mathbb{E}(Z_{ij}^{(k)^2}|x_{i}^{(k)}, \bm{\hat{\Theta}}^{(m)}, D)$ for $j = 1,\ldots,p$ using (\ref{eq:z})
        \ELSE
        \STATE Update 
        $\mathbb{E}(Z_{ij}^{(k)}|x_{i}^{(k)}, \bm{\hat{\Theta}}^{(m)}, D)$ for $j = 1,\ldots,p$ using (\ref{eq:z2})
        \STATE Set $\mathbb{E}(Z_{ij}^{(k)}Z_{ij'}^{(k)}|x_{i}^{(k)}, \bm{\hat{\Theta}}^{(m)}, D) = \mathbb{E}(Z_{ij}^{(k)}|x_{i}^{(k)}, \bm{\hat{\Theta}}^{(m)}, D)\mathbb{E}(Z_{ij'}^{(k)}|x_{i}^{(k)}, \bm{\hat{\Theta}}^{(m)}, D)$ for $1 \leq j \neq j' \leq p$
        \ENDIF
        \ENDFOR
          \STATE Compute $\tilde{r}_{jj'}^{(k)} = \frac{1}{n_{k}}\sum_{i = 1}^{n_{k}}\mathbb{E}(Z_{ij}^{(k)}Z_{ij'}^{(k)}|x_{i}^{(k)}, \bm{\hat{\Theta}}^{(m)}, D)$ for $1 \leq j, j' \leq p$
       \ENDFOR
  \end{algorithmic}
\end{algorithm}

\noindent After obtaining an estimate for all $\bar{R}^{(k)}$, using either the Gibbs method or the approximate method, the M-step commences which consists of maximizing (\ref{eq:e-step}) with respect to the precision matrices, subject to the imposed penalties of (\ref{eq:loglikpen}): 
\begin{equation*}
\hat{\bm{\Theta}}_{\lambda_1,\lambda_2}^{(m + 1)} = \argmax_{\bm{\Theta}}\{\frac{1}{2}\sum_{k = 1}^K n_{k}\left[\log(|\Theta^{(k)}|) - \text{tr}(\Theta^{(k)}\bar{R}^{(k)})\right] - \lambda_1\sum_{k=1}^K||\Theta^{(k)}||_1 - \lambda_2\sum_{k<k'}||\Theta^{(k)} - \Theta^{(k')}||_1\},
\end{equation*}
 which is done using the fused graphical lasso by Danaher et al.\ (\citeyear{danaher2014joint}).

\subsection{Model selection}
Instead of one penalty parameter, as is typical for graphical models, the copula graphical model for heterogeneous data requires the selection of two penalty parameters. In a predictive setting, the AIC (Akaike, \citeyear{akaike1973second}) penalty and cross-validation approaches are commonly applied, whereas the BIC (Schwarz, \citeyear{schwarz1978estimating}), EBIC (Chen \& Chen, \citeyear{chen2012extended}) and StARS (Liu et al., \citeyear{liu2010stability}) approaches are designed for graph identification (Vujačić et al., \citeyear{vujavcic2015computationally}). When considering a grid of $\lambda_1 \times \lambda_2$ combinations of penalty parameters, computational cost becomes crucial. Therefore, unless a coarse grid structure for the penalty parameters is used or if researchers are willing to spend a substantial amount of time waiting for the ‘‘optimal’’ combination of penalty parameters, information criteria are preferred over computationally intensive methods such as cross-validation and StARS. Two of these, the AIC and EBIC are given for heterogeneous data:
\begin{gather}
    \text{AIC}(\lambda_1,\lambda_2) = \sum_{k=1}^K\left[n_{k} \text{tr}(S^{(k)} \hat{\Theta}_{\lambda_1,\lambda_2}^{(k)})-n_{k} \log(|\hat{\Theta}_{\lambda_1,\lambda_2}^{(k)}|) + 2 \nu^{(k)}_{\lambda_1,\lambda_2}\right]\\
    \text{EBIC}(\lambda_1,\lambda_2) = \sum_{k=1}^K\left[n_{k} \text{tr}(S^{(k)} \hat{\Theta}_{\lambda_1,\lambda_2}^{(k)})-n_{k} \log(|\hat{\Theta}_{\lambda_1,\lambda_2}^{(k)}|) + \log(n_{k}) \nu^{(k)}_{\lambda_1,\lambda_2} + 4\gamma\log(p) \nu^{(k)}_{\lambda_1,\lambda_2}\right] \label{eq:EBIC}
\end{gather}
where the degrees of freedom $\nu^{(k)} = \text{card}\{(i,j): i < j, \hat{\theta}_{ij}^{(k)} \neq 0\}$ and $0 \leq \gamma \leq 1$ is a penalty parameter, commonly set to 0.5. The EBIC tends to lead to sparser networks than the AIC penalty (Vujačić et al., \citeyear{vujavcic2015computationally}). 

\section{Simulation studies} \label{Simulation study}
To demonstrate the added value of the proposed copula graphical model for heterogeneous mixed data, a simulation study is undertaken using a variety of settings. In this simulation, relevant parameter values are known a-priori. Both the Gibbs-based and approximation-based copula graphical models will be evaluated, together with the following models; the fused graphical lasso (FGL) by Danaher et al.\ (\citeyear{danaher2014joint}) and the graphical lasso (GLASSO) method (Friedman et al., \citeyear{friedman2008sparse}), where the networks are fitted seperately for each group. Whilst the joint mixed learning method by Jia and Liang (\citeyear{jia2020joint}) is a method that aims to analyse similar data to the copula graphical model for heterogeneous data, their R package \texttt{equSA} (Jia et al., \citeyear{equSA}) is no longer supported, which, at the time of writing, resulted in constant crashes of the R software when running the joint mixed learning method. 

As an aside, it should be noted that both the FGL and GLASSO methods assume Gaussian data. However, the data used to compare the methods is of mixed-type. Even though Poisson data can be normalised using for instance a logarithmic base-10 transformation, ordinal and binary data cannot. For this reason, the data was not normalised. 
\\
\\
For each combination of network-type, 25 datasets are generated consisting of $p = \{50,100\}$, $n_{k} = \{10, 50, 100, 500\}$ and $K = 3$ groups. The combinations of these values for $n_k$ and $p$ result in both high- and low-dimensional scenarios. Moreover, the choice of $p$ is pursuant to the typical number of variables in a production-ecological dataset and the $K$ with the number of seasons or environments analysed in such data. The networks used in this simulation study are a cluster network, a scale-free network and a random network according to the Erdős-Rényi model (Erdős \& Rényi, \citeyear{erdos1959random}). The choice for the first network is motivated by the fact that for production-ecological data, when group-membership variables exist, such as environments or seasons, we expect clusters consisting of a large number of within-cluster edges compared to the number of between-cluster edges to arise. The scale-free network represents the opposite of a cluster network, consisting of a relatively high number of edges between clusters and a low number of edges within clusters. As model performance under opposite conditions is also relevant to evaluate, the scale-free network was chosen. The last network choice; the random network, allows the evaluation of the proposed copula graphical model under unspecified graph structure, where edge connection probability $p_e$ results in sparse or dense graphs, depending on whether the probability is close to 0 or 1, respectively. Having a model that performs well under assumed sparsity without additional structural network assumptions is useful, as it is not always known a-priori what the underlying graph should look like. The data is simulated in the following way:
\begin{enumerate}
    \item Generate graph $G$ and (initial) shared precision matrix $\Theta^{(s)}$ according to the type of network: cluster, scale-free or random, by setting values in $\Theta$ to $[-1, 0.5] \cup [0.5, 1]$ the correspond to edges in $G$
    \item Create different precision matrices $\Theta^{(k)}$ for $k = 1,\ldots,K$ by randomly filling in $\lfloor\rho M\rfloor$ zero elements of $\Theta^{(k)}$, where $\rho$ is a dissimilarity parameter and $M$ is the number of nonzero elements in the lower diagonal of $\Theta^{(s)}$.
    \item Set diag$(\Theta^{(k)})$ = 0
    \item Ensure that the $\Theta^{(k)}$ are positive definite by setting diag$(\Theta^{(k)}) = |\lambda_{\min}(\Theta^{(k)})| + \epsilon$, where $\lambda_{\min}(\Theta^{(k)})$ are the smallest eigenvalues of $\Theta^{(k)}$ and $\epsilon > 0$.
    \item Compute covariance matrix $\Sigma^{(k)} = \Theta^{(k)^{-1}}$
    \item Turn covariance matrix into correlation matrix $\Sigma^{(k)} = \text{diag}(\Sigma^{(k)})^{-\frac{1}{2}}\Sigma^{(k)}\text{diag}(\Sigma^{(k)})^{-\frac{1}{2}}$
    \item Set $(\Theta^{(k)}) = \Sigma^{(k)^{-1}}$ 
    \item From $V = \{1,\ldots,p\}$ sample $s_b = \lfloor\gamma_b p\rfloor$, $s_o = \lfloor\gamma_o p\rfloor$, $s_p = \lfloor\gamma_p p\rfloor$ and $s_g = V - s_b - s_o - s_p$ columns without replacement for the binomial, ordinal, Poisson and Gaussian variables respectively, where $\gamma_i, i \in I = \{b,o,p,g\}$ represents the proportion of that variable occurring in the data. Then, $\bigcup_{i \in I}s_i = V$ and $\bigcap_{i \in I}s_i = \emptyset$. 
    \item Sample latent data $Z \sim N(0, \Sigma^{(k)})$
    \item Generate observed data $X_{s_i}^{(k)} = F_{s_i}^{(k)^{-1}}(\Phi_i(Z_{s_i}^{(k)}))$
\end{enumerate}

\noindent In the simulations, we set distribution proportions $\gamma_b$ to 0.1, $\gamma_o$ to 0.5, $\gamma_p$ to 0.2 and $\gamma_g$ to 0.2. Moreover, the success parameter for the binomial marginals is set at 0.5, the rate parameter of the Poisson marginals is set at 10 and the number of categories for ordinal variables is set at 6. The edge connection probability $p_e$ for the random network is set at 0.05, resulting in sparse random graphs. The number of clusters in the cluster network is set at 3. Finally, we considered $\rho = 0.25$ and $1$, resulting in respectively similar and different graphs for each group. This results in a total of $3\times4\times2\times2$ = 48 unique combinations of settings. Each combination is used to sample 25 different datasets to minimise the effect of randomness on the results.
\\
\\
To evaluate the performance of the models, ROC curves are drawn, consisting of the false positive rate (FPR) and true positive rate (TPR), which respectively represent the rate of false and true edges selected by the model. These are defined as:
\begin{equation*}
    \begin{gathered}
    \text{FPR} = \frac{1}{K}\sum_{k = 1}^K \frac{\sum_{i < j}\mathbb{I}(\theta_{ij}^{(k)} = 0, \hat{\theta}_{ij}^{(k)}\neq 0)}{\mathbb{I}(\theta_{ij}^{(k)} = 0)}\\
    \text{TPR} = \frac{1}{K}\sum_{k = 1}^K \frac{\sum_{i < j}\mathbb{I}(\theta_{ij}^{(k)} \neq 0, \hat{\theta}_{ij}^{(k)}\neq 0)}{\mathbb{I}(\theta_{ij}^{(k)} \neq 0)},
    \end{gathered}
\end{equation*}
where $\mathbb{I}()$ is an indicator function. The FPR and TPR are computed by varying $\lambda_1$ over $[0,1]$ with a step size of 0.05, whilst fixing $\lambda_2$ to either 0, 0.1 or 1, resulting in 3 ROC curves per plot (one for each value of $\lambda_2$), per dataset per model. Correspondingly, AUC scores are computed for these curves. In addition, the Frobenius and entropy loss are computed and averaged over the groups:
\begin{equation*}
    \begin{gathered}
        \text{FL} = \frac{1}{K}\sum_{k=1}^K \frac{||\Theta^{(k)} - \hat{\Theta}^{(k)}||_F^2}{||\Theta^{(k)}||_F^2} \\
        \text{EL} = \frac{1}{K}\sum_{k=1}^K \text{tr}(\Theta^{(k)^{-1}}\hat{\Theta}^{(k)}) - \log(\det(\Theta^{(k)^{-1}}\hat{\Theta}^{(k)})) - p
    \end{gathered}
\end{equation*}

\noindent Aside from the averaged performance measures across the values of $\lambda_2$, the best choice performance measure is given: the value of the performance measure the attained the best score, either highest or lowest depending on the measure, for a particular choice of $\lambda_2$. 
\\
\\
Results for the random network are given in Table \ref{tab:simrandom} and Figure \ref{fig:roccurves}, whereas the results for the cluster and scale-free networks are given in Appendix \ref{Additional simulation results} in respectively Table \ref{tab:simcluster} and Figure \ref{fig:roccurves_cl} and Table \ref{tab:simscalefree} and Figure \ref{fig:roccurves_sf}. 

\begin{table}[H]
\centering
  \begin{threeparttable}
  \caption{Simulation results for random networks, where AUC stands for area under the curve, FL stands for Frobenius loss, EL stands for entropy loss and the bc suffix stands for best choice, i.e. the best result of that respective metric (highest for AUC and lowest for EL and FL) for a particular value of $\lambda_2$. The value corresponding to the winning method is written in bold.}
  \label{tab:simrandom}
     \begin{tabular}{lcccccc}
        \toprule
         \multicolumn{4}{r}{Gibbs method/Approximate method} 
         &
         \multicolumn{3}{c}{Fused graphical lasso/GLASSO}\\
        \midrule
         \textbf{$n, p$ $\rho$} & \textbf{AUC} &\textbf{FL} & \textbf{EL}& \textbf{AUC} & \textbf{FL} & \textbf{EL}\\ \midrule
$10, 50, 0.25$ & \textbf{0.60}/0.60 & \textbf{0.58}/0.64 & \textbf{12.29}/13.59 & 0.56/0.56 & 3.15/1.23 & 48.30/22.09\\ 
$50, 50, 0.25$ & \textbf{0.83}/0.83 & \textbf{0.18}/0.18 & \textbf{5.42}/5.43 & 0.74/0.76 & 0.94/0.34 & 34.93/10.05\\ 
$100, 50, 0.25$ & \textbf{0.89}/0.89 & \textbf{0.17}/0.17 & \textbf{5.05}/5.06 & 0.81/0.86 & 0.91/0.32 & 35.53/9.44\\ 
$500, 50, 0.25$ & 0.94/0.94 & \textbf{0.17}/0.17 & \textbf{4.92}/4.94 & 0.91/\textbf{0.98} & 0.90/0.32 & 35.55/9.22\\ 
$10, 100, 0.25$ & \textbf{0.56}/0.56 & \textbf{0.79}/0.95 & \textbf{34.19}/40.47 & 0.54/0.53 & 4.01/1.51 & 112.07/55.05\\ 
$50, 100, 0.25$ & \textbf{0.77}/0.77 & \textbf{0.22}/0.22 & \textbf{14.49}/14.62 & 0.69/0.67 & 0.94/0.39 & 72.79/24.57\\ 
$100, 100, 0.25$ & \textbf{0.84}/0.84 & \textbf{0.20}/0.20 & \textbf{13.17}/13.20 & 0.77/0.78 & 0.88/0.35 & 72.70/22.11\\ 
$500, 100, 0.25$ & 0.93/0.93 & \textbf{0.19}/0.19 & \textbf{12.69}/12.72 & 0.89/\textbf{0.95} & 0.87/0.35 & 73.63/21.27\\ 
$10, 50, 1$ & \textbf{0.56}/0.56 & \textbf{0.58}/0.63 & \textbf{12.76}/14.06 & 0.54/0.54 & 3.05/1.19 & 49.03/22.43\\ 
$50, 50, 1$ & \textbf{0.72}/0.72 & \textbf{0.19}/0.19 & \textbf{5.96}/5.97 & 0.66/0.68 & 0.93/0.35 & 35.63/10.52\\ 
$100, 50, 1$ & \textbf{0.77}/0.77 & \textbf{0.18}/0.18 & \textbf{5.62}/5.63 & 0.72/0.78 & 0.89/0.33 & 35.76/9.94\\ 
$500, 50, 1$ & 0.85/0.85 & \textbf{0.18}/0.18 & \textbf{5.48}/5.50 & 0.83/\textbf{0.94} & 0.89/0.33 & 35.93/9.74\\ 
$10, 100, 1$ & \textbf{0.54}/0.54 & \textbf{0.77}/0.92 & \textbf{35.09}/41.27 & 0.52/0.52 & 3.97/1.47 & 114.34/55.70\\ 
$50, 100, 1$ & \textbf{0.67}/0.67 & \textbf{0.23}/0.23 & \textbf{15.51}/15.64 & 0.62/0.62 & 0.93/0.40 & 73.43/25.44\\ 
$100, 100, 1$ & \textbf{0.73}/0.73 & \textbf{0.21}/0.21 & \textbf{14.23}/14.26 & 0.68/0.70 & 0.87/0.36 & 74.06/23.05\\ 
$500, 100, 1$ & 0.82/0.82 & \textbf{0.20}/0.21 & \textbf{13.78}/13.81 & 0.79/\textbf{0.89} & 0.86/0.36 & 75.03/22.27\\ 
        \midrule
         \multicolumn{4}{r}{Gibbs method/Approximate method} 
         &
         \multicolumn{3}{c}{Fused graphical lasso}\\
        \midrule
         \textbf{$n, p$ $\rho$} & \textbf{AUC bc} & \textbf{FL bc} & \textbf{EL bc} & \textbf{AUC bc} & \textbf{FL bc} & \textbf{EL bc}\\ \midrule
$10, 50, 0.25$ & 0.63/\textbf{0.64} & \textbf{0.28}/0.29 & \textbf{7.99}/8.22 & 0.59 & 1.45 & 37.35 \\ 
$50, 50, 0.25$ & \textbf{0.88}/0.87 & \textbf{0.17}/0.17 & \textbf{5.09}/5.10 & 0.79 & 0.91 & 34.42 \\ 
$100, 50, 0.25$ & \textbf{0.91}/0.91 & \textbf{0.17}/0.17 & \textbf{4.96}/4.98 & 0.84 & 0.90 & 35.42 \\ 
$500, 50, 0.25$ & \textbf{0.98}/0.98 & \textbf{0.17}/0.17 & \textbf{4.91}/4.93 & 0.91 & 0.90 & 35.50 \\ 
$10, 100, 0.25$ & \textbf{0.58}/0.58 & \textbf{0.39}/0.43 & \textbf{23.02}/24.82 & 0.56 & 1.88 & 84.83 \\ 
$50, 100, 0.25$ & \textbf{0.82}/0.82 & \textbf{0.20}/0.20 & \textbf{13.28}/13.30 & 0.74 & 0.87 & 70.54 \\ 
$100, 100, 0.25$ & \textbf{0.88}/0.88 & \textbf{0.19}/0.19 & \textbf{12.84}/12.86 & 0.81 & 0.87 & 72.25 \\ 
$500, 100, 0.25$ & \textbf{0.95}/0.95 & \textbf{0.19}/0.19 & \textbf{12.68}/12.72 & 0.89 & 0.87 & 73.47 \\ 
$10, 50, 1$ & \textbf{0.58}/0.58 & \textbf{0.29}/0.29 & \textbf{8.52}/8.75 & 0.56 & 1.41 & 38.21 \\ 
$50, 50, 1$ & \textbf{0.74}/0.74 & \textbf{0.18}/0.18 & \textbf{5.67}/5.67 & 0.68 & 0.90 & 35.12 \\ 
$100, 50, 1$ & \textbf{0.78}/0.78 & \textbf{0.18}/0.18 & \textbf{5.55}/5.57 & 0.73 & 0.89 & 35.62 \\ 
$500, 50, 1$ & \textbf{0.94}/0.94 & \textbf{0.18}/0.18 & \textbf{5.44}/5.45 & 0.87 & 0.89 & 35.82 \\ 
$10, 100, 1$ & \textbf{0.55}/0.55 & \textbf{0.39}/0.43 & \textbf{24.18}/25.92 & 0.53 & 1.87 & 86.84 \\ 
$50, 100, 1$ & \textbf{0.69}/0.69 & \textbf{0.21}/0.21 & \textbf{14.39}/14.40 & 0.65 & 0.87 & 71.20 \\ 
$100, 100, 1$ & \textbf{0.74}/0.74 & \textbf{0.21}/0.21 & \textbf{13.95}/13.97 & 0.69 & 0.86 & 73.61 \\ 
$500, 100, 1$ & \textbf{0.89}/0.88 & \textbf{0.20}/0.20 & \textbf{13.71}/13.74 & 0.82 & 0.86 & 74.74 \\
        \bottomrule
     \end{tabular}
  \end{threeparttable}
\end{table}

\begin{figure}[H]
\centering
\text{Random network, $p = 50$, $\rho = 0.25$}\\
\includegraphics[width=0.25\textwidth]{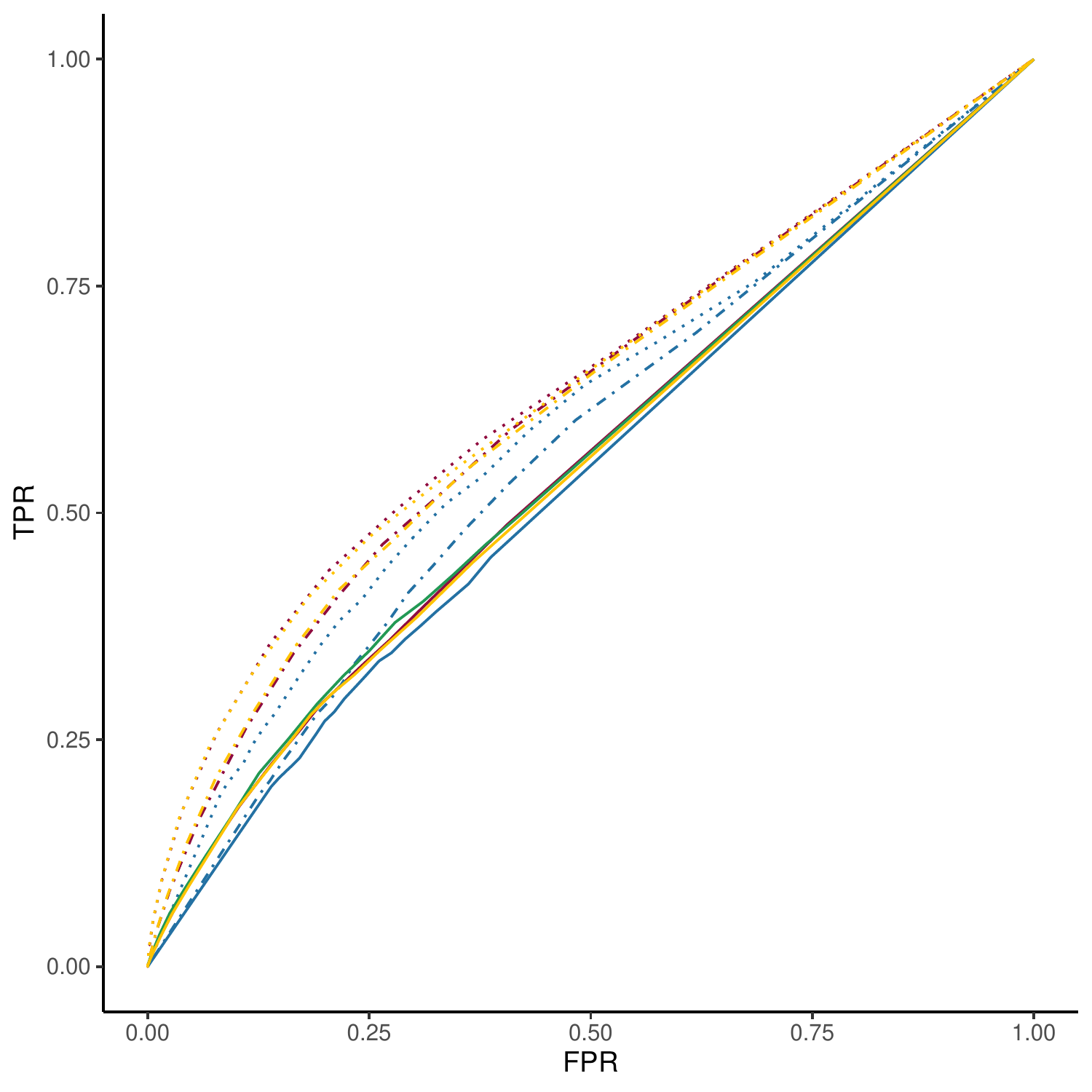}\hfill
\includegraphics[width=0.25\textwidth]{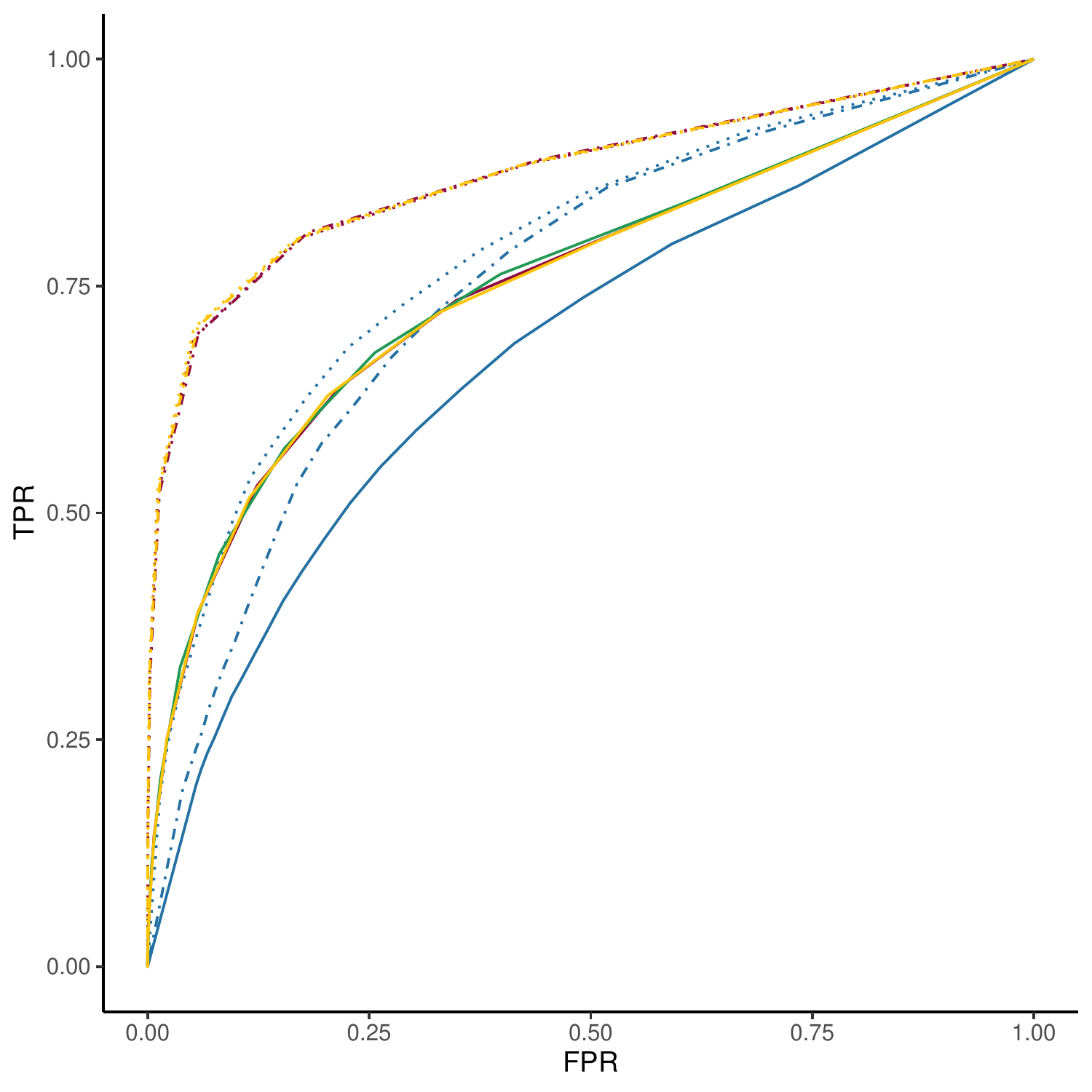}\hfill
\includegraphics[width=0.25\textwidth]{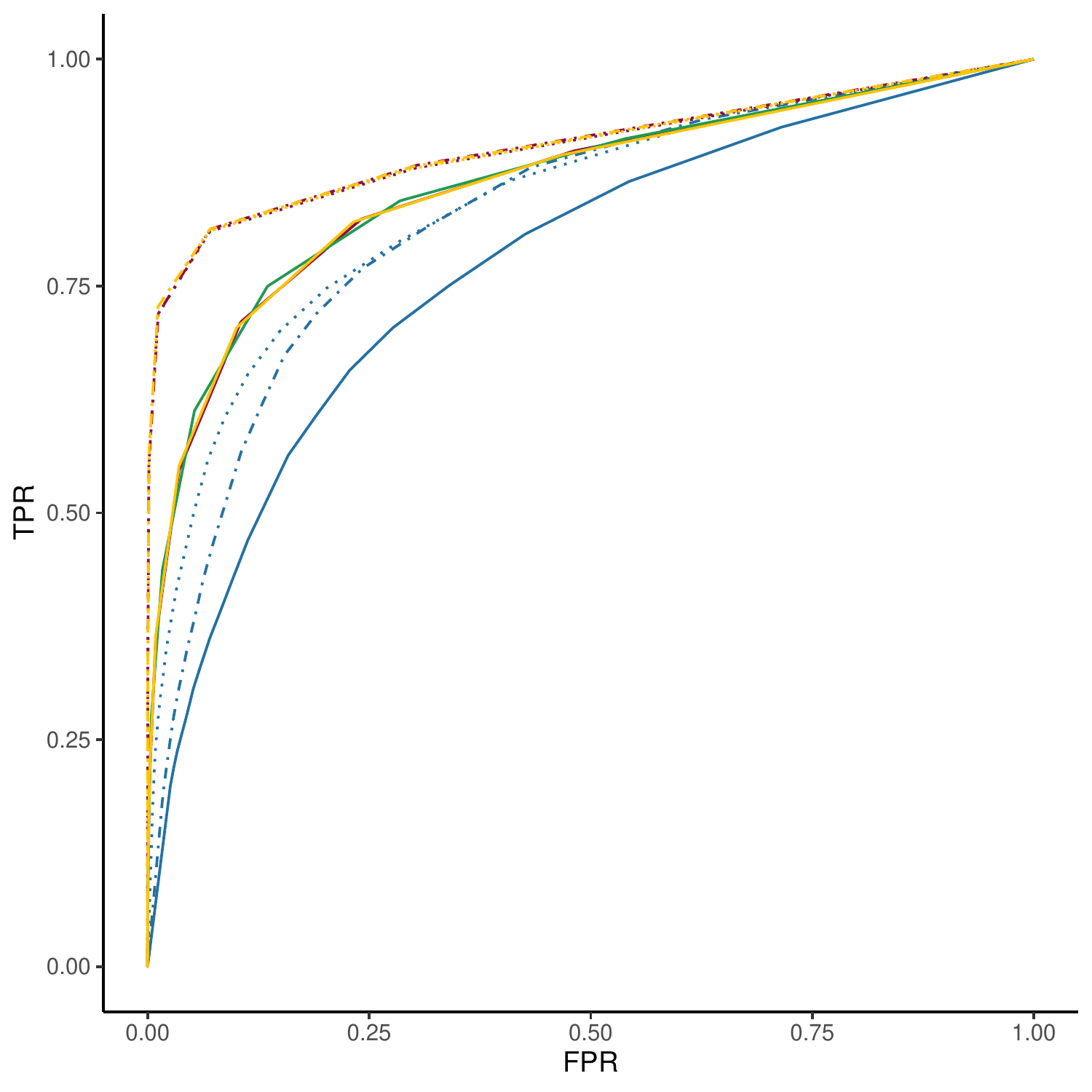}\hfill
\includegraphics[width=0.25\textwidth]{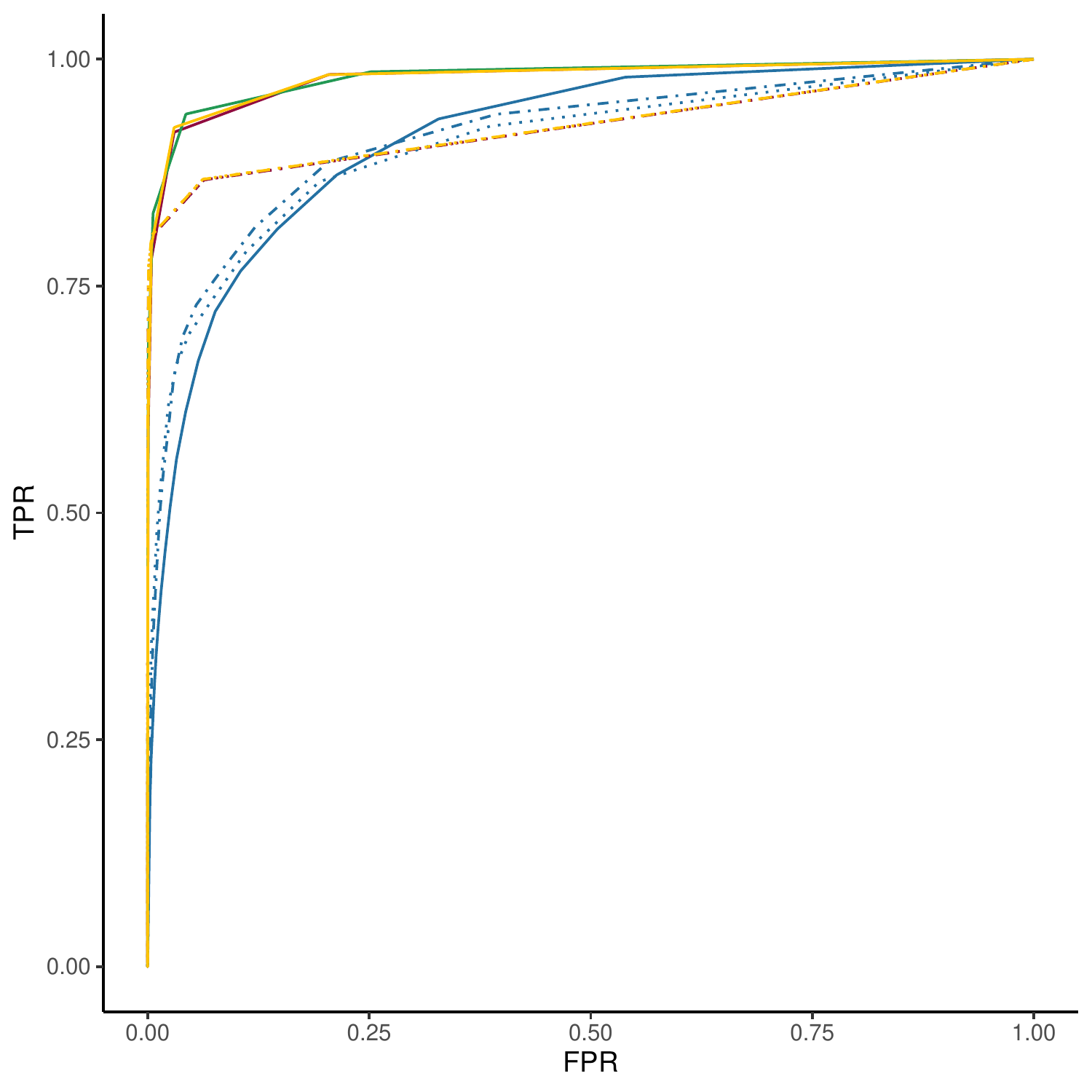}\hfill
\text{Random network, $p = 100$, $\rho = 0.25$}\\
\includegraphics[width=0.25\textwidth]{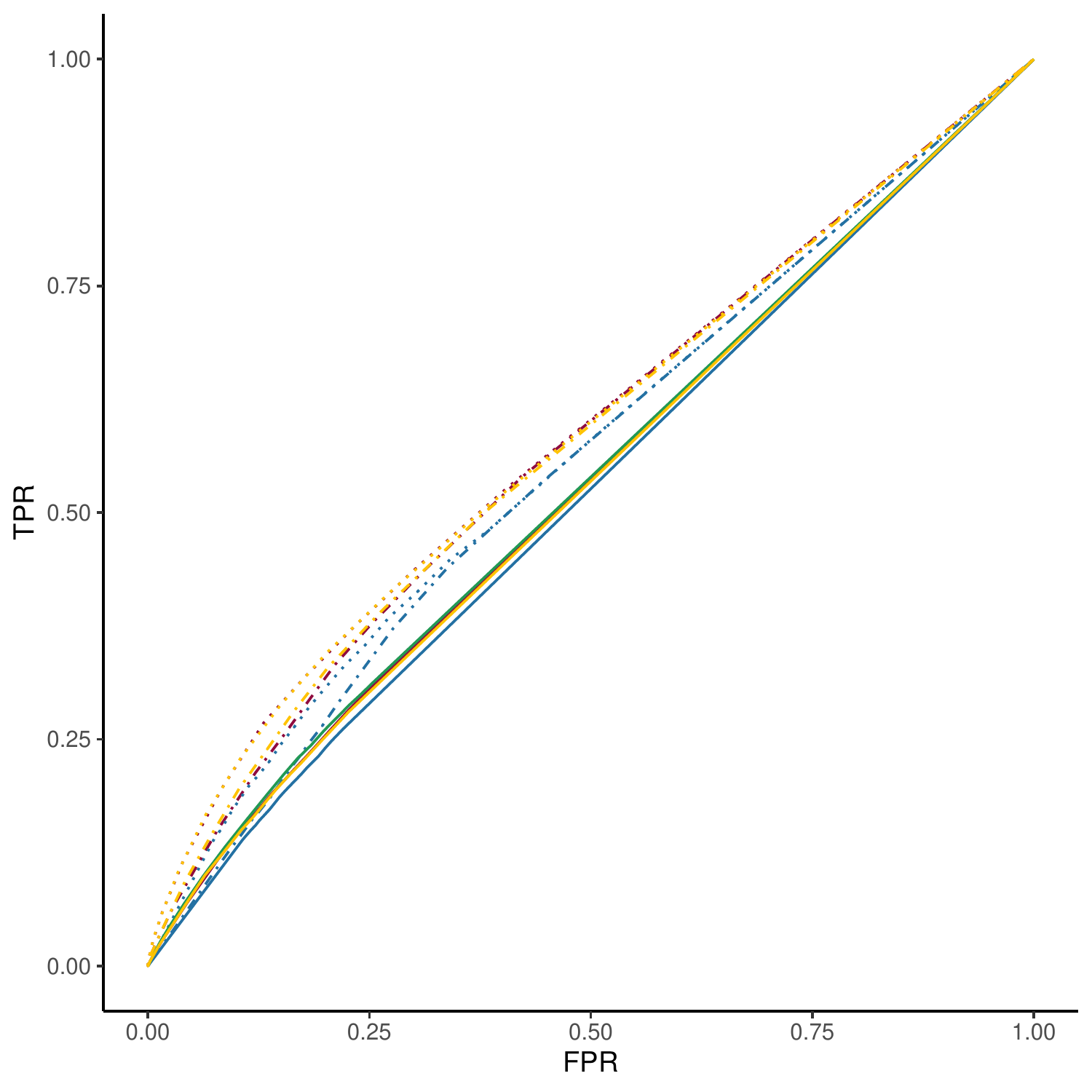}\hfill
\includegraphics[width=0.25\textwidth]{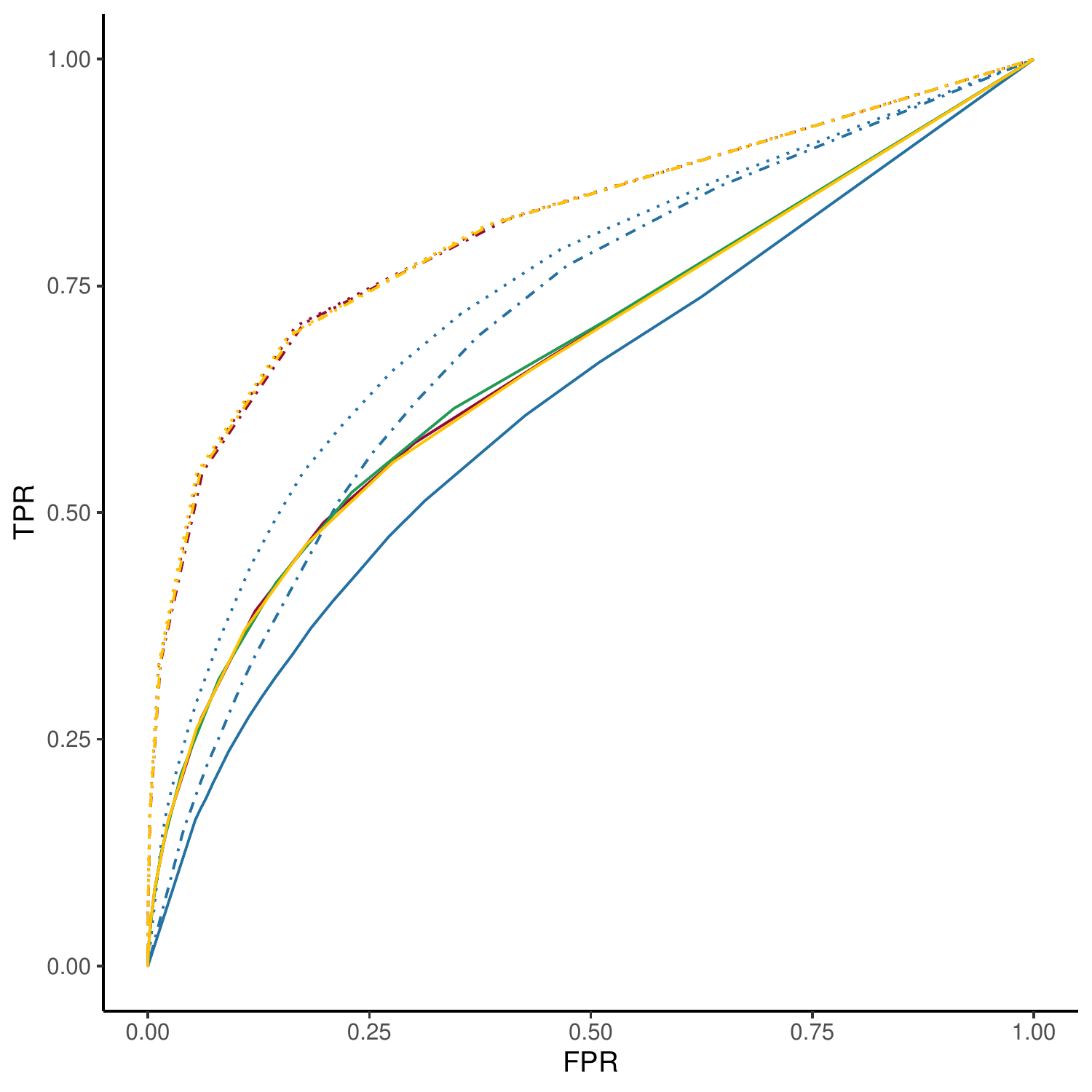}\hfill
\includegraphics[width=0.25\textwidth]{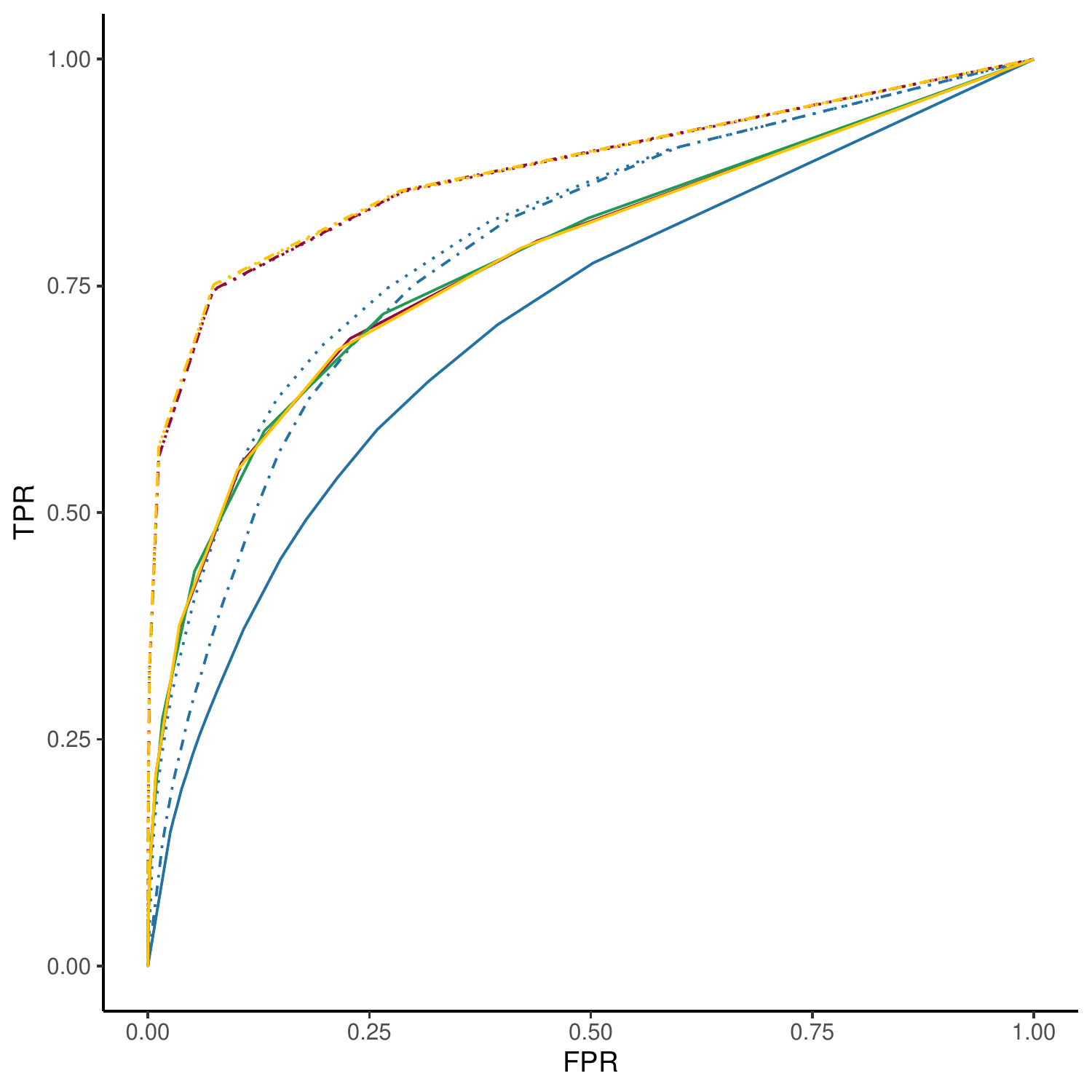}\hfill
\includegraphics[width=0.25\textwidth]{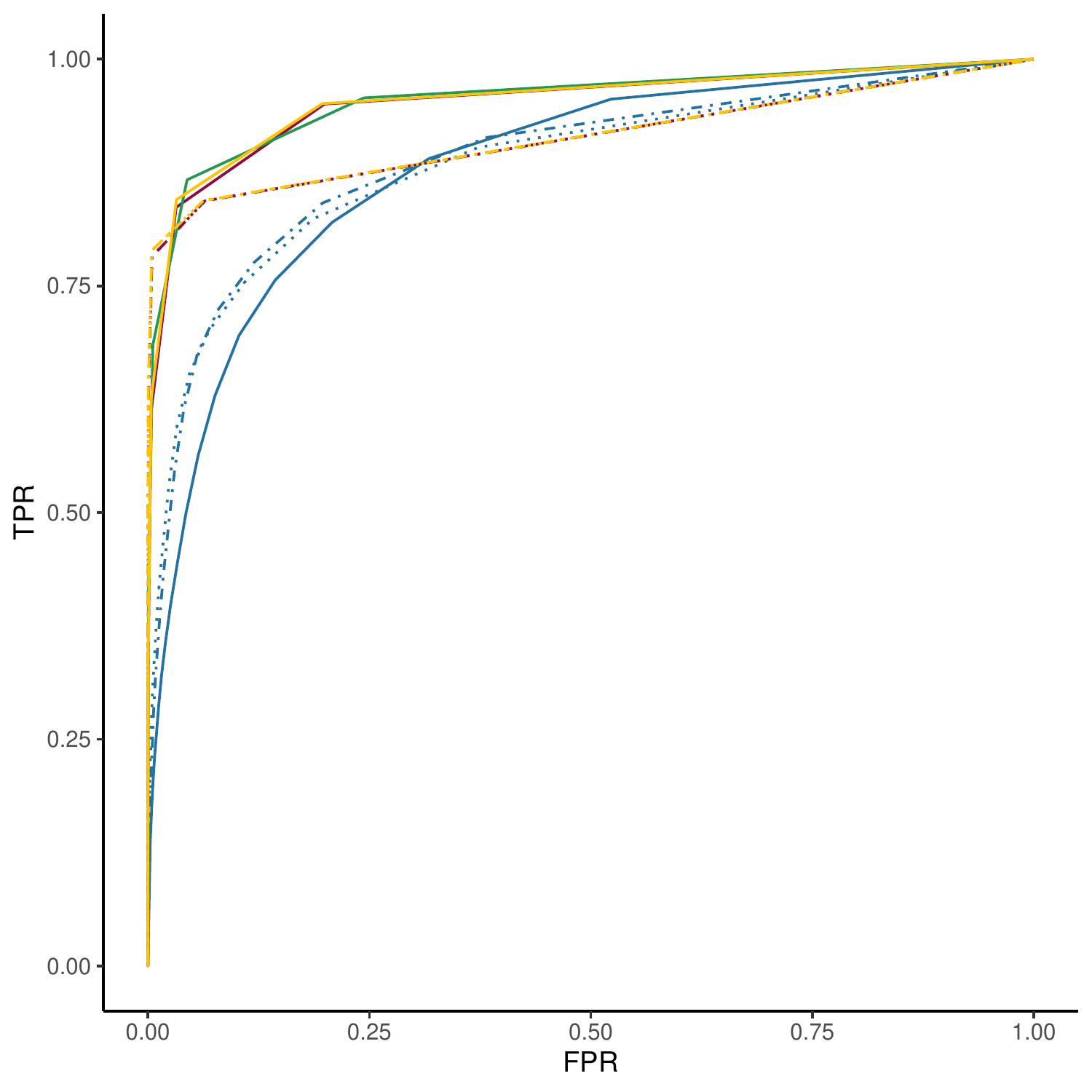}\hfill
\text{Random network, $p = 50$, $\rho = 1$}\\
\includegraphics[width=0.25\textwidth]{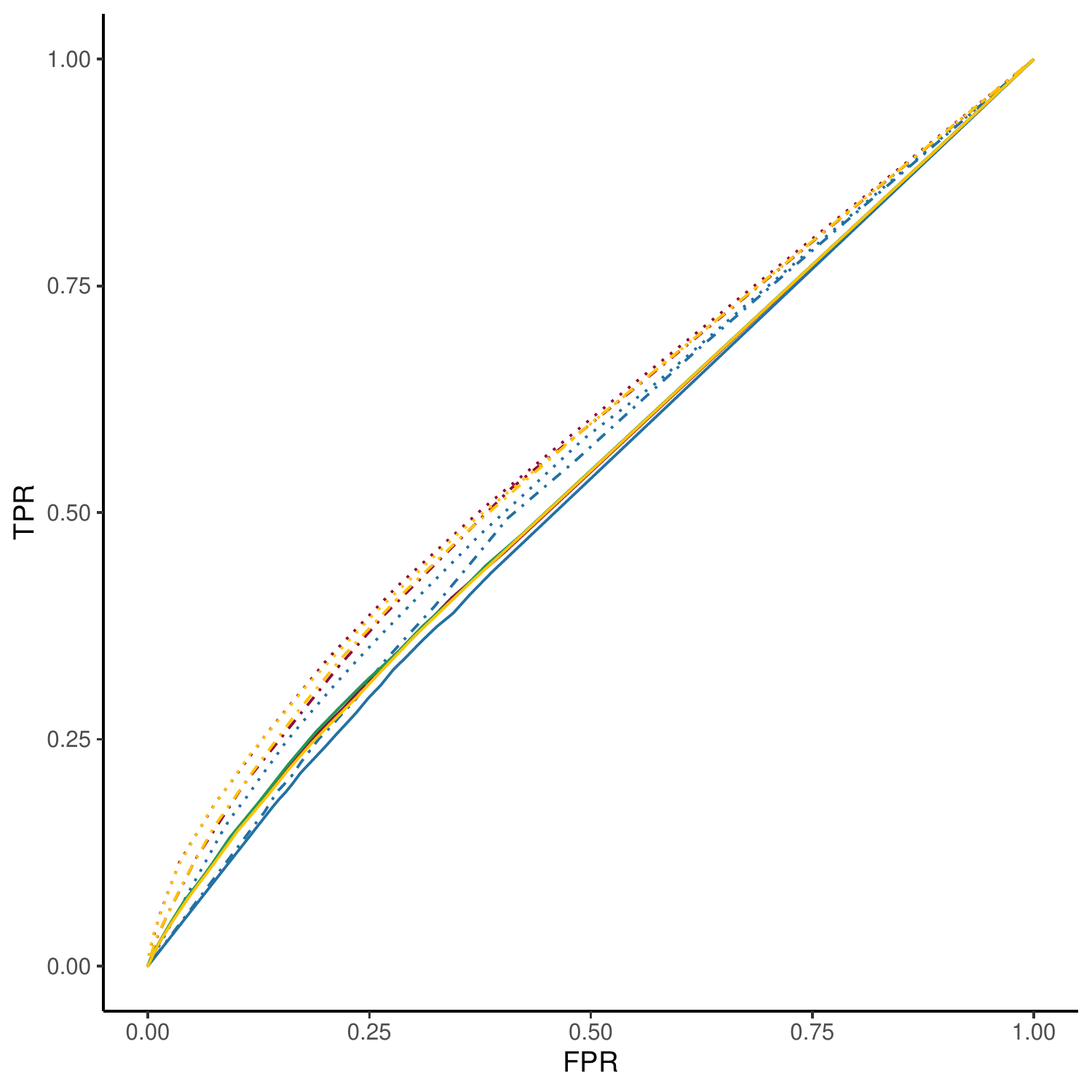}\hfill
\includegraphics[width=0.25\textwidth]{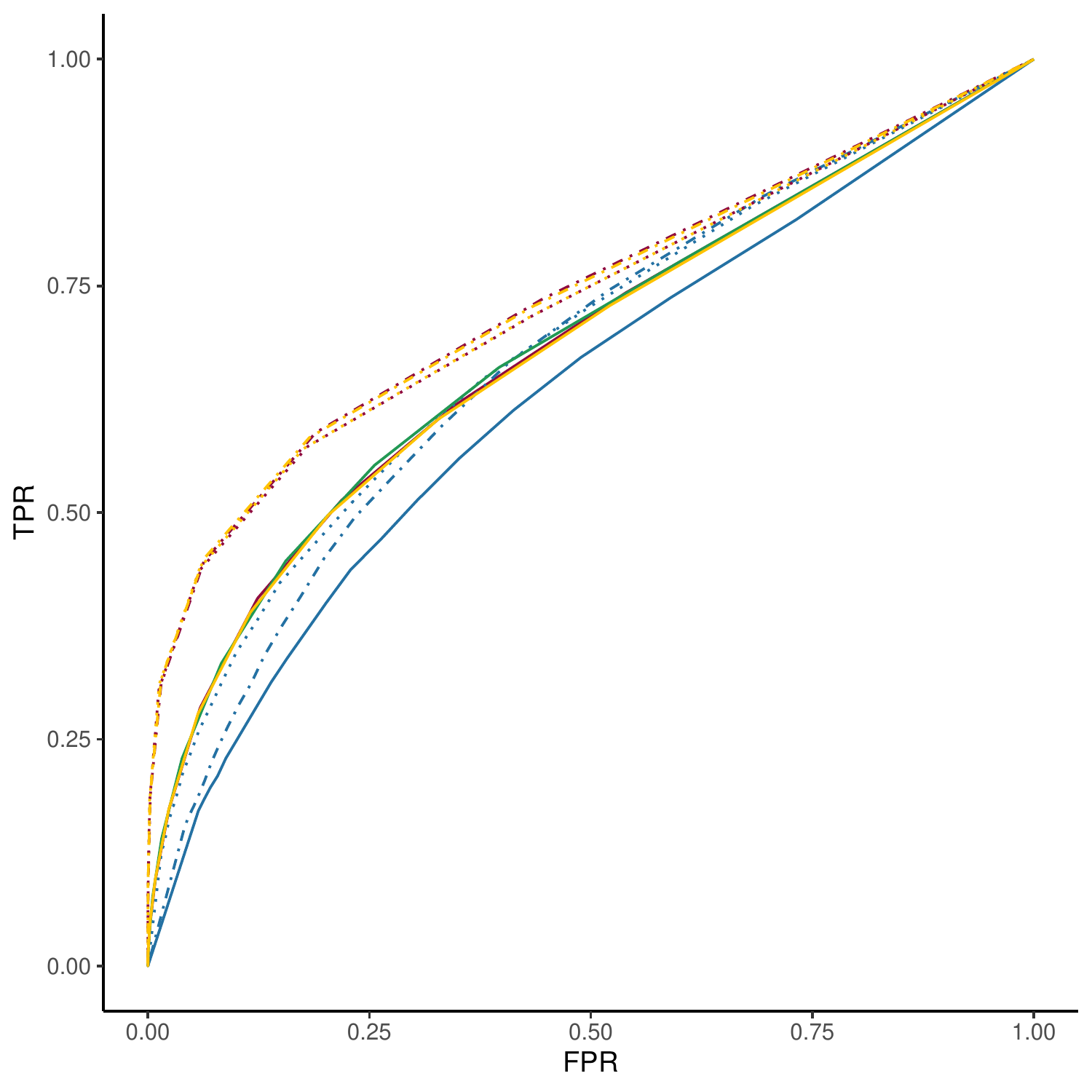}\hfill
\includegraphics[width=0.25\textwidth]{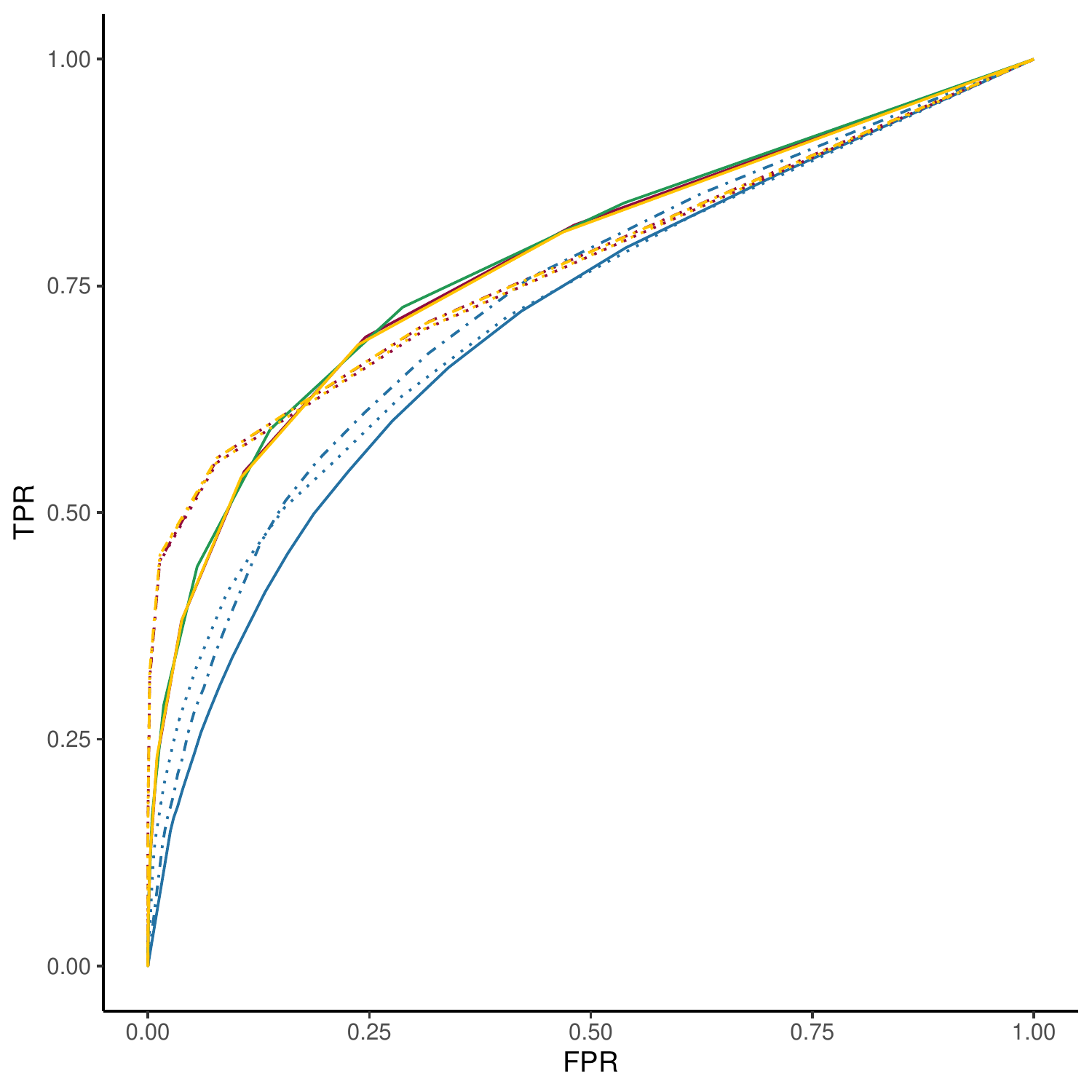}\hfill
\includegraphics[width=0.25\textwidth]{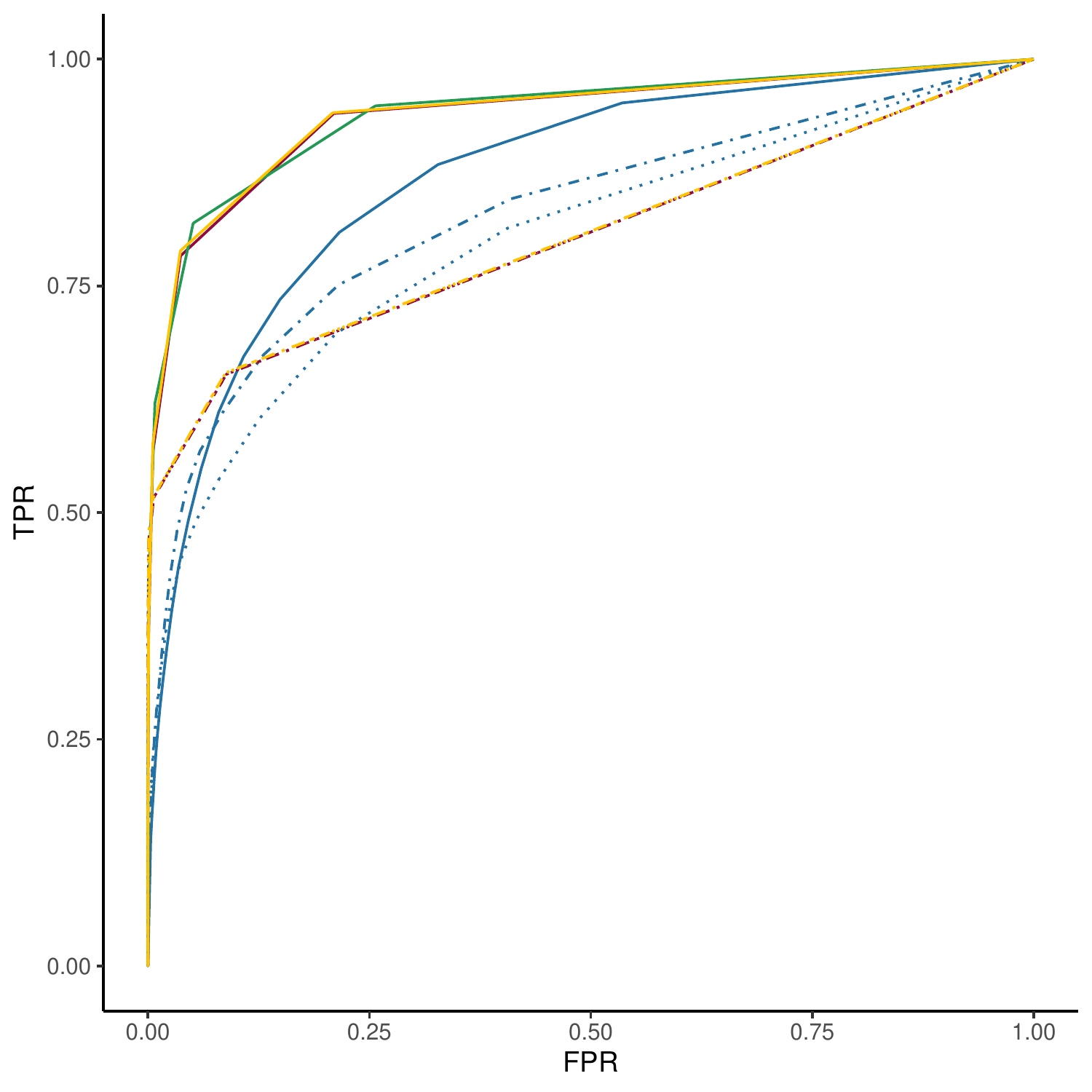}\hfill
\text{Random network, $p = 100$, $\rho = 1$}\\
\includegraphics[width=0.25\textwidth]{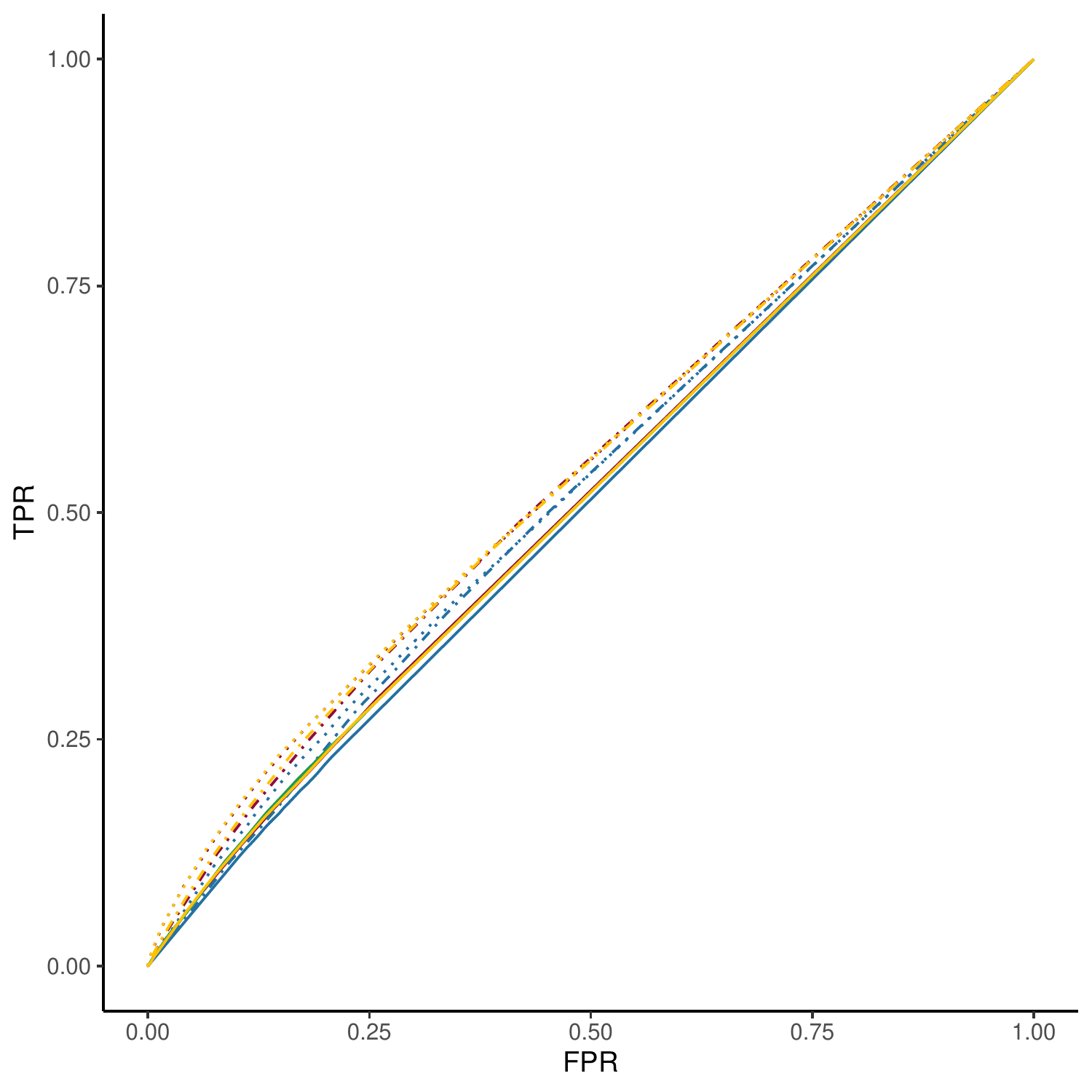}\hfill
\includegraphics[width=0.25\textwidth]{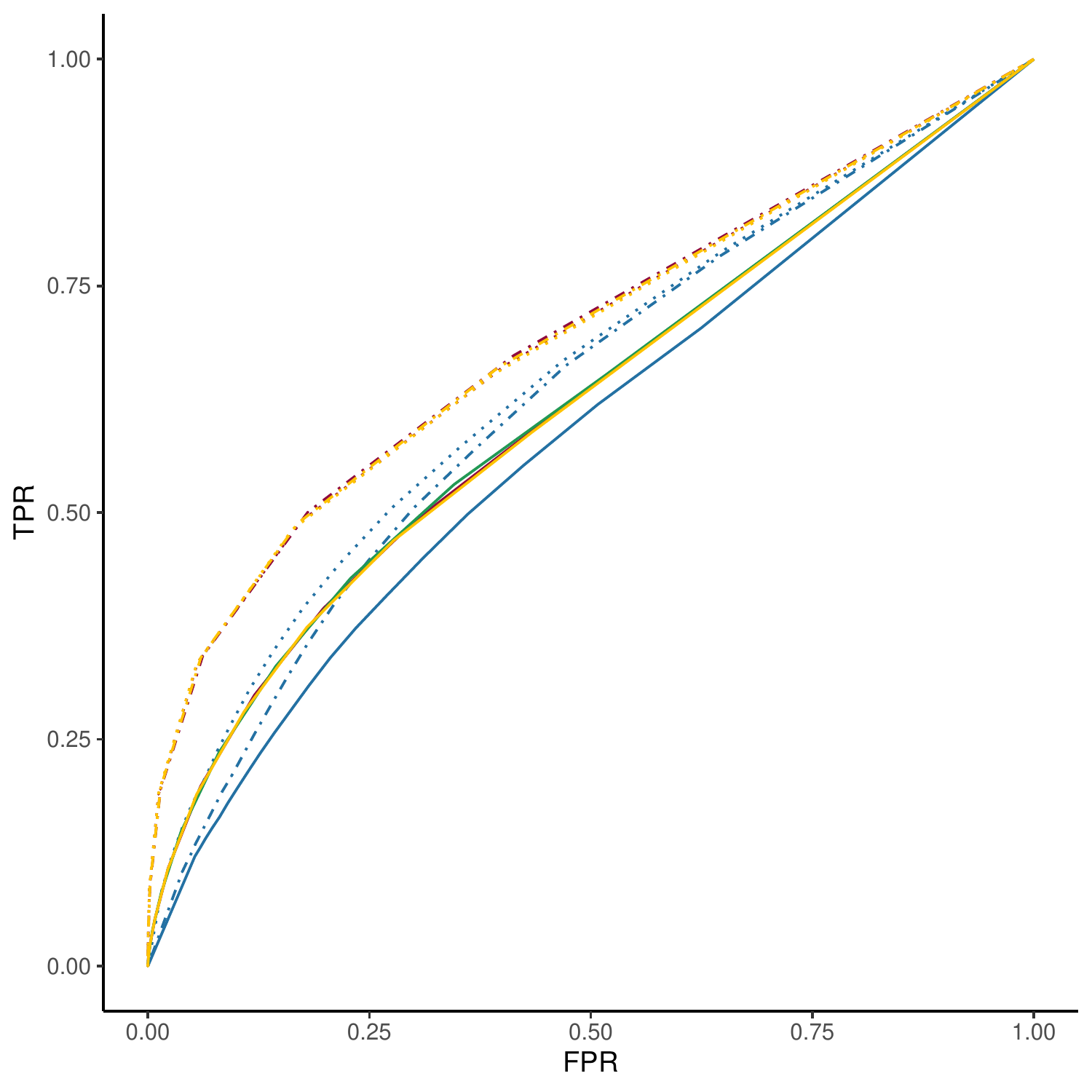}\hfill
\includegraphics[width=0.25\textwidth]{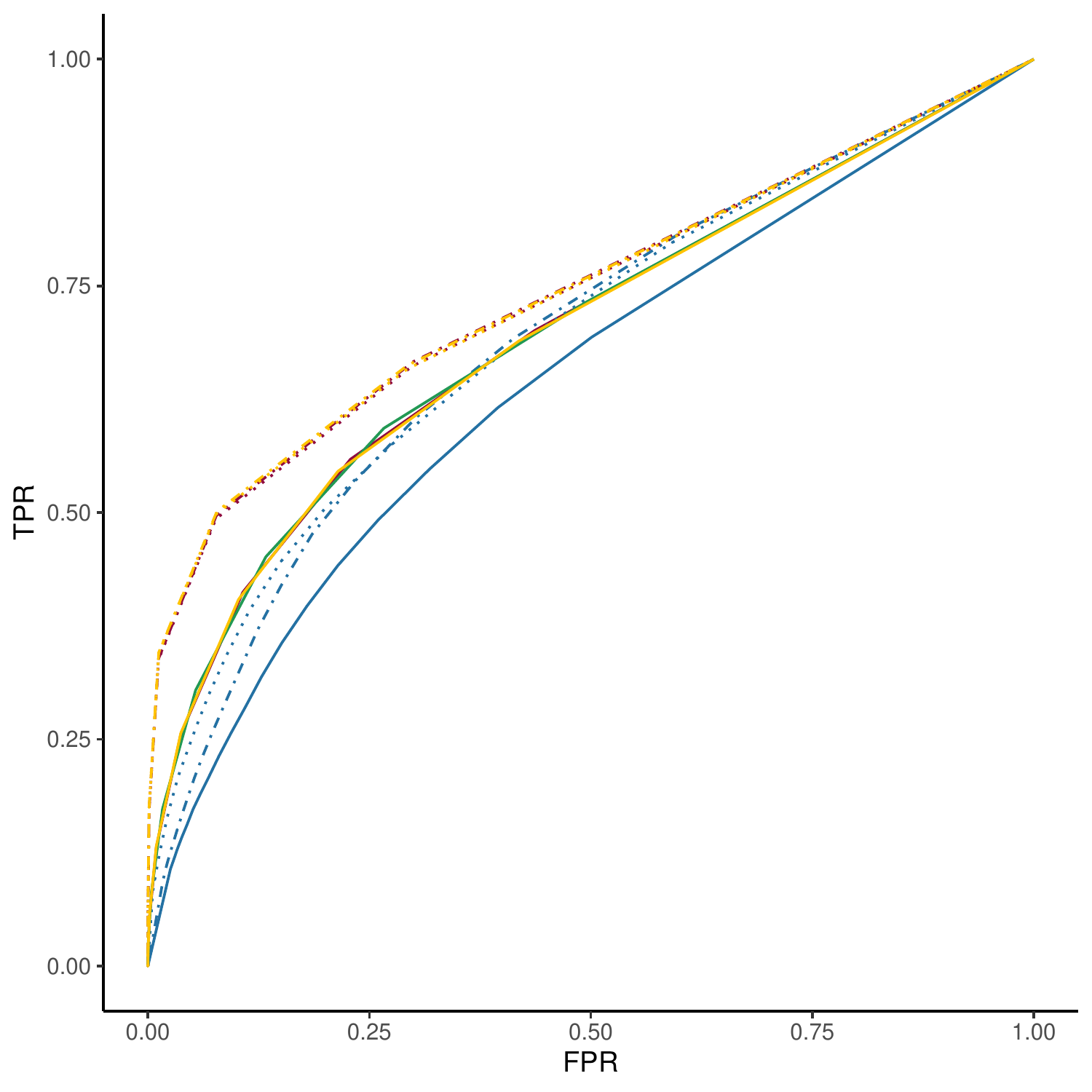}\hfill
\includegraphics[width=0.25\textwidth]{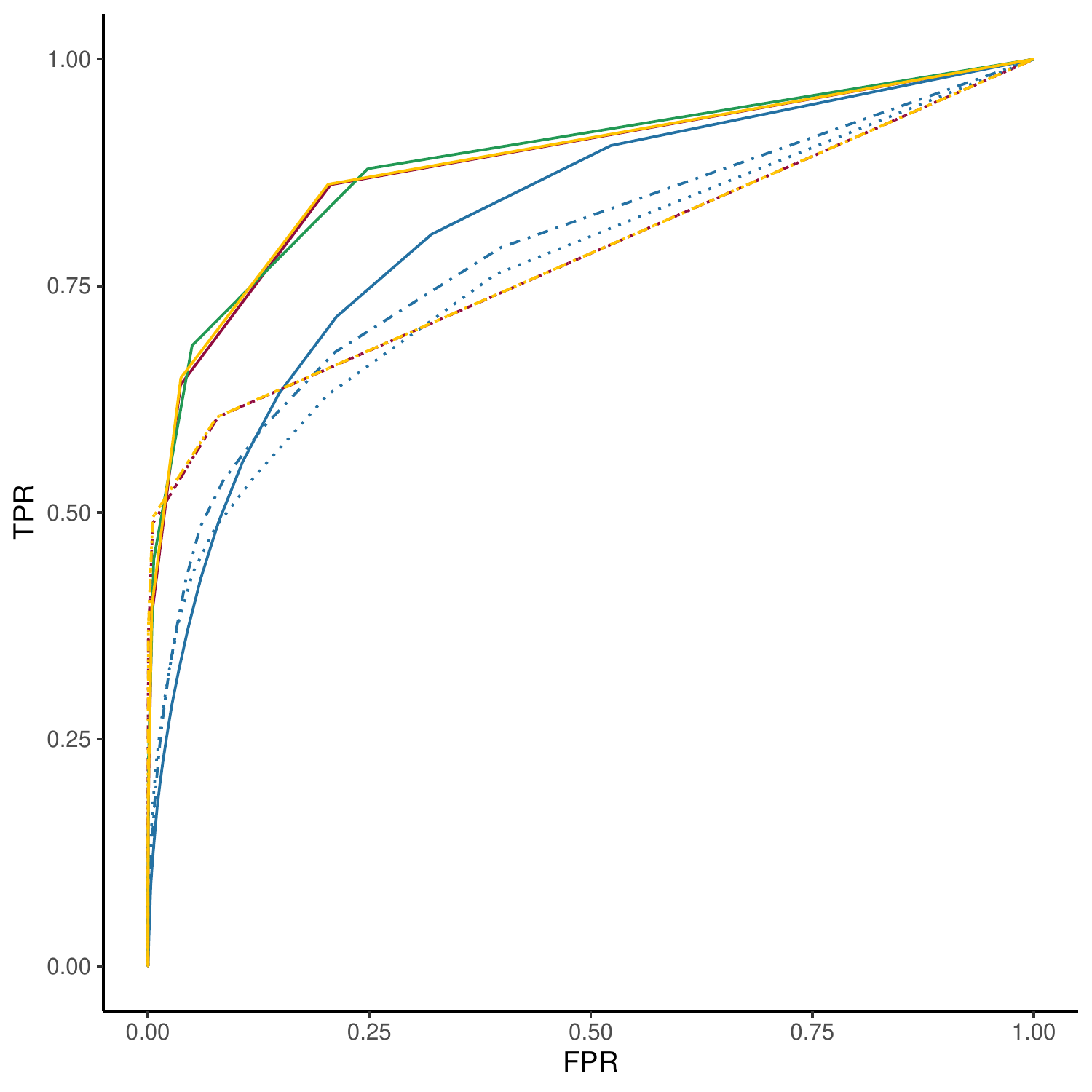}\hfill
\caption[]{ROC curves plotted over values of $\lambda_1$. The line type denotes the penalty for $\lambda_2$: $\lambda_2 = 0$ \begin{tikzpicture}\draw [thick] (0,2) -- (1,2);\end{tikzpicture}, $\lambda_2 = 0.1$  \begin{tikzpicture}\draw [dashdotted] (0,2) -- (1,2);\end{tikzpicture} and $\lambda_2 = 1$ \begin{tikzpicture}\draw [dotted] (0,2) -- (1,2);\end{tikzpicture}. The colors represent the method used: Gibbs method \begin{tikzpicture}\draw [colorGibbs, thick] (0,2) -- (1,2);\end{tikzpicture}, Approximate method \begin{tikzpicture}\draw [colorApprox, thick] (0,2) -- (1,2);\end{tikzpicture}, Fused graphical lasso \begin{tikzpicture}\draw [colorFGL, thick] (0,2) -- (1,2);\end{tikzpicture} and GLASSO \begin{tikzpicture}\draw [colorGLASSO, thick] (0,2) -- (1,2);\end{tikzpicture}. From left to right the values of $n$ are respectively 10, 50, 100 and 500.}
\label{fig:roccurves}
\end{figure}

\noindent The simulation results shown in Table \ref{tab:simrandom} and Figure \ref{fig:roccurves} indicate that the proposed copula graphical model for heterogeneous mixed data in general outperforms the alternative models. Only under very low-dimensional settings does the GLASSO approach attain a better AUC that the proposed method, which is not a setting that typically occurs in real-world data. In high dimensional settings, the performance of the proposed model is substantially better than that of the alternative models. When groups become more dissimilar, by increasing dissimilarity parameter $\rho$, the relative advantage of the proposed method becomes less substantial as compared to the GLASSO method, but this is to be expected, as there is less common information to borrow across the groups. Moreover, in low-dimensional settings, enforcing the precision matrices to be equal across groups ($\lambda_2 = 1$) results in better performance than when relatively little information is borrowed across groups. Conversely, the opposite phenomenon is observed for simulations when $n = p$ and $n < p$. When values for $\lambda_2$ are high, more information is borrowed across graphs, which is of added value when the number of samples per group is low. Conversely, when groups contain enough observations, such as in the rightmost figures, setting $\lambda_2 > 0$ unnecessarily restricts the graphs, as each group contain enough information for individual estimation. Differences between the Gibbs and approximate methods are also observed: even though both methods select approximately the same number of true and false positive edges, marked differences are present when inspecting the entropy loss results, indicating differences in edge values. The Gibbs method near-consistently outperforms the approximate method, which matches the results obtained by Behrouzi and Wit (\citeyear{behrouzi2019detecting}). Therefore, the Gibbs method is recommended over the approximate method when accuracy is the primary concern of the researcher.

\section{Application to production-ecological data} \label{Real world data example}
\begin{figure}[H]
\centering
\includegraphics[width=0.35\textwidth]{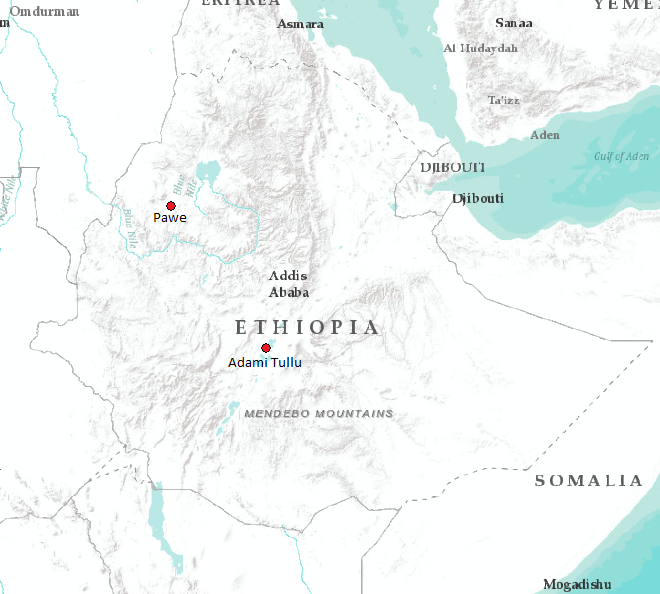}\hfill
\caption{Map showing the farm locations in Ethiopia}
\label{fig:map}
\end{figure}

The Gibbs version of the copula graphical model for heterogeneous mixed data was applied to a real world production-ecological data set.  The data contained variables pertaining to soil properties (e.g.\ clay content, silt content and total nitrogen), weather (e.g.\ average mean temperature and number of frost days), management influences (e.g.\ pesticide use and weeding frequency), external stresses (pests, diseases) and maize yield. The data were collected in Ethiopia by the Ethiopian Institute of Agricultural Research (EIAR) in collaboration with the International Maize and Wheat Improvement Center (CIMMYT) and was used as part of a larger study (\citeyear{VascoSilva_forthcoming}) aimed at explaining crop yield variability.
\\
\\
The maize data used to illustrate the proposed model consists of measurements taken on two different locations: Pawe in northwestern Ethiopia and Adami Tullu in central Ethiopia, see Figure \ref{fig:map}. These two locations have many properties that are not present in the data, but can influence the underlying relationships, such as soil water level, air pressure, wind, drainage, etc, as the two locations have different climatic properties and altitude levels (Abate et al., \citeyear{abate2015factors}). For this reason, 4 groups were created: data from farms in Pawe in 2010 (group 1) and 2013 (group 2) and data data from farms in Adami Tullu in 2010 (group 3) and 2013 (group 4). Each group consists of measurements across 63 variables with sample sizes 82, 82, 129 and 132 for Pawe 2010/2013 and Adamui Tullu 2010/2013 respectively. The data comprises 26 continuous, 22 count, 3 ordinal and 12 binomial variables. 
\\
\\
Crop yield tends to be the principal variable of interest in production ecology and will be the focus of this analysis. Both reported yield (\texttt{yield}) and simulated, water-limited, yield (\texttt{yield\_theoretical}) were present in the data. The difference between these two quantities (i.e. the yield gap) tends to be substantial in Ethiopia which makes unraveling the factors determining their relationship of interest to production ecologists (van Dijk et al., \citeyear{van2020reducing}; Assefa et al., \citeyear{assefa2020unravelling}; Kassie et al., \citeyear{kassie2014climate}; \citeyear{silva2019labour}; Getnet et al., \citeyear{getnet2016yield}). The literature (Rabbinge et al., \citeyear{rabbinge1990theoretical}; Connor et al., \citeyear{connor2011crop}) frequently identifies the following factors as potentially important for determining yield: total rainfall, planting density, soil fertility, use of intercropping, crop residue, amount of labour used, maize variety, plot size and fertiliser use. However, establishing the relative importance of these factors under different conditions is not trivial, since many of these factors may interact in complex ways depending on time and place.  

\begin{figure}[H]
\centering
\text{Network for Pawe in 2010 (left) and 2013 (right)}\\
\includegraphics[width=0.5\textwidth]{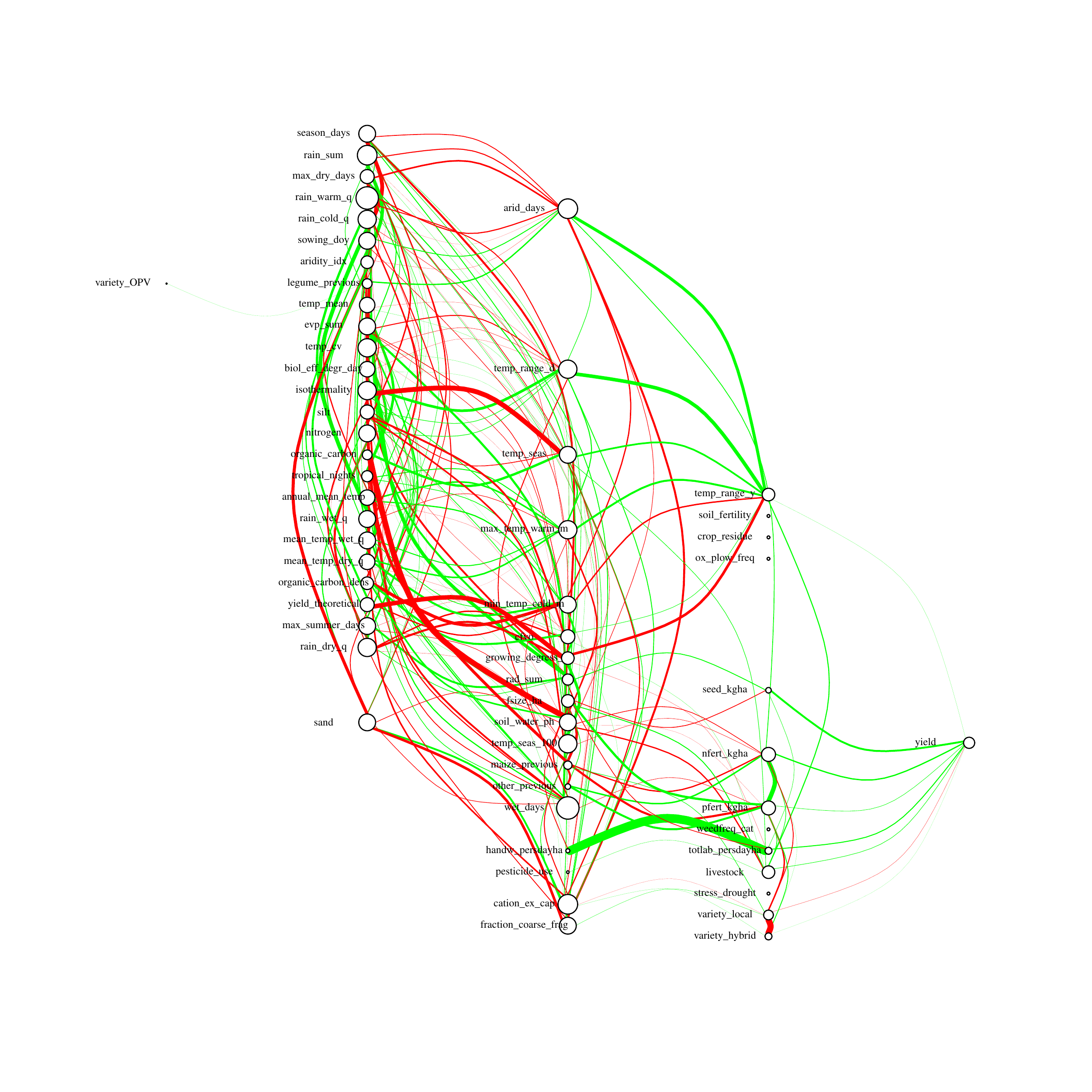}\hfill
\includegraphics[width=0.5\textwidth]{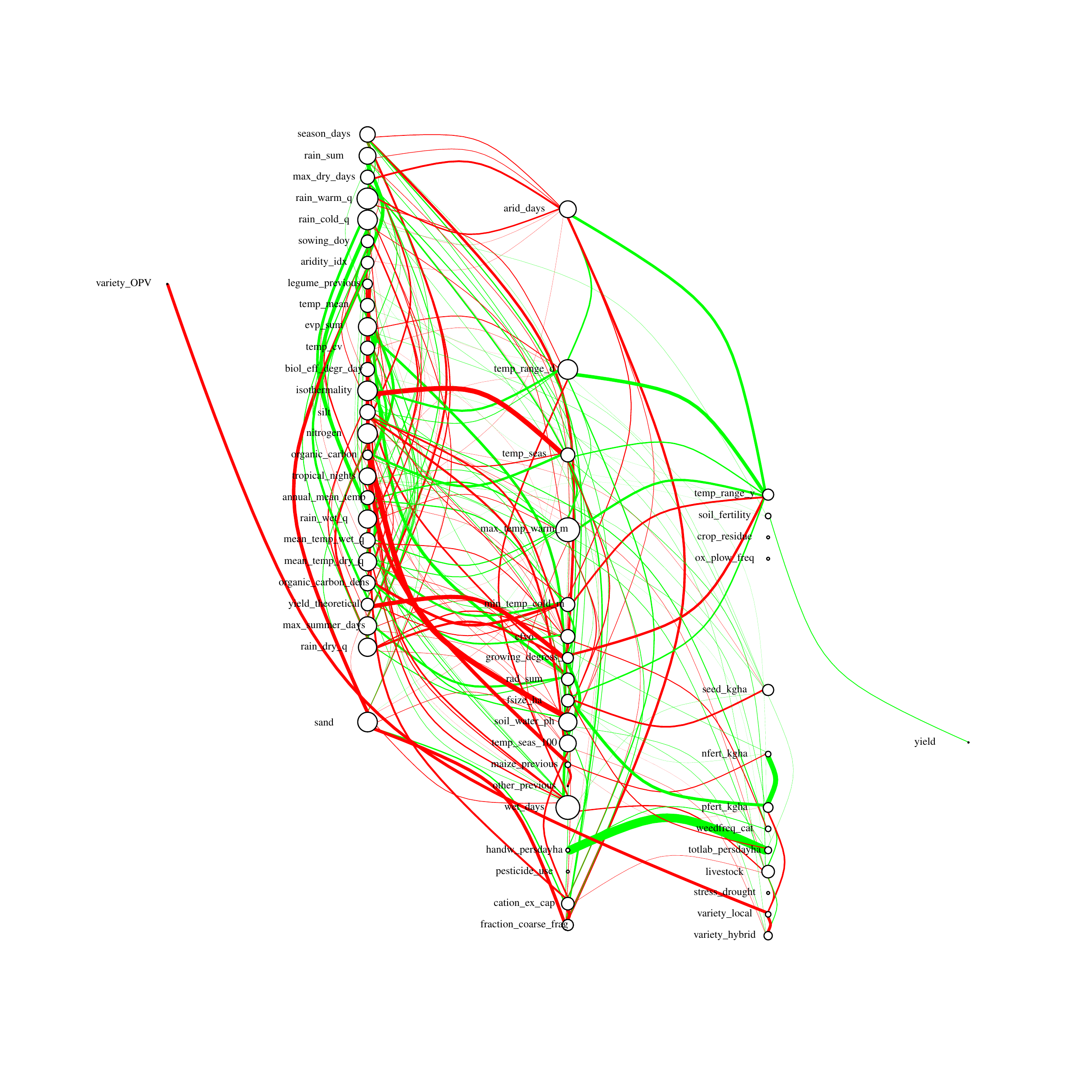}\hfill
\text{Network for Adami Tullu in 2010 (left) and 2013 (right)}\\
\includegraphics[width=0.5\textwidth]{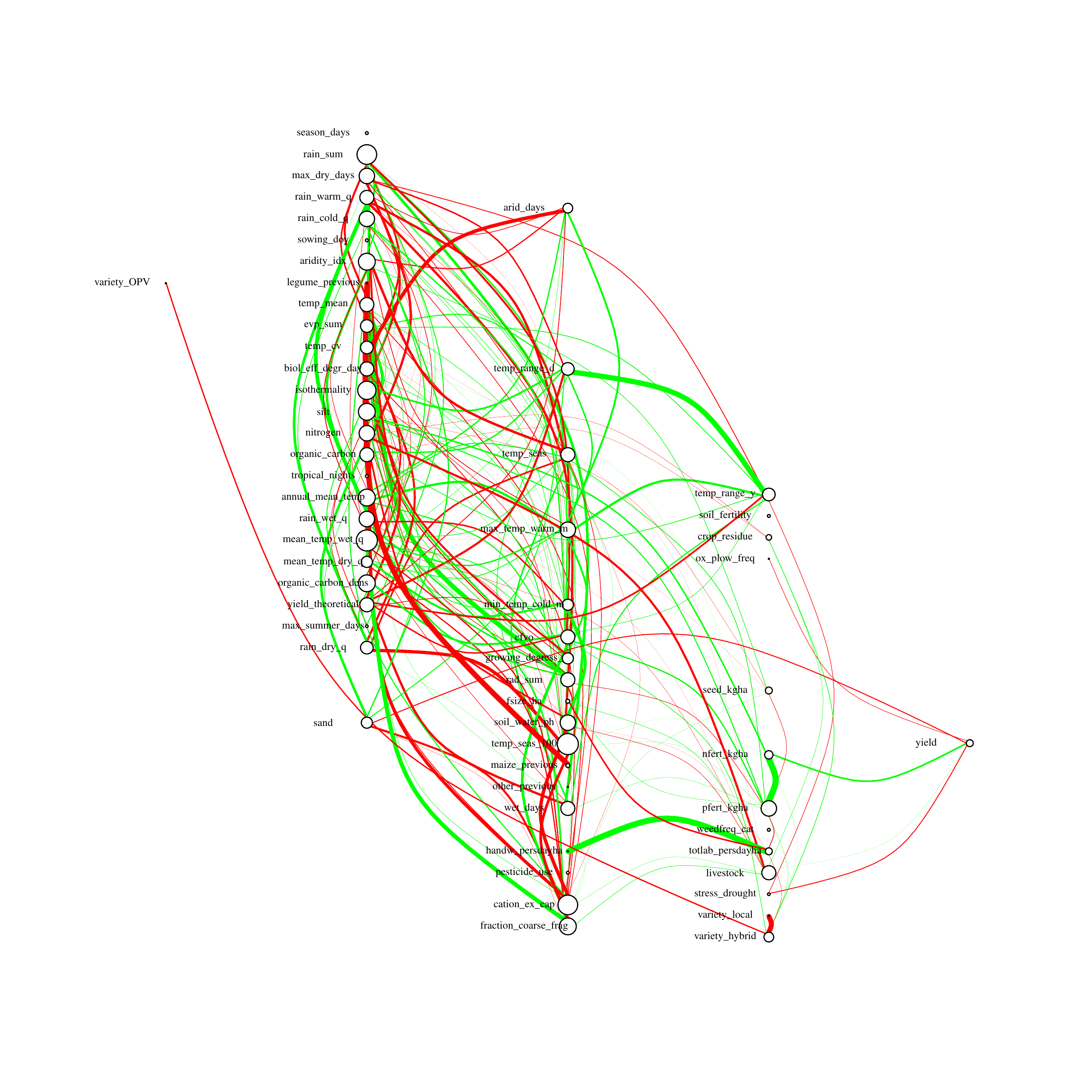}\hfill
\includegraphics[width=0.5\textwidth]{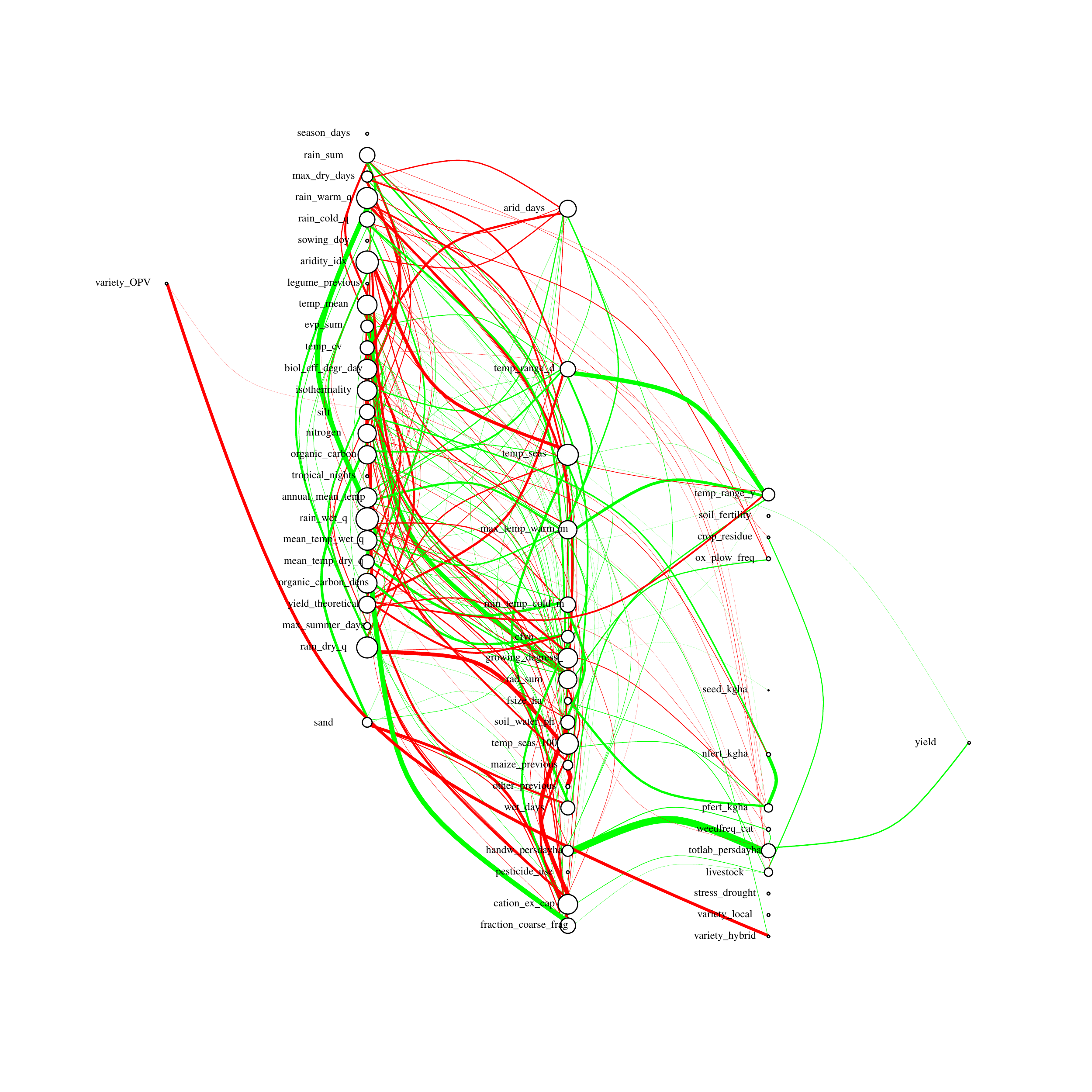}\hfill
\caption{Results for the 4 networks, with penalty parameters set to $\lambda_1 = 0.2$, $\lambda_2 = 0$ as chosen by the EBIC with $\gamma = 0.5$. Node size is indicative of the node degree, edge width reflects the absolute value of the partial correlation and color is used to make visual distinction between positive (green) and negative (red) partial correlations (green).}
\label{fig:realdata}
\end{figure}

\begin{figure}[H]
\centering
\text{Yield network for Pawe in 2010 (left) and 2013 (right)}\\
\includegraphics[width=0.5\textwidth]{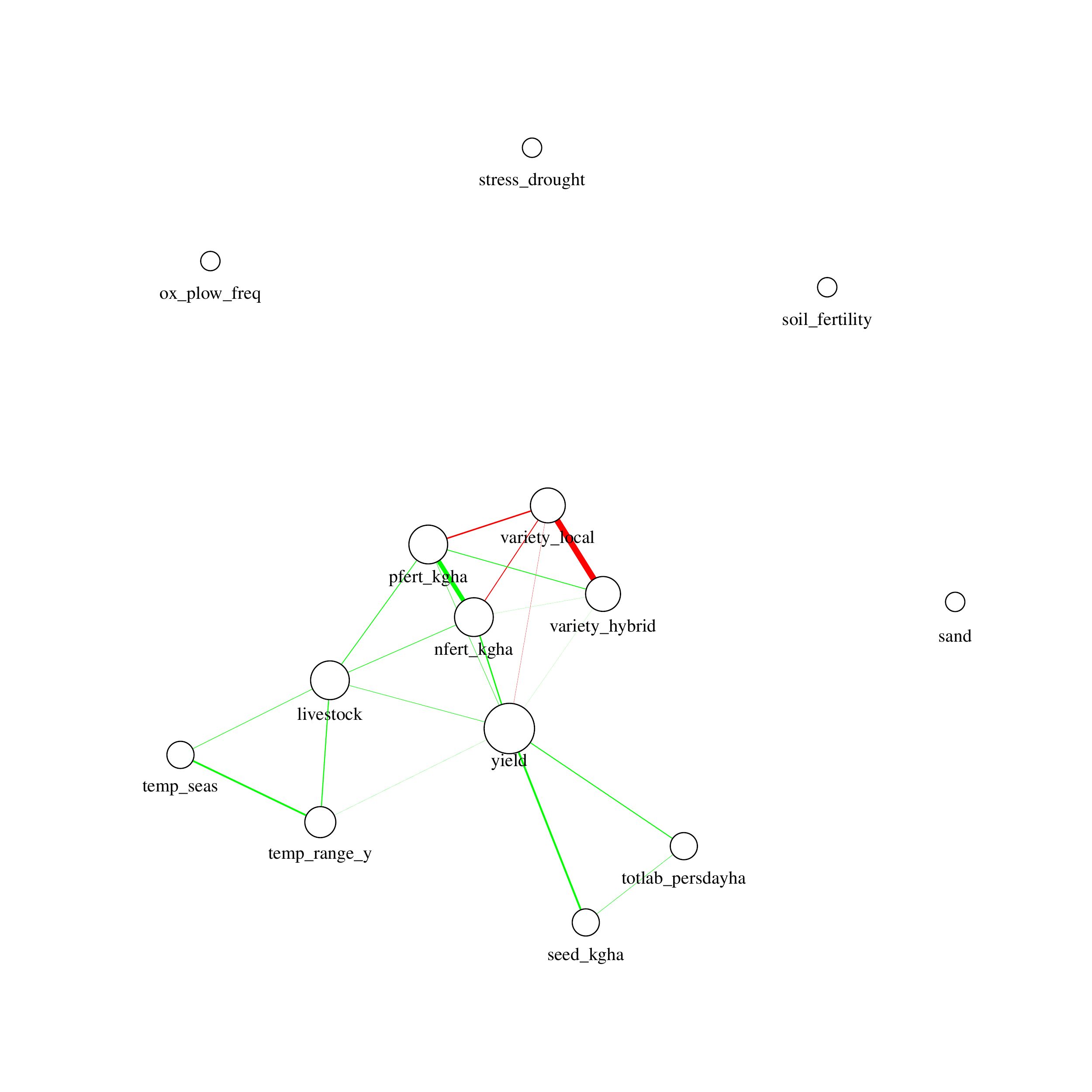}\hfill
\includegraphics[width=0.5\textwidth]{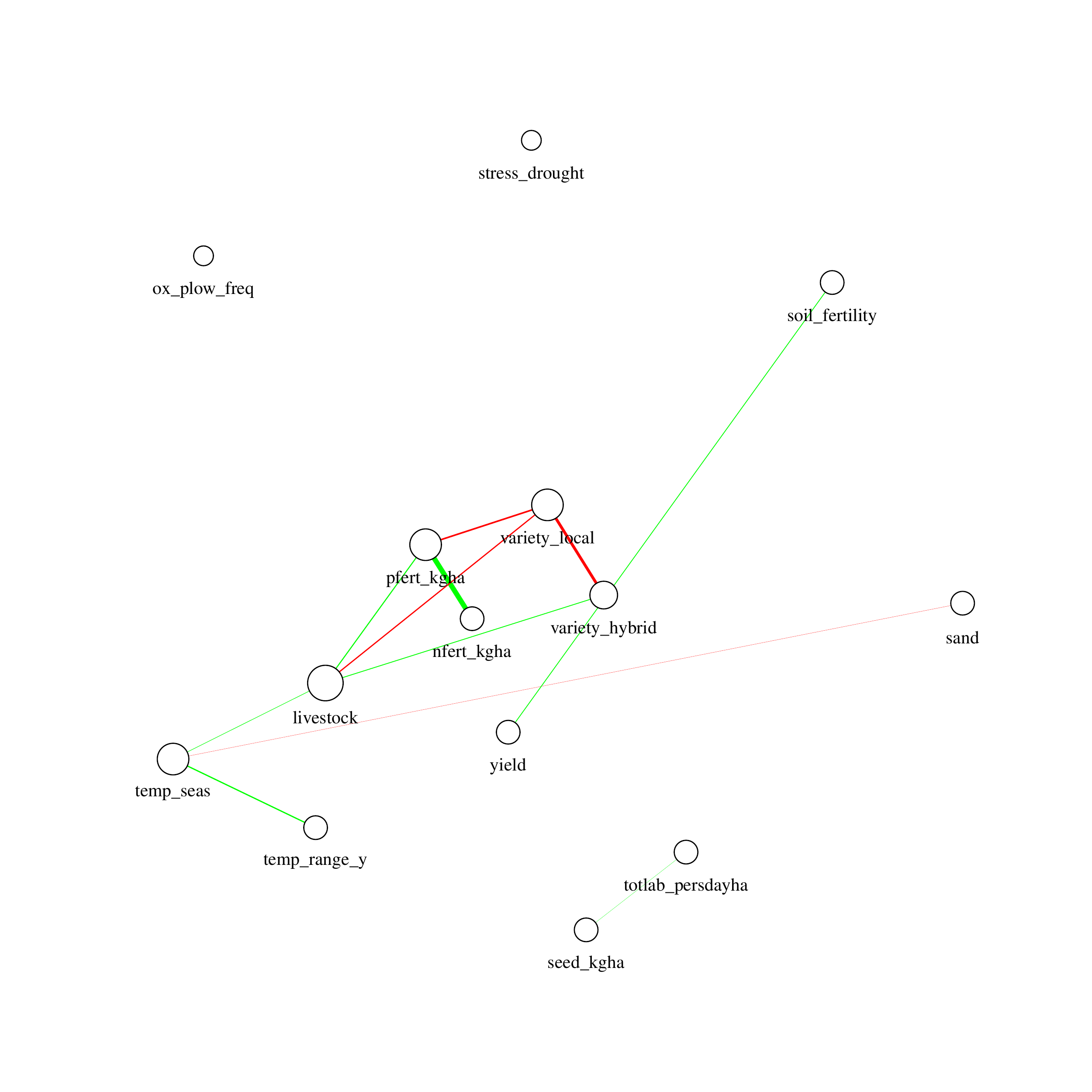}\hfill
\text{Yield network for Adami Tullu in 2010 (left) and 2013 (right)}\\
\includegraphics[width=0.5\textwidth]{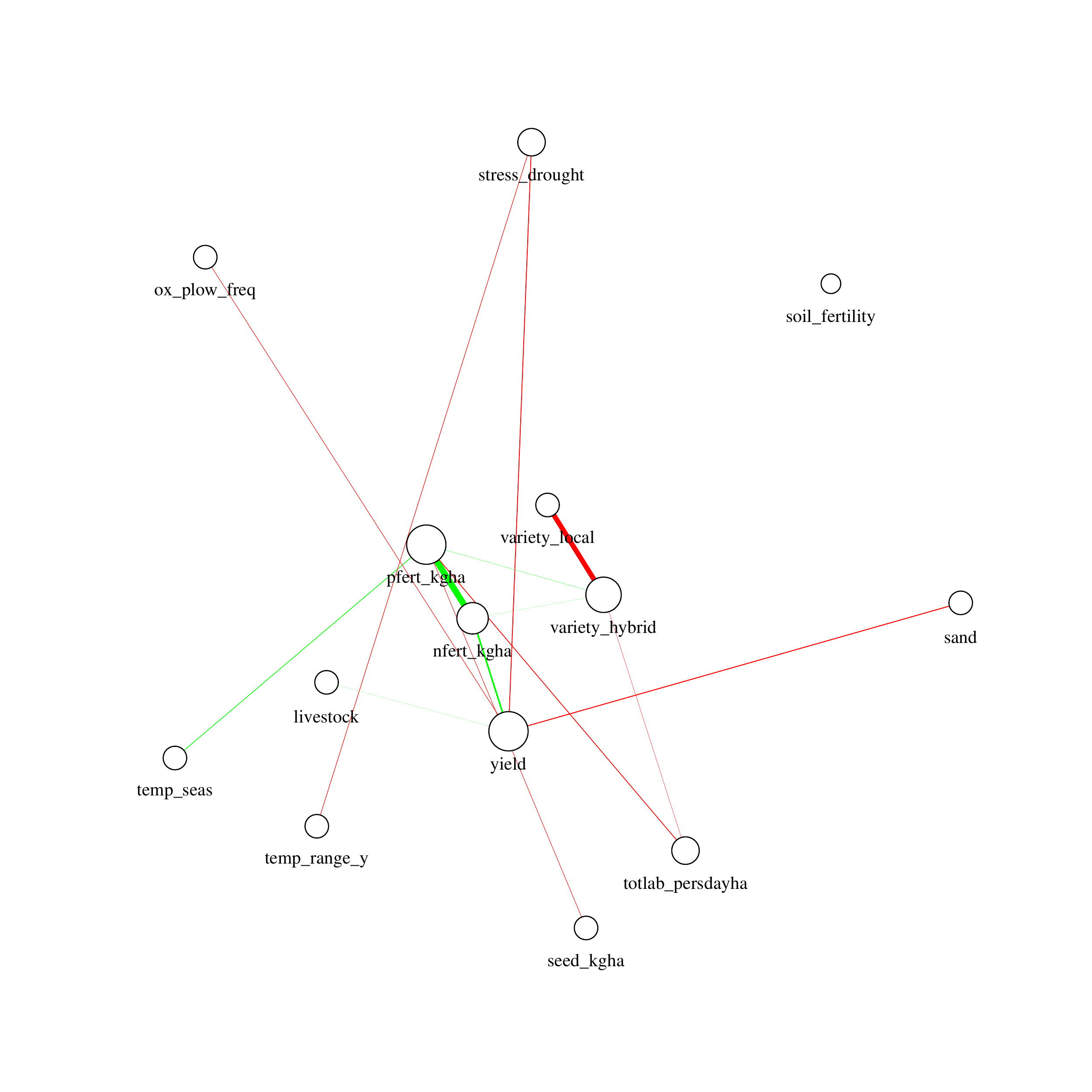}\hfill
\includegraphics[width=0.5\textwidth]{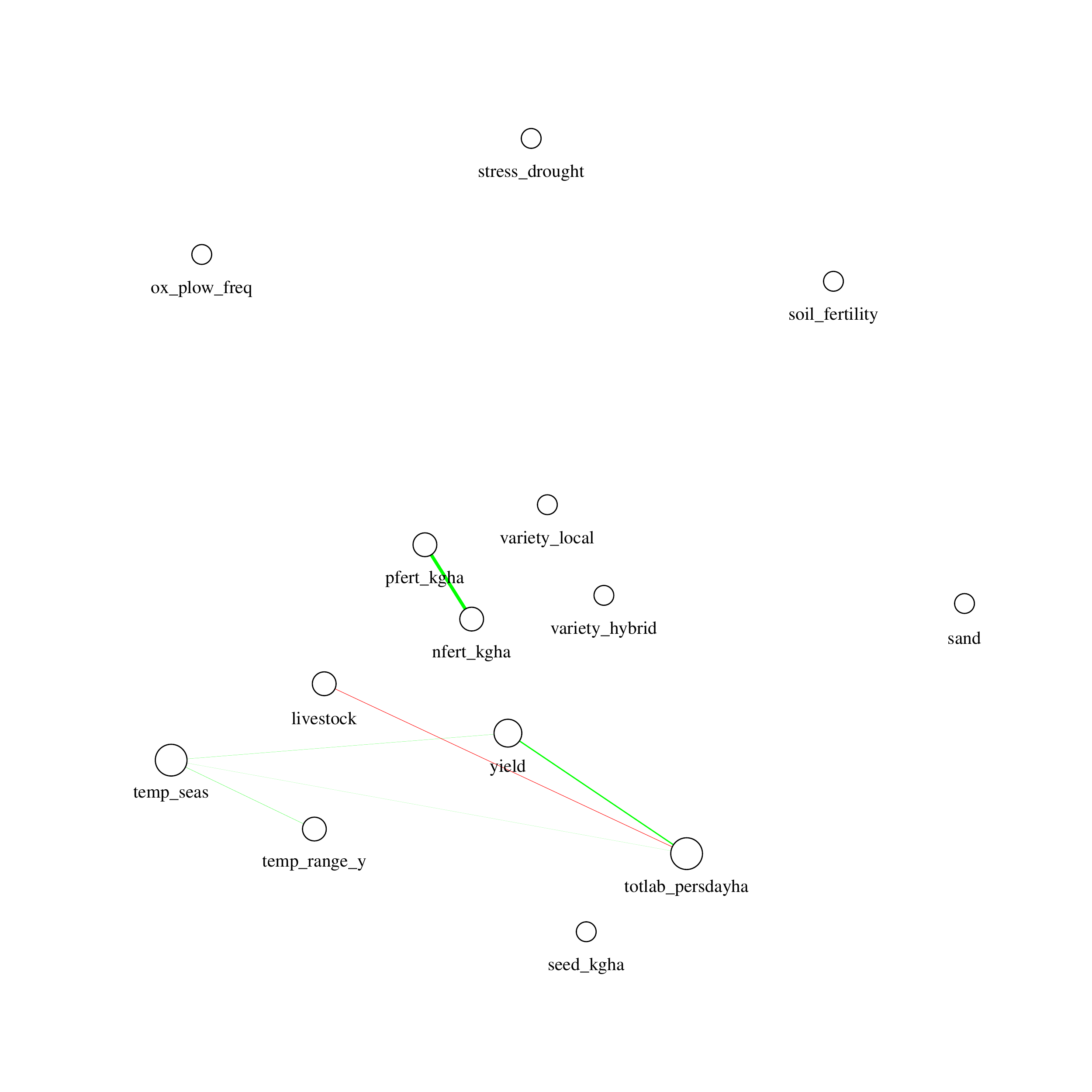}\hfill
\caption{Results for the 4 yield networks obtained by applying the Local Markov Property to the yield variable. All variables that share an edge with yield in at least one of the four networks are included in all four networks. Node size is indicative of the node degree, edge width reflects the absolute value of the partial correlation and color is used to make visual distinction between positive (green) and negative (red) partial correlations.}
\label{fig:yield}
\end{figure}

To gain more insight into the relations underlying yield variation, we apply the proposed copula graphical model for heterogeneous mixed data. The model is fitted using a grid of 11$\times$11 combinations for $\lambda_1$ and $\lambda_2$, from which the combination is selected that minimises the EBIC (\ref{eq:EBIC}) with $\gamma = 0.5$.  Some general graph properties of the full graphs as seen in Figure \ref{fig:realdata} are discussed, followed by a discussion of the yield graphs seen in Figure \ref{fig:yield}.

\noindent Topologically, the graphs consist of a dense cluster of variables mainly consisting of soil and weather properties with some management variables mixed in, a less dense cluster of yield with its neighbours and some conditionally independent management and external-stress variables. The graphs consist respectively (by increasing group order) of 271, 261, 220 and 253 edges, with average node degrees of 8.60, 8.29, 6.98 and 8.03. However, in order to gain insights into which variables are conditionally dependent on yield, the full graphs are impractical. Therefore, the subgraphs containing the yield variable and its neighbours are given in Figure \ref{fig:yield} below. By the Local Markov Property (cf.\ Lauritzen, \citeyear{lauritzen1996graphical}), yield is conditionally independent of all variables not shown in the graph given its neighbours, resulting in Figure \ref{fig:yield} being sufficient to infer all conditional dependencies of the yield variable. 

\noindent Central in these graphs is the actual, reported yield. Whereas this variable has many direct relationships in the graph of Pawe and Adami Tullu in 2010, this is not the case for the other groups, including Pawe in 2013, possibly indicating a temporal interaction. Most results presented in Figure \ref{fig:yield} are not unexpected, such as the positive effects of the use of (\texttt{variety\_hybrid}) seed, and consequently the negative effect of planting a local variety (\texttt{variety\_local}) (Assefa et al., \citeyear{assefa2020unravelling}), the benefit from the application of N fertiliser and labour input.  More surprising  perhaps was the negative relationship with the amount of seed  found in Adami Tullu in 2010, but this may be a true indication that optimum densities were relatively low for that location and year. The direct relation found in  Pawe and Adami Tullu in 2010 between \texttt{livestock} (ownership) and yield is also interesting and may reflect either beneficial effects of their use as draft animals or positive effects on soil fertility, either directly through manure, or through greater resource endowment in general, for which livestock ownership is an indicator (Silva et al., \citeyear{silva2019labour}). In this regard, the negative effect of livestock on the labour input per person, which in turn has a positive effect on yield could also reflect the beneficial effect of the use of animal power over manual labour. With respect to labour per se, both graphs for Pawe in 2010 and Adami Tullu in 2013 contain an edge between yield and total labour use (\texttt{totlab\_persdayha}), indicating conditional dependence. Whilst the presence of the edges, and the corresponding positive partial correlations are not surprising per se, the fact that this relation is only found in Pawe in 2010 and Adami Tullu in 2013 is unusual. As labour is highly seasonal (Silva et al., \citeyear{silva2019labour}), and labour seasonality can vary with location, this edge is expected either for Pawe in 2010 and 2013 or for Adami Tullu in 2010 and 2013. This is an example of a conditional dependence that can be explored further by production ecologists.
\\
\\
The present analysis is an example of how graphical models can aid in the exploration and understanding of fundamental production-ecological relations. Once other researchers take note of this novel application, methodologies can be tailored which will further understanding of how yield is influenced.  

\subsection{Goodness of fit}
\noindent In order to evaluate the stability of the selected network, we apply a bootstrapping procedure to the data, where 200 (row) permutations of the original data are created, the proposed model is fitted over all values for $\lambda_1$ and $\lambda_2$, and an optimal model is selected using the the EBIC ($\gamma = 0.5$). For each graph $G_k, k\in 1\ldots,4$, edges that occur over $90\%$ across the 200 bootstrapped $G_k$ graphs (hereinafter referred to as the acceptance ratio) are compared to the pre-bootstrap fitted graphs, indicating how stable the model performance is across random permutations of the data. The choice of an acceptance ratio of $90\%$ was based on the fact that we are primarily interested in whether the results obtained by the proposed model are stable across small perturbations of the data, and using a high threshold only takes into account edges that can be considered part of the underlying graph.

\begin{figure}[H]
\centering
\text{Results for Pawe in 2010 (left) and 2013 (right)}\\
\includegraphics[width=0.5\textwidth]{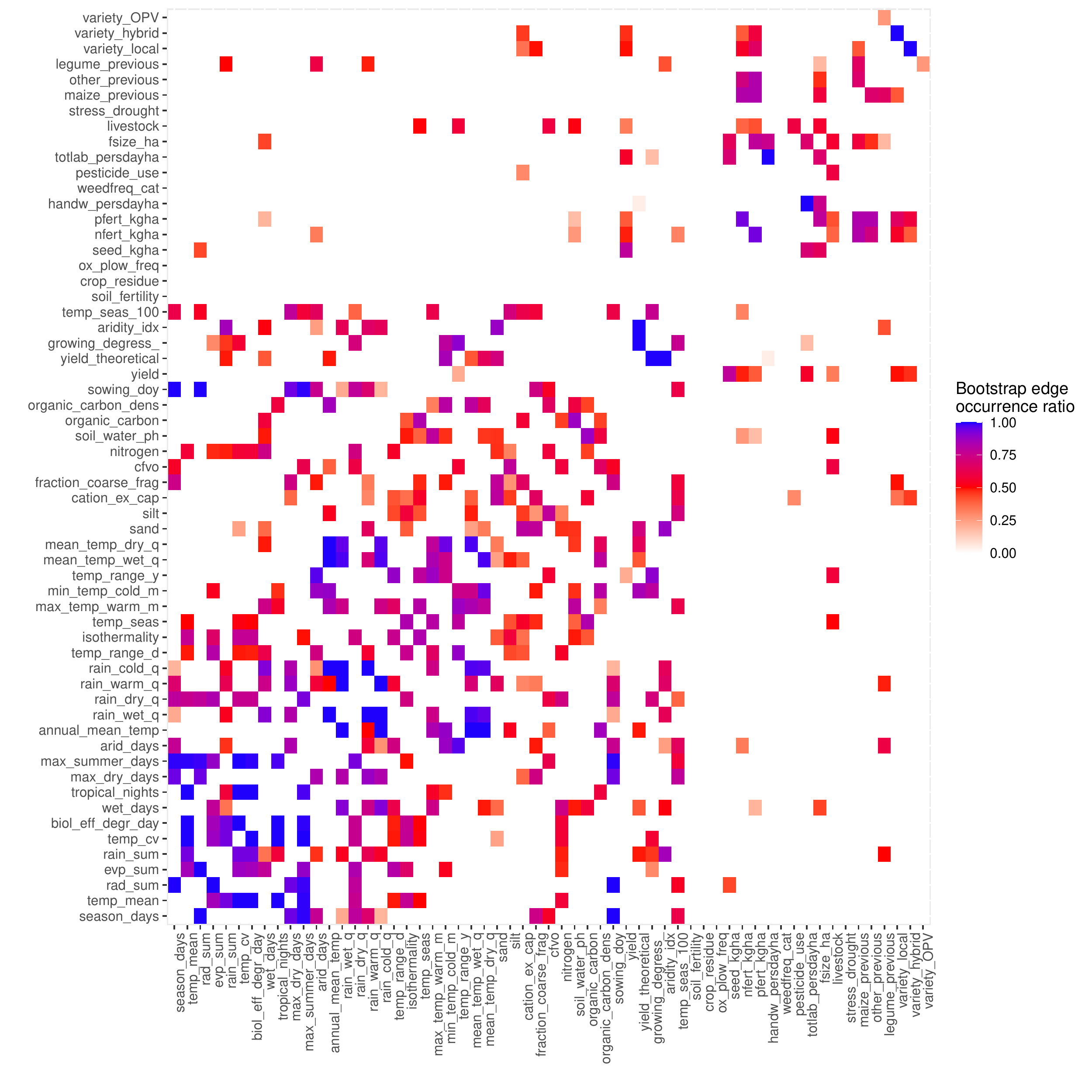}\hfill
\includegraphics[width=0.5\textwidth]{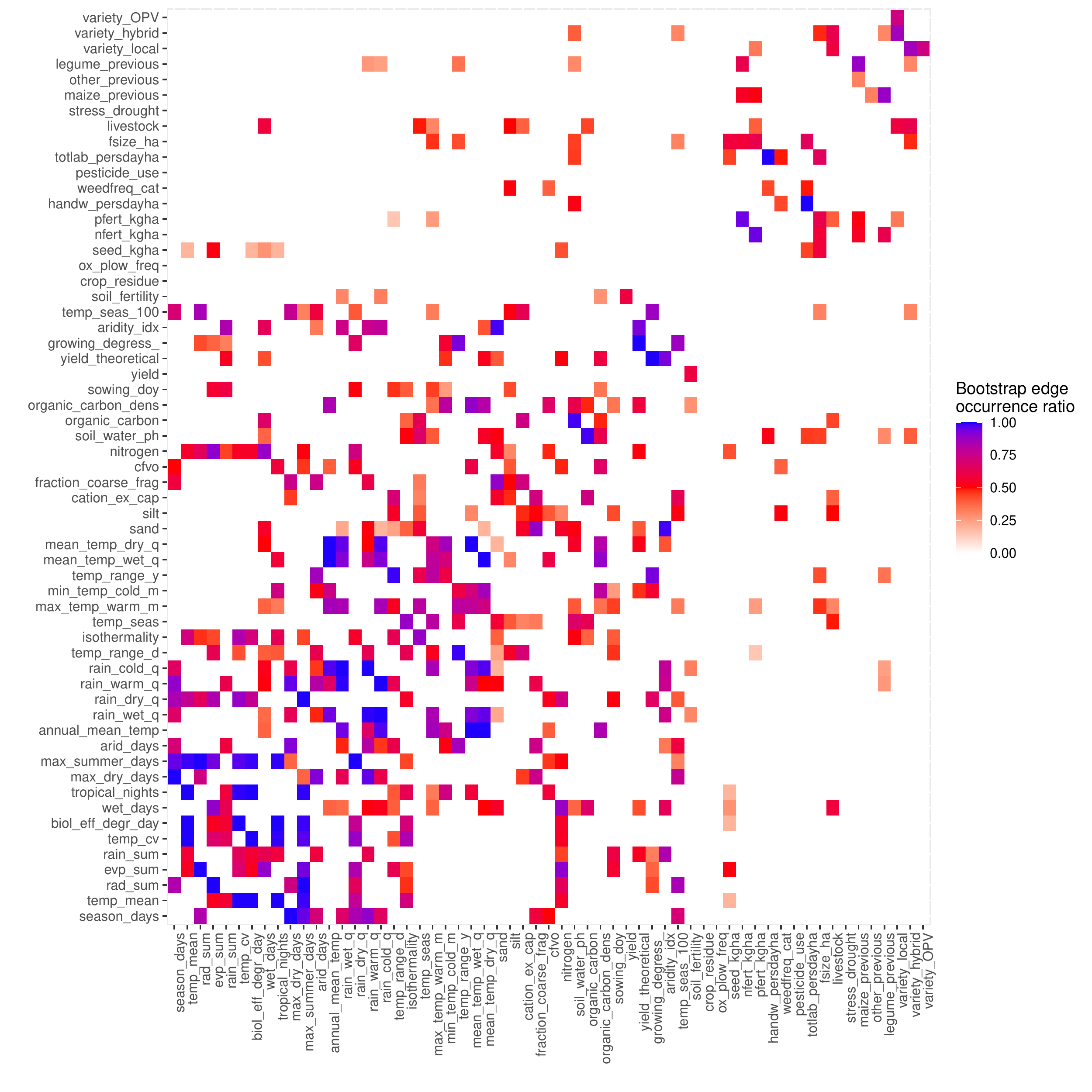}\hfill
\text{Results for Adami Tullu in 2010 (left) and 2013 (right)}\\
\includegraphics[width=0.5\textwidth]{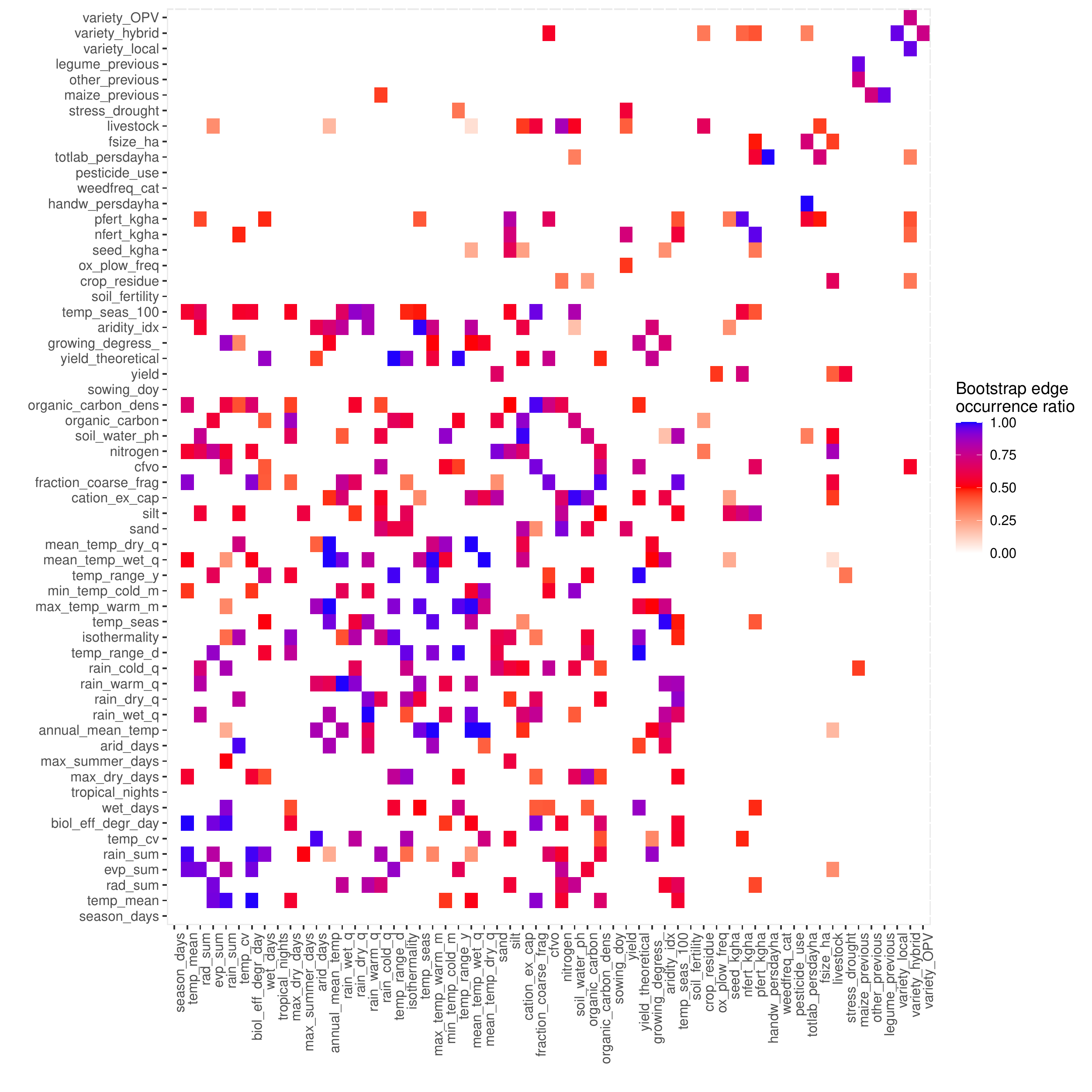}\hfill
\includegraphics[width=0.5\textwidth]{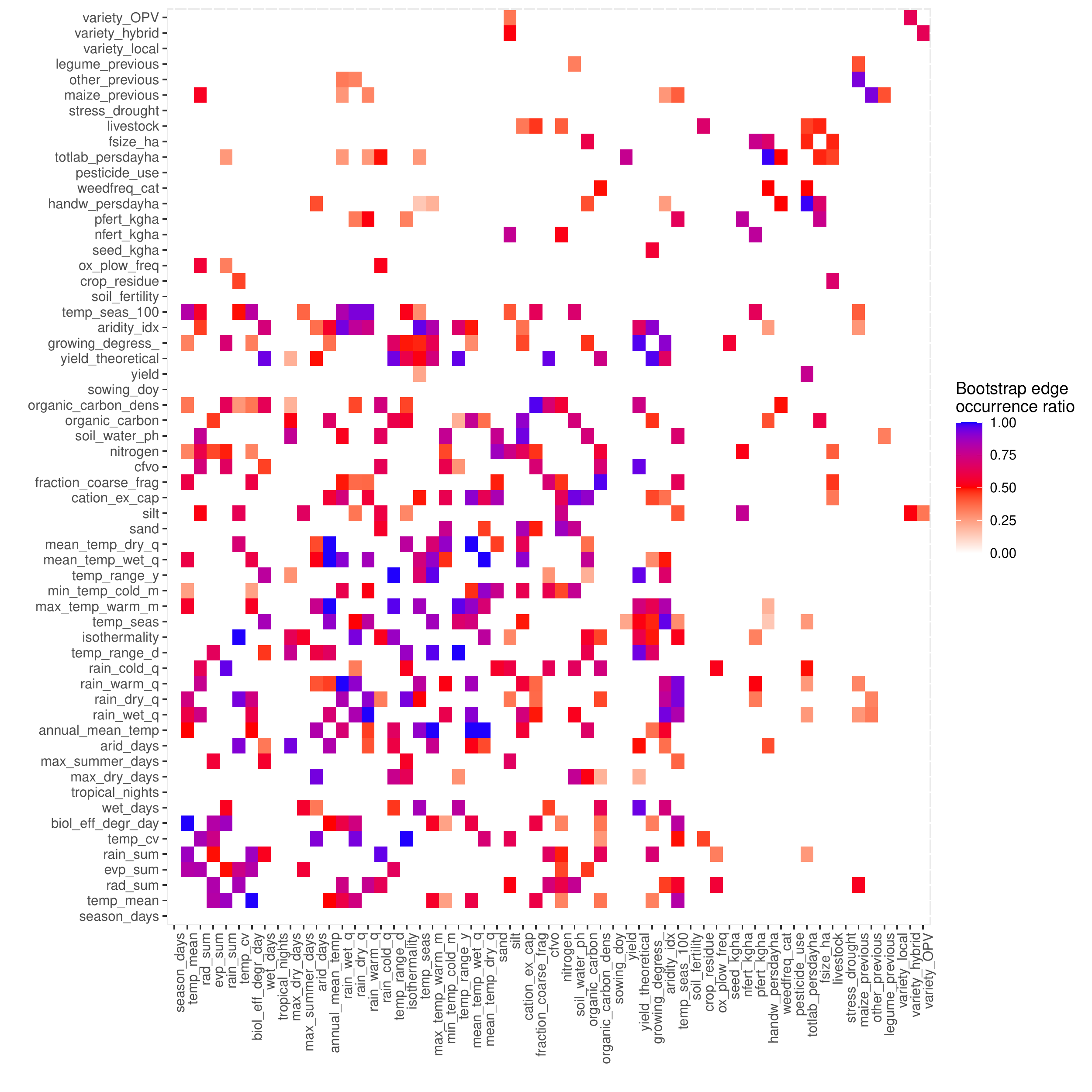}\hfill
\caption{Edge stability results for the 4 networks, as determined by the number of times the edges from the fitted graphs occurred in the 200 bootstrapped graphs.}
\label{fig:bootstrapedge}
\end{figure}

\noindent Using an acceptance ratio of $90\%$ for the edges obtained from bootstrapped data, the fitted model described above with $\lambda_1 = 0.2$ and $\lambda_2 = 0$ is able to infer all fundamental relations (the discovery rate). This perfect discovery rate remains up until the acceptance ratio is decreased to $70\%$, where across all four graphs, the discovery rate decreases to $98.71\%$. Even when we identify those edges that appear in $50\%$ of the graphs obtained from the bootstrap as fundamental, the discovery rate remains $89.53\%$. When the acceptance ratio is set even lower, non-fundamental relations are more likely to be included, which is not of interest here. In addition, judging by Figure \ref{fig:bootstrapedge}, most edges in the fitted model occur frequently in the bootstrapped graphs, as judged by the low amount of very light red edges. Of particular interest are the edges surrounding the yield variable, which are stable. Therefore, using the EBIC results in stable model selection where fundamental relations are satisfactorily recovered. 

\section{Conclusion} \label{Conclusion}
Responding to the need of production ecologists for a statistical technique that can shed light into fundamental relations underlying plant productivity in agricultural systems, this article introduces the copula graphical model for mixed heterogeneous data: a novel statistical method that can be used to obtain estimate simultaneous interactions amongst multi-group mixed-type datasets. The proposed model can be seen as the fusion of two different models: the copula graphical model and the fused graphical lasso. The former extends the Gaussian graphical model with non-Gaussian variables, whereas the latter extends the graphical model to a multi-group setting and enforces similarity between similar groups. The model performs competitively for a myriad of graph structures underlying very different datasets, thereby extending its use beyond production-ecological data, to any mixed-type heterogeneous data. Moreover, the proposed method was applied to a production-ecological dataset consisting of 4 groups, reflecting spatial and temporal heterogeneity, as is typical of production-ecological data. Aside from yield relationships that are typically identified in production-ecological research, the model also found some peculiar relationships, giving motivation for future research. In terms of future statistical research, one recommendation that we give is model selection for multi-group graphical models, in order to facilitate applied researchers. Current approaches do not give theoretical guarantees for these types of models and model selection remains an important part of any statistical application. Another possible research direction is to extend the proposed method by allowing for unordered categorical (nominal) data, which is one of the shortcomings of the copula. Finally, we urge statisticians to develop methodologies that make optimal use of the intricacies that production-ecological data offer and is in line with the goals of production ecologists. 

\clearpage
\bibliographystyle{Chicago}
\bibliography{library}

\begin{appendices}
\section{Approximate method details} \label{Approximate method details}
As computing the first and second moments of a truncated normal distribution is computationally expensive even for moderate $p$, Guo et al.\ (\citeyear{guo2015graphical}) proposed an approximation method which ensures feasibility of the copula graphical model even for high $p$. The mathematical derivations and theory behind it is outlined in this section. 
\\
\\
In order to approximate the first and second moments of the latent variables arising from a truncated normal distribution, we apply mean field theory:

  \begin{gather}
    \mathbb{E}(Z_{ij}^{(k)}|x_{i}^{(k)}, \bm{\hat{\Theta}}^{(m)}, D) \approx \mathbb{E}[\mathbb{E}(Z_{ij}^{(k)}|z_{i-j}^{(k)}, x_{ij}^{(k)}, \bm{\hat{\Theta}}^{(m)}, D)|x_{i}^{(k)}, \bm{\hat{\Theta}}^{(m)}, D ]\label{eq:mom1} \\
    \mathbb{E}(Z_{ij}^{(k)^2}|x_{i}^{(k)}, \bm{\hat{\Theta}}^{(m)}, D) \approx \mathbb{E}[\mathbb{E}(Z_{ij}^{(k)^2}|z_{i-j}^{(k)}, x_{ij}^{(k)}, \bm{\hat{\Theta}}^{(m)}, D)|x_{i}^{(k)}, \bm{\hat{\Theta}}^{(m)}, D ], \label{eq:mom2} 
  \end{gather}
where $z_{i -j} = (z_{i1}^{(k)},\ldots,z_{i j-1}^{(k)},z_{i j+1}^{(k)}, z_{i p}^{(k)})$. We have that $z_{ij}^{(k)}|z_{i-j}^{(k)}, x_{ij}^{(k)} \sim TN(\mu_{ij}^{(k)}, \sigma^{(k)^2}_{ij}, t_{x_{ij}^{(k)} j}^{(k)}, t_{x_{ij}^{(k)}+1 j}^{(k)})$, where $\mu_{ij}^{(k)} = \hat{\Sigma}^{(k)}_{j-j}\hat{\Sigma}^{(k)^{-1}}_{-j-j}z_{i-j}^{(k)^T}$ and $\sigma^{(k)^2}_{ij} = 1-\hat{\Sigma}^{(k)}_{j-j}\hat{\Sigma}^{(k)^{-1}}_{-j-j}\hat{\Sigma}^{(k)}_{-j-j}$. The $(j,j')$th element of sample correlation matrix $\bar{R}^{(k)}: \bar{r}^{(k)}_{j j'}$ is compute as $\frac{1}{n_{k}}\sum_{i = 1}^{n_{k}}\mathbb{E}(Z_{ij}^{(k)}Z_{ij'}^{(k)}|x_{i}^{(k)}, \bm{\hat{\Theta}}^{(m)}, D)$. If $j = j'$ we use that $\mathbb{E}(Z_{ij}^{(k)}Z_{ij'}^{(k)^T}|x_{i}^{(k)}, \bm{\hat{\Theta}}^{(m)}, D) = \mathbb{E}(Z_{ij}^{(k)^2}|x_{i}^{(k)}, \bm{\hat{\Theta}}^{(m)}, D)$ and if $j \neq j'$ we approximate the quantity as

$\mathbb{E}(Z_{ij}^{(k)}Z_{ij'}^{(k)^T}|x_{i}^{(k)}, \bm{\hat{\Theta}}^{(m)}, D) \approx \mathbb{E}(Z_{ij}^{(k)}|x_{i}^{(k)}, \bm{\hat{\Theta}}^{(m)}, D)\mathbb{E}(Z_{ij'}^{(k)}|x_{i}^{(k)}, \bm{\hat{\Theta}}^{(m)}, D)$.
\\
\\
For the remainder, we need some theory on truncated normal distributions; for a random variable $X\sim N(\mu_0, \sigma_0)$  with lower and upper truncation points $a$ and $b$ respectively, $X|a \leq X \leq b$ follows a truncated normal distribution. Define $\epsilon_a = \frac{(a - \mu_0)}{\sigma_0}$ and $\epsilon_b = \frac{(b - \mu_0)}{\sigma_0}$; then, the first and second order moments are 
\begin{equation*}
    \begin{gathered}
    \mathbb{E}(X|a \leq X \leq b) = \mu_0 + \frac{\phi(\epsilon_a) - \phi(\epsilon_b)}{\Phi(\epsilon_a) - \Phi(\epsilon_b)}\sigma_0\\
    \mathbb{E}(X^2|a \leq X \leq b) = \mu_0^2 + \sigma_{0}^2 + 2\frac{\phi(\epsilon_a) - \phi(\epsilon_b)}{\Phi(\epsilon_a) - \Phi(\epsilon_b)}\mu_0\sigma_0 + \frac{\epsilon_a\phi(\epsilon_a) - \epsilon_b\phi(\epsilon_b)}{\Phi(\epsilon_a) - \Phi(\epsilon_b)}\sigma_0
    \end{gathered}
\end{equation*}

Therefore, we have that
\begin{equation}
\label{eq:z}
    \mathbb{E}(Z_{ij}^{(k)}|x_{i}^{(k)}; \bm{\hat{\Theta}}^{(m)}, D) = \hat{\Sigma}^{(k)}_{j-j}\hat{\Sigma}^{(k)^{-1}}_{-j-j}\mathbb{E}(Z_{i-j}^{(k)^T}|x_{i}^{(k)}; \bm{\hat{\Theta}}^{(m)}, D)\frac{\phi\left(\tilde{\delta}^{(k)}_{i_{x_{ij}^{(k)}j}}\right) - \phi\left(\tilde{\delta}^{(k)}_{i_{x_{ij}^{(k)}+1 j}}\right)}{\Phi\left(\tilde{\delta}^{(k)}_{i_{x_{ij}^{(k)}+1 j}}\right) - \Phi\left(\tilde{\delta}^{(k)}_{i_{x_{ij}^{(k)}j}}\right)}\tilde{\sigma}_{ij}^{(k)} 
\end{equation}

\begin{equation}
    \begin{gathered}
        \mathbb{E}(Z_{ij}^{(k)^2}|x_{i}^{(k)}; \bm{\hat{\Theta}}^{(m)}, D) = \hat{\Sigma}^{(k)}_{j-j}\hat{\Sigma}^{(k)^{-1}}_{-j-j}\mathbb{E}(Z_{i-j}^{(k)^T}Z_{i-j}^{(k)}|x_{i}^{(k)}; \bm{\hat{\Theta}}^{(m)}, D)\hat{\Sigma}^{(k)^{-1}}_{-j-j}\hat{\Sigma}^{(k)^{T}}_{j-j} + \tilde{\sigma}_{ij}^{(k)^2}\\
        + 2\frac{\phi\left(\tilde{\delta}^{(k)}_{i_{x_{ij}^{(k)}j}}\right) - \phi\left(\tilde{\delta}^{(k)}_{i_{x_{ij}^{(k)}+1 j}}\right)}{\Phi\left(\tilde{\delta}^{(k)}_{i_{x_{ij}^{(k)}+1 j}}\right) - \Phi\left(\tilde{\delta}^{(k)}_{i_{x_{ij}^{(k)}j}}\right)}[\hat{\Sigma}^{(k)^{-1}}_{j-j}\hat{\Sigma}^{(k)^T}_{-j-j}\mathbb{E}(Z_{i-j}^{(k)^T}|x_{i}^{(k)}; \bm{\hat{\Theta}}^{(m)}, D)]\tilde{\sigma}_{ij}^{(k)}\\
        + \frac{\tilde{\delta}^{(k)}_{i_{x_{ij}^{(k)}j}}\phi\left(\tilde{\delta}^{(k)}_{i_{x_{ij}^{(k)}j}}\right) - \tilde{\delta}^{(k)}_{i_{x_{ij}^{(k)}+1 j}}\phi\left(\tilde{\delta}^{(k)}_{i_{x_{ij}^{(k)}+1 j}}\right)}{\Phi\left(\tilde{\delta}^{(k)}_{i_{x_{ij}^{(k)}+1 j}}\right) - \Phi\left(\tilde{\delta}^{(k)}_{i_{x_{ij}^{(k)}j}}\right)}\tilde{\sigma}_{ij}^{(k)^2},
        \label{eq:z2}
    \end{gathered}
\end{equation}
where $Z_{i -j} = (Z_{i1}^{(k)},\ldots,Z_{i j-1}^{(k)},Z_{i j+1}^{(k)}, Z_{i p}^{(k)})$ and $\tilde{\delta}^{(k)}_{i_{x_{ij}^{(k)}j}} = \frac{t_{ij}^{(k)} - \mathbb{E}(\tilde{\mu}_{ij}^{k}|x_{i}^{(k)}; \bm{\hat{\Theta}}^{(m)}, D)}{\tilde{\sigma_{ij}^{(k)}}}$

\section{Additional simulation results} \label{Additional simulation results}

\begin{table}[H]
\centering
  \begin{threeparttable}
  \caption{Simulation results for cluster networks, where AUC stands for area under the curve, FL stands for Frobenius loss, EL stands for entropy loss and the bc suffix stands for best choice, i.e. the best result of that respective metric (highest for AUC and lowest for EL and FL) for a particular value of $\lambda_2$. The value corresponding to the winning method is written in bold.}
  \label{tab:simcluster}
     \begin{tabular}{lcccccc}
        \toprule
         \multicolumn{4}{r}{Gibbs method/Approximate method} 
         &
         \multicolumn{3}{c}{Fused graphical lasso/GLASSO}\\
        \midrule
         \textbf{$n, p$ $\rho$} & \textbf{AUC} &\textbf{FL} & \textbf{EL}& \textbf{AUC} & \textbf{FL} & \textbf{EL}\\ \midrule
$10, 50, 0.25$ & \textbf{0.64}/0.64 & \textbf{0.71}/0.80 & \textbf{9.67}/11.08 & 0.60/0.58 & 4.24/1.57 & 45.71/19.64\\ 
$50, 50, 0.25$ & \textbf{0.86}/0.86 & \textbf{0.11}/0.11 & \textbf{2.54}/2.55 & 0.77/0.81 & 1.12/0.26 & 32.44/7.20\\ 
$100, 50, 0.25$ & \textbf{0.92}/0.92 & \textbf{0.09}/0.09 & \textbf{2.13}/2.14 & 0.85/0.91 & 1.08/0.23 & 32.60/6.51\\ 
$500, 50, 0.25$ & 0.97/0.97 & \textbf{0.09}/0.09 & \textbf{1.99}/2.00 & 0.95/\textbf{1.00} & 1.05/0.23 & 32.91/6.25\\ 
$10, 100, 0.25$ & \textbf{0.59}/0.60 & \textbf{1.01}/1.25 & \textbf{27.91}/34.49 & 0.56/0.55 & 5.65/1.99 & 107.44/49.01\\ 
$50, 100, 0.25$ & \textbf{0.81}/0.81 & \textbf{0.15}/0.15 & \textbf{7.77}/7.92 & 0.73/0.73 & 1.13/0.33 & 66.19/18.09\\ 
$100, 100, 0.25$ & \textbf{0.89}/0.89 & \textbf{0.12}/0.12 & \textbf{6.32}/6.36 & 0.81/0.85 & 1.03/0.27 & 66.50/15.34\\ 
$500, 100, 0.25$ & 0.96/0.96 & \textbf{0.11}/0.11 & \textbf{5.78}/5.83 & 0.92/\textbf{0.99} & 1.01/0.25 & 67.67/14.33\\ 
$10, 50, 1$ & \textbf{0.60}/0.60 & \textbf{0.69}/0.78 & \textbf{10.06}/11.46 & 0.56/0.56 & 4.00/1.52 & 45.52/19.95\\ 
$50, 50, 1$ & \textbf{0.77}/0.77 & \textbf{0.12}/0.12 & \textbf{2.95}/2.96 & 0.71/0.75 & 1.11/0.27 & 32.82/7.58\\ 
$100, 50, 1$ & 0.83/0.83 & \textbf{0.10}/0.10 & \textbf{2.57}/2.58 & 0.78/\textbf{0.86} & 1.05/0.24 & 32.68/6.91\\ 
$500, 50, 1$ & 0.92/0.92 & \textbf{0.10}/0.10 & \textbf{2.44}/2.45 & 0.90/\textbf{0.99} & 1.03/0.24 & 33.44/6.66\\ 
$10, 100, 1$ & \textbf{0.56}/0.56 & \textbf{0.97}/1.20 & \textbf{28.94}/35.58 & 0.54/0.54 & 5.31/1.90 & 107.66/49.93\\ 
$50, 100, 1$ & \textbf{0.71}/0.72 & \textbf{0.16}/0.16 & \textbf{8.88}/9.03 & 0.66/0.68 & 1.10/0.34 & 67.86/19.07\\ 
$100, 100, 1$ & 0.78/0.78 & \textbf{0.13}/0.13 & \textbf{7.46}/7.49 & 0.72/\textbf{0.79} & 1.00/0.28 & 67.85/16.38\\ 
$500, 100, 1$ & 0.89/0.89 & \textbf{0.12}/0.12 & \textbf{6.94}/6.98 & 0.86/\textbf{0.97} & 0.99/0.27 & 68.54/15.40\\
        \midrule
         \multicolumn{4}{r}{Gibbs method/Approximate method} 
         &
         \multicolumn{3}{c}{Fused graphical lasso}\\
        \midrule
         \textbf{$n, p$ $\rho$} & \textbf{AUC bc} & \textbf{FL bc} & \textbf{EL bc} & \textbf{AUC bc} & \textbf{FL bc} & \textbf{EL bc}\\ \midrule
$10, 50, 0.25$ & 0.68/\textbf{0.69} & \textbf{0.26}/0.27 & \textbf{5.23}/5.50 & 0.65 & 1.87 & 34.83 \\
$50, 50, 0.25$ & \textbf{0.90}/0.90 & \textbf{0.09}/0.09 & \textbf{2.17}/2.18 & 0.81 & 1.08 & 31.90 \\ 
$100, 50, 0.25$ & \textbf{0.93}/0.93 & \textbf{0.09}/0.09 & \textbf{2.03}/2.04 & 0.87 & 1.07 & 32.48 \\ 
$500, 50, 0.25$ & \textbf{1.00}/1.00 & \textbf{0.09}/0.09 & \textbf{1.98}/2.00 & 0.95 & 1.05 & 32.88 \\ 
$10, 100, 0.25$ & 0.62/\textbf{0.63} & \textbf{0.42}/0.48 & \textbf{16.67}/18.61 & 0.58 & 2.53 & 79.72 \\ 
$50, 100, 0.25$ & \textbf{0.85}/0.85 & \textbf{0.12}/0.12 & \textbf{6.44}/6.47 & 0.78 & 1.03 & 63.82 \\ 
$100, 100, 0.25$ & \textbf{0.91}/0.91 & \textbf{0.11}/0.11 & \textbf{5.94}/5.97 & 0.84 & 1.01 & 66.02 \\
$500, 100, 0.25$ & \textbf{0.99}/0.99 & \textbf{0.11}/0.11 & \textbf{5.78}/5.82 & 0.93 & 1.01 & 67.55 \\
$10, 50, 1$ & \textbf{0.63}/0.63 & \textbf{0.26}/0.28 & \textbf{5.65}/5.93 & 0.58 & 1.75 & 34.74 \\ 
$50, 50, 1$ & \textbf{0.78}/0.78 & \textbf{0.10}/0.10 & \textbf{2.61}/2.62 & 0.72 & 1.07 & 32.28 \\ 
$100, 50, 1$ & \textbf{0.86}/0.86 & \textbf{0.10}/0.10 & \textbf{2.49}/2.50 & 0.79 & 1.04 & 32.56 \\ 
$500, 50, 1$ & \textbf{0.99}/0.99 & \textbf{0.10}/0.10 & \textbf{2.40}/2.41 & 0.93 & 1.03 & 33.35 \\ 
$10, 100, 1$ & \textbf{0.57}/0.57 & \textbf{0.41}/0.47 & \textbf{17.71}/19.70 & 0.55 & 2.41 & 80.47 \\ 
$50, 100, 1$ & \textbf{0.74}/0.74 & \textbf{0.13}/0.13 & \textbf{7.61}/7.63 & 0.68 & 1.00 & 65.50 \\ 
$100, 100, 1$ & \textbf{0.79}/0.79 & \textbf{0.13}/0.13 & \textbf{7.12}/7.14 & 0.73 & 0.99 & 67.36 \\ 
$500, 100, 1$ & \textbf{0.97}/0.97 & \textbf{0.12}/0.12 & \textbf{6.87}/6.91 & 0.90 & 0.98 & 68.30 \\
        \bottomrule
     \end{tabular}
  \end{threeparttable}
\end{table}

\begin{figure}[H]
\centering
\text{Cluster network, $p = 50$, $\rho = 0.25$}\\
\includegraphics[width=0.25\textwidth]{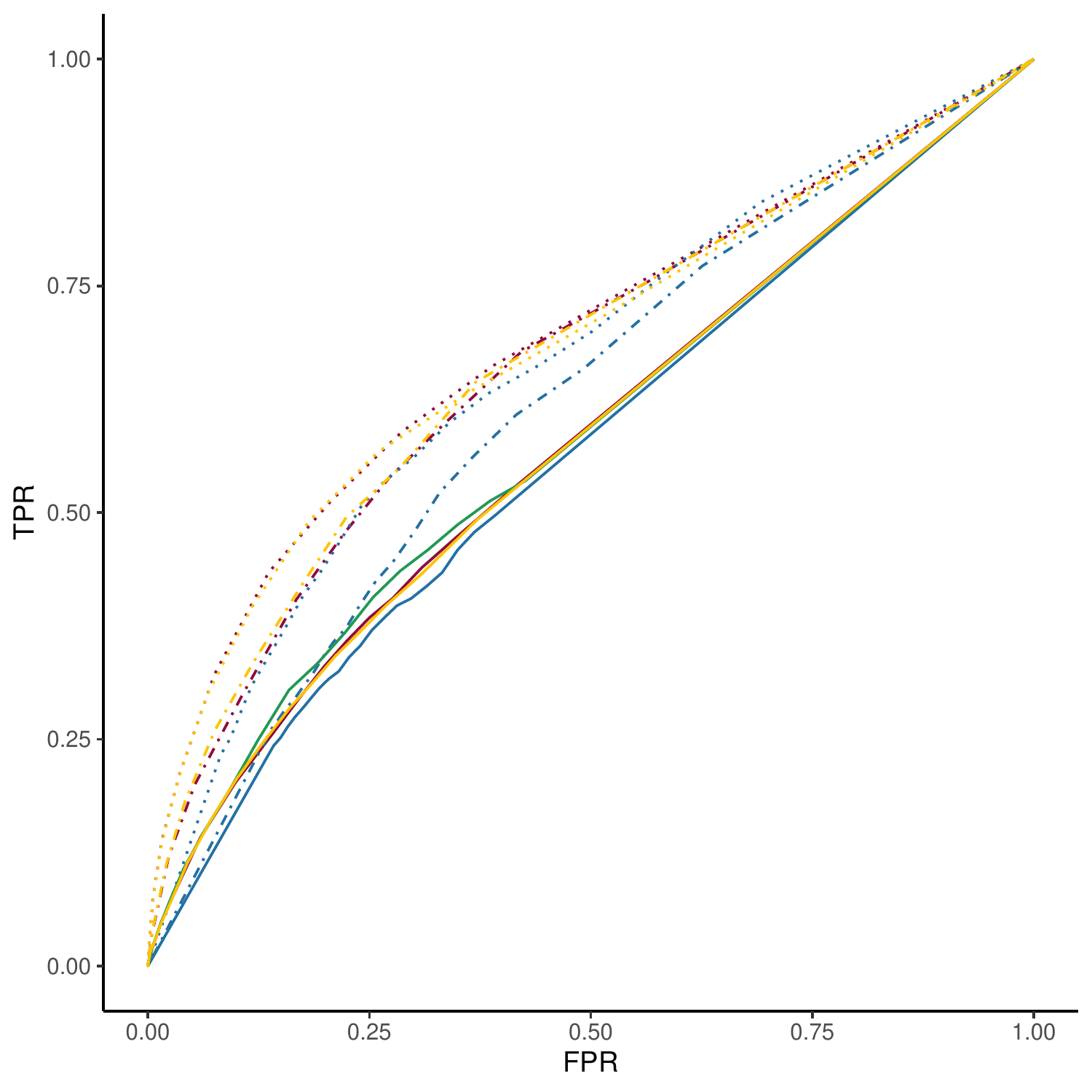}\hfill
\includegraphics[width=0.25\textwidth]{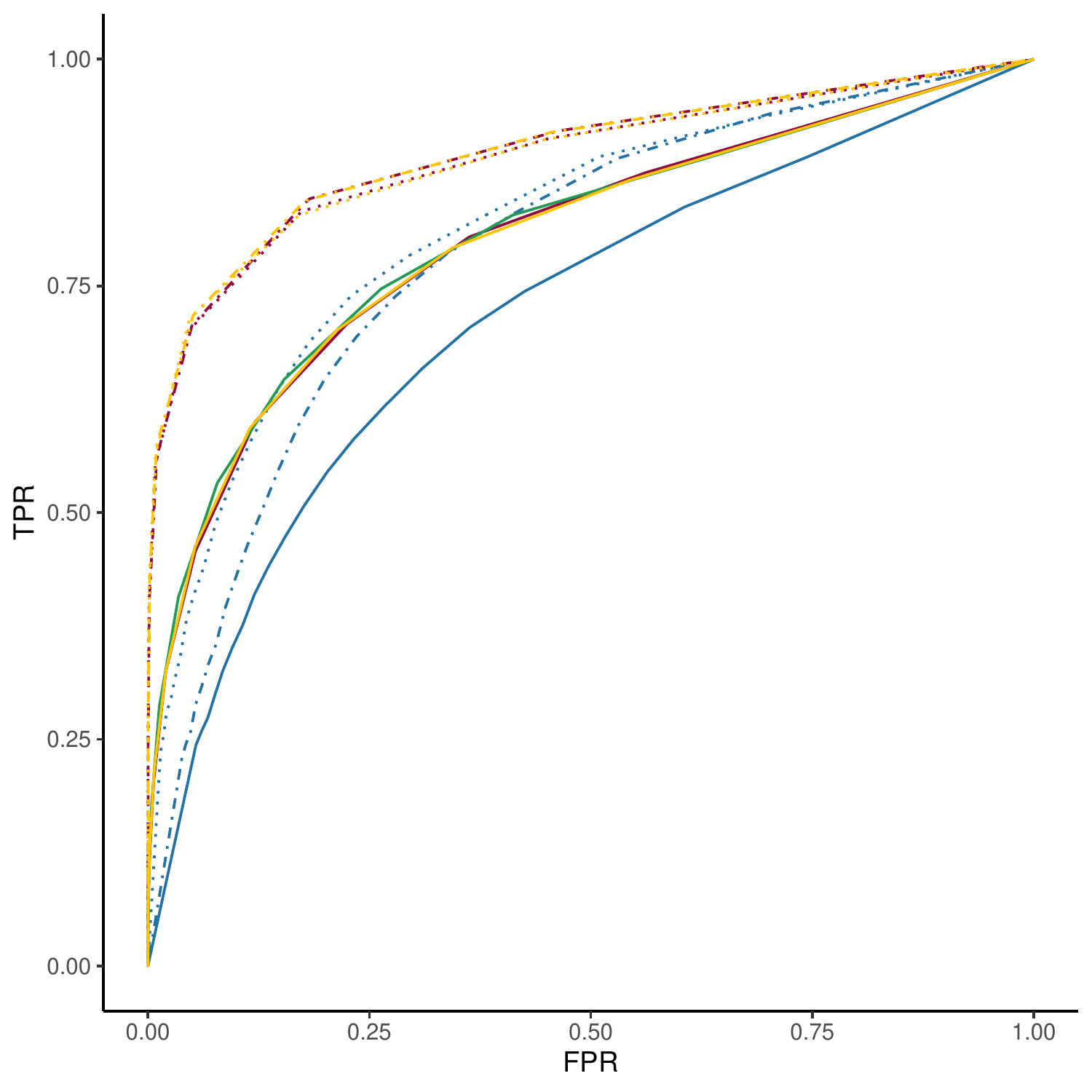}\hfill
\includegraphics[width=0.25\textwidth]{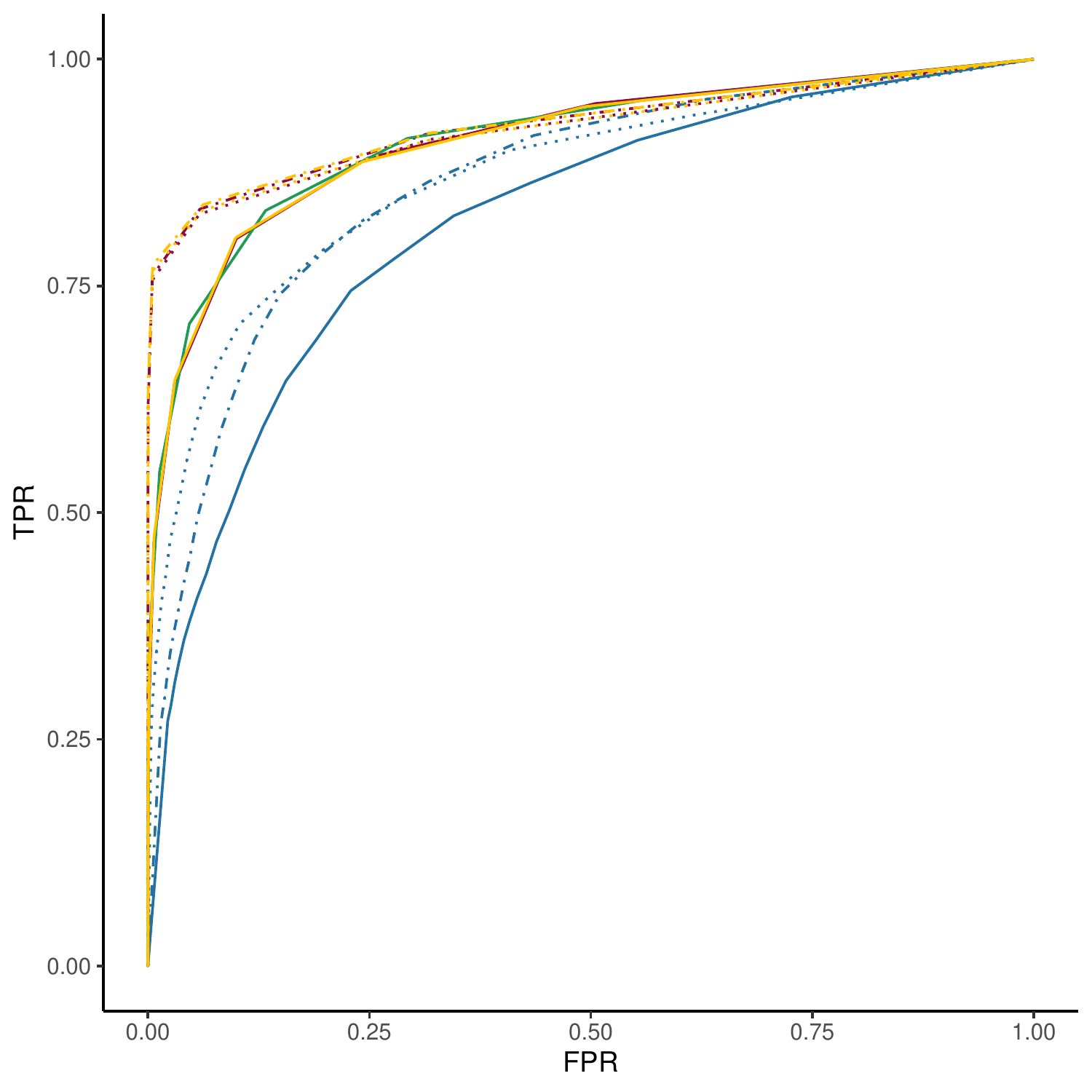}\hfill
\includegraphics[width=0.25\textwidth]{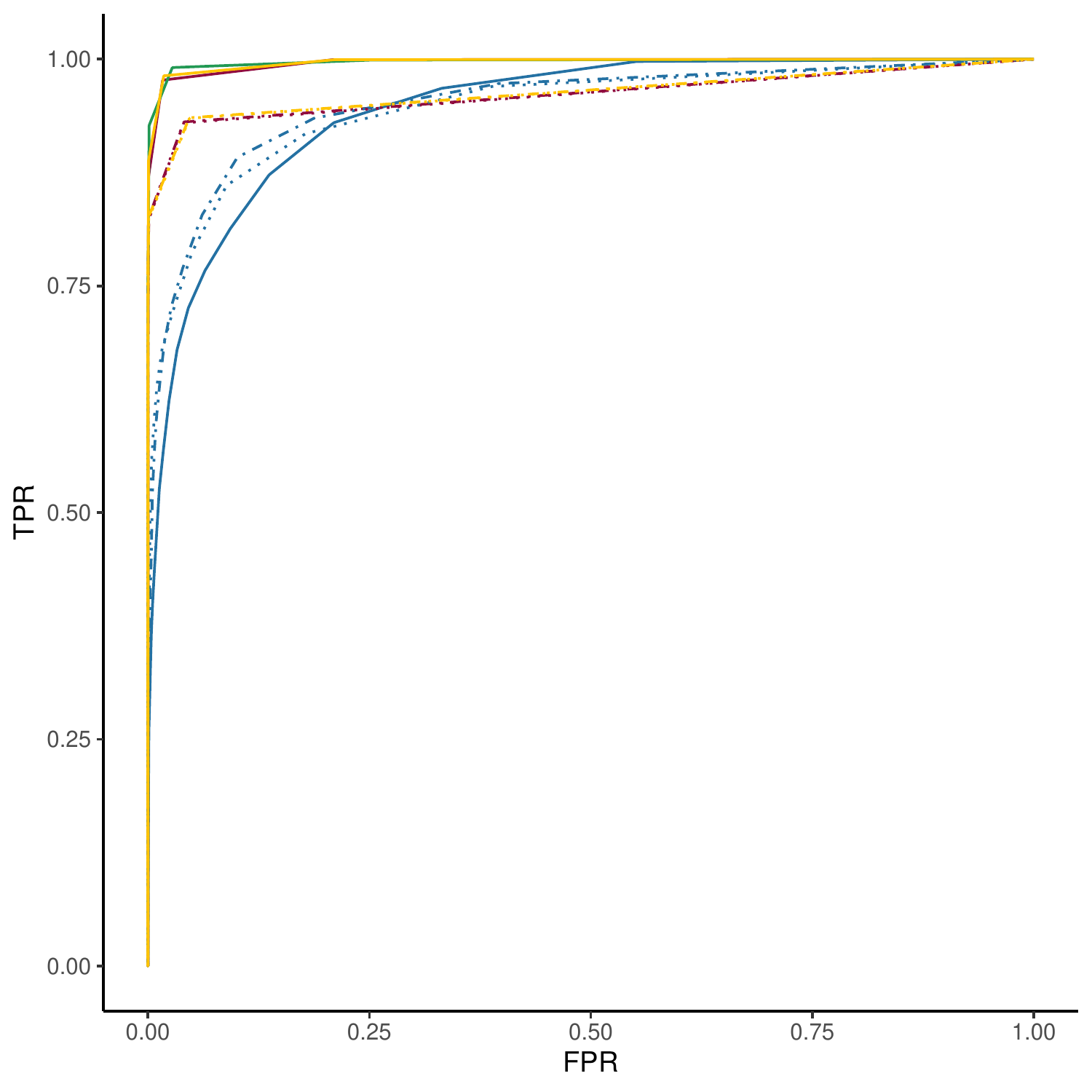}\hfill
\text{Cluster network, $p = 100$, $\rho = 0.25$}\\
\includegraphics[width=0.25\textwidth]{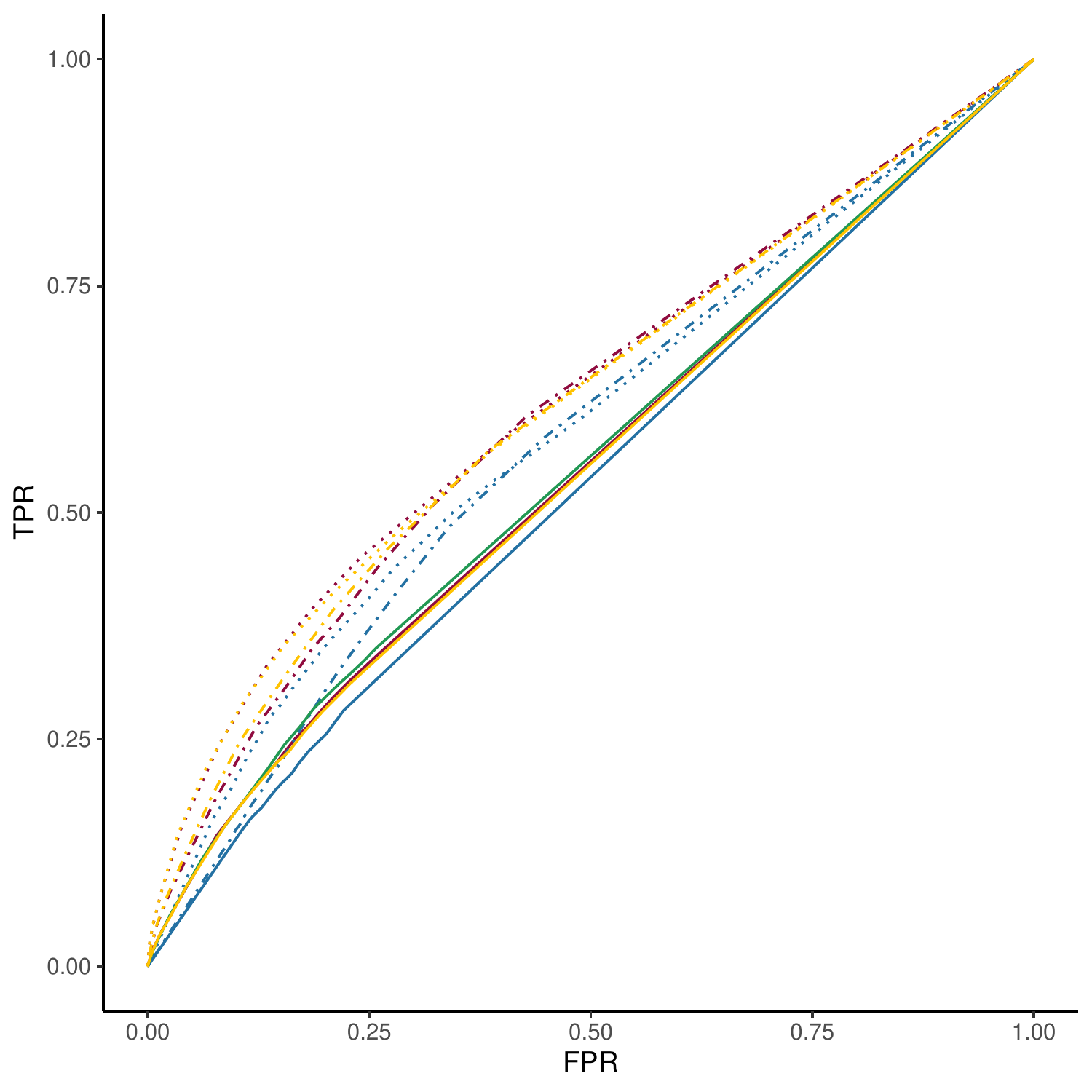}\hfill
\includegraphics[width=0.25\textwidth]{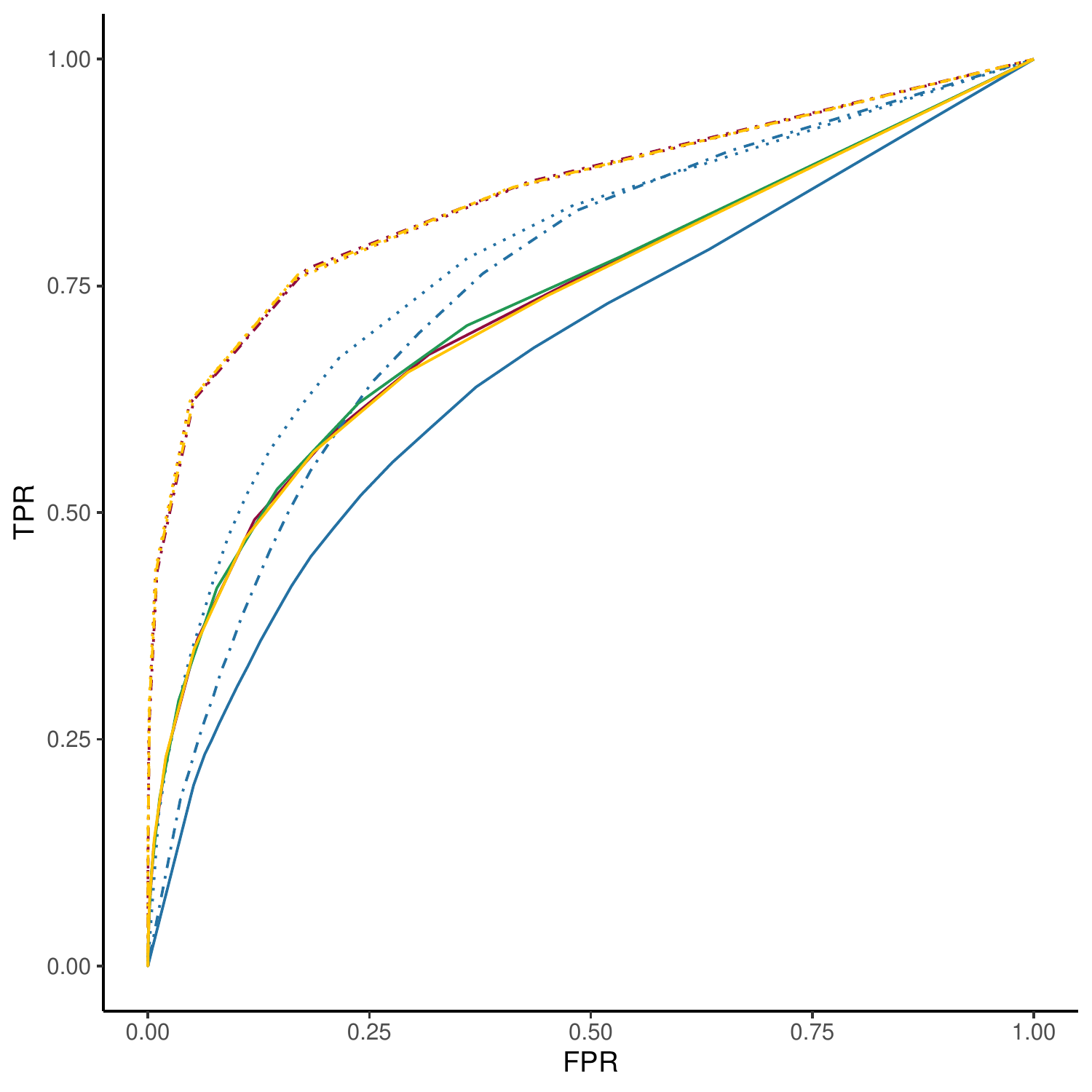}\hfill
\includegraphics[width=0.25\textwidth]{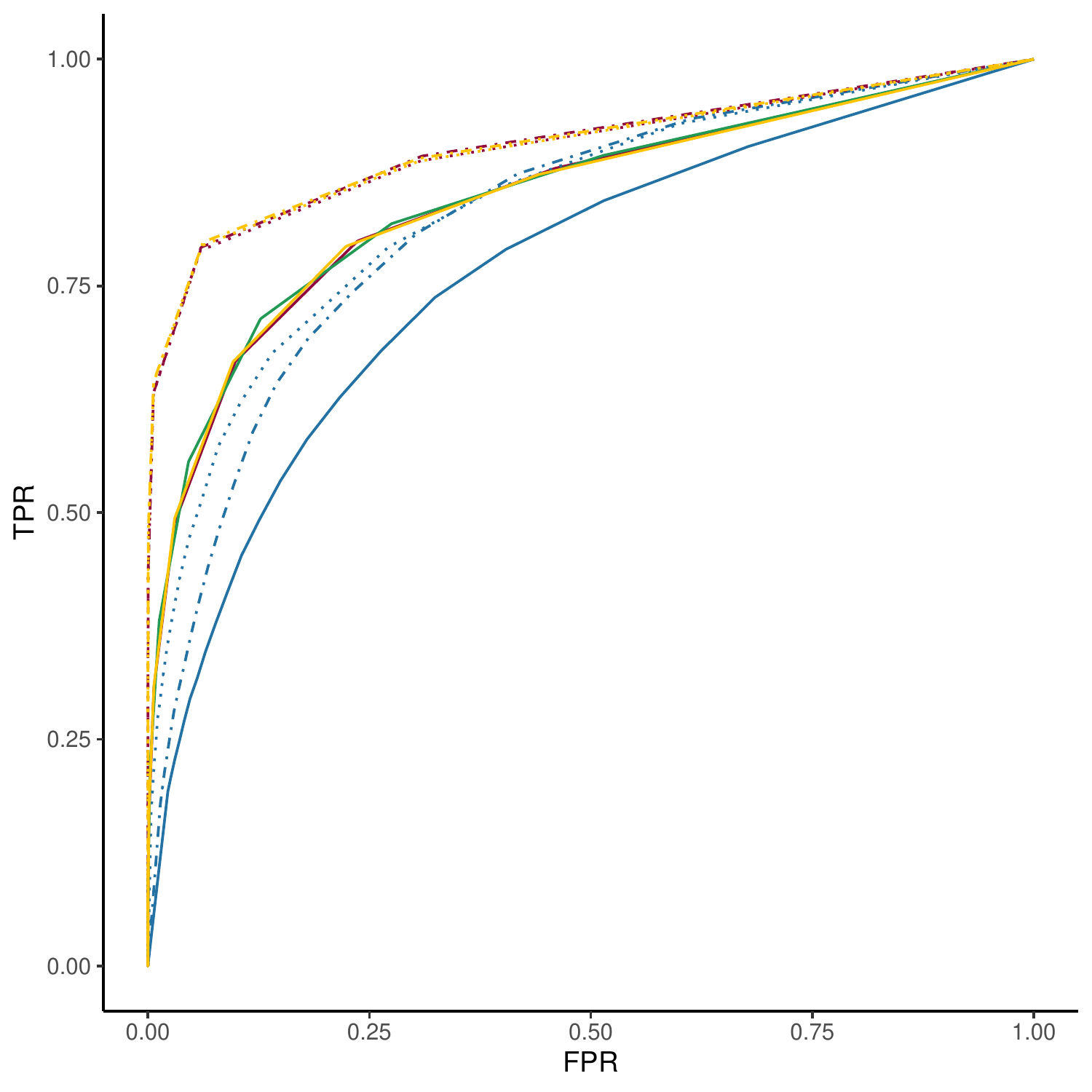}\hfill
\includegraphics[width=0.25\textwidth]{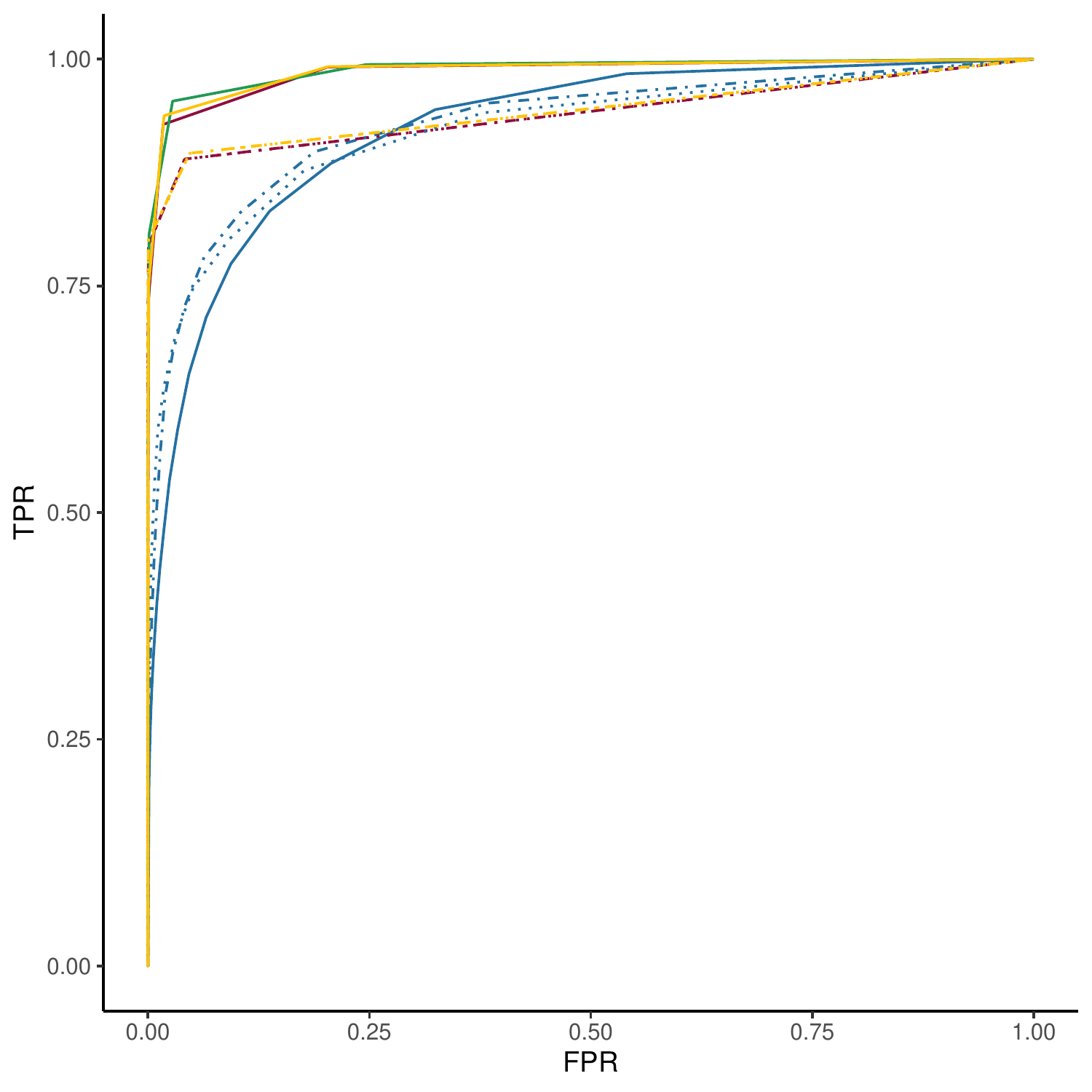}\hfill
\text{Cluster network, $p = 50$, $\rho = 1$}\\
\includegraphics[width=0.25\textwidth]{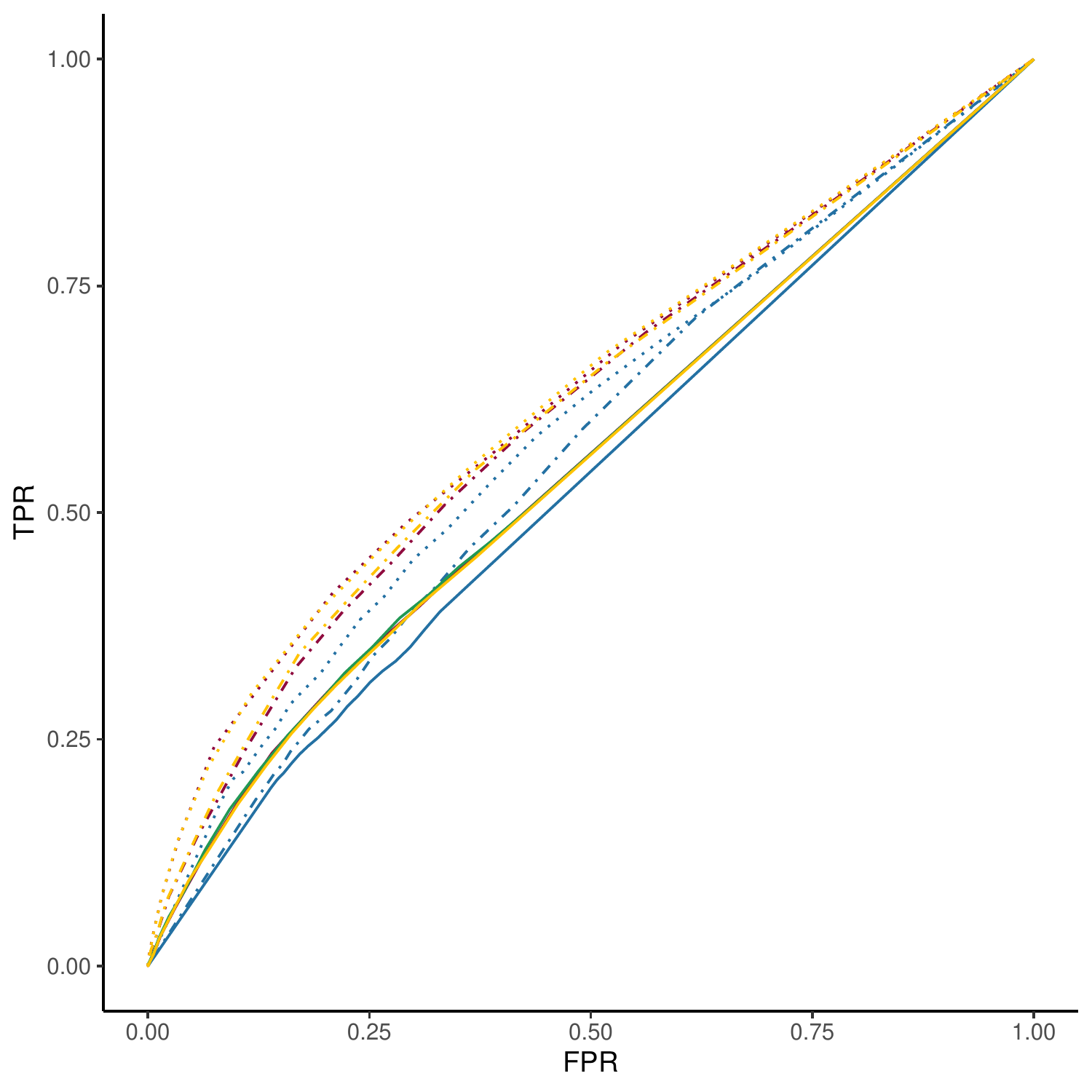}\hfill
\includegraphics[width=0.25\textwidth]{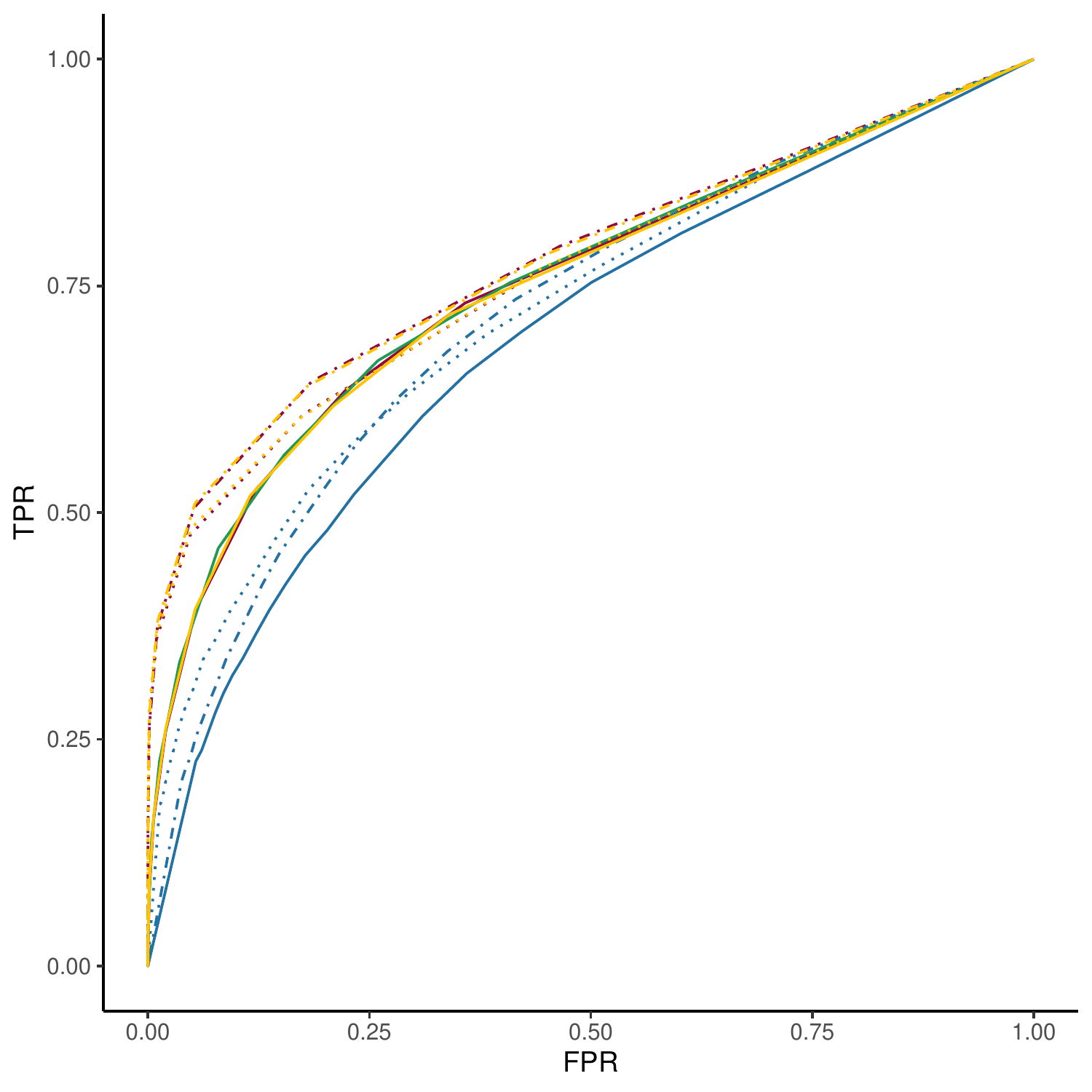}\hfill
\includegraphics[width=0.25\textwidth]{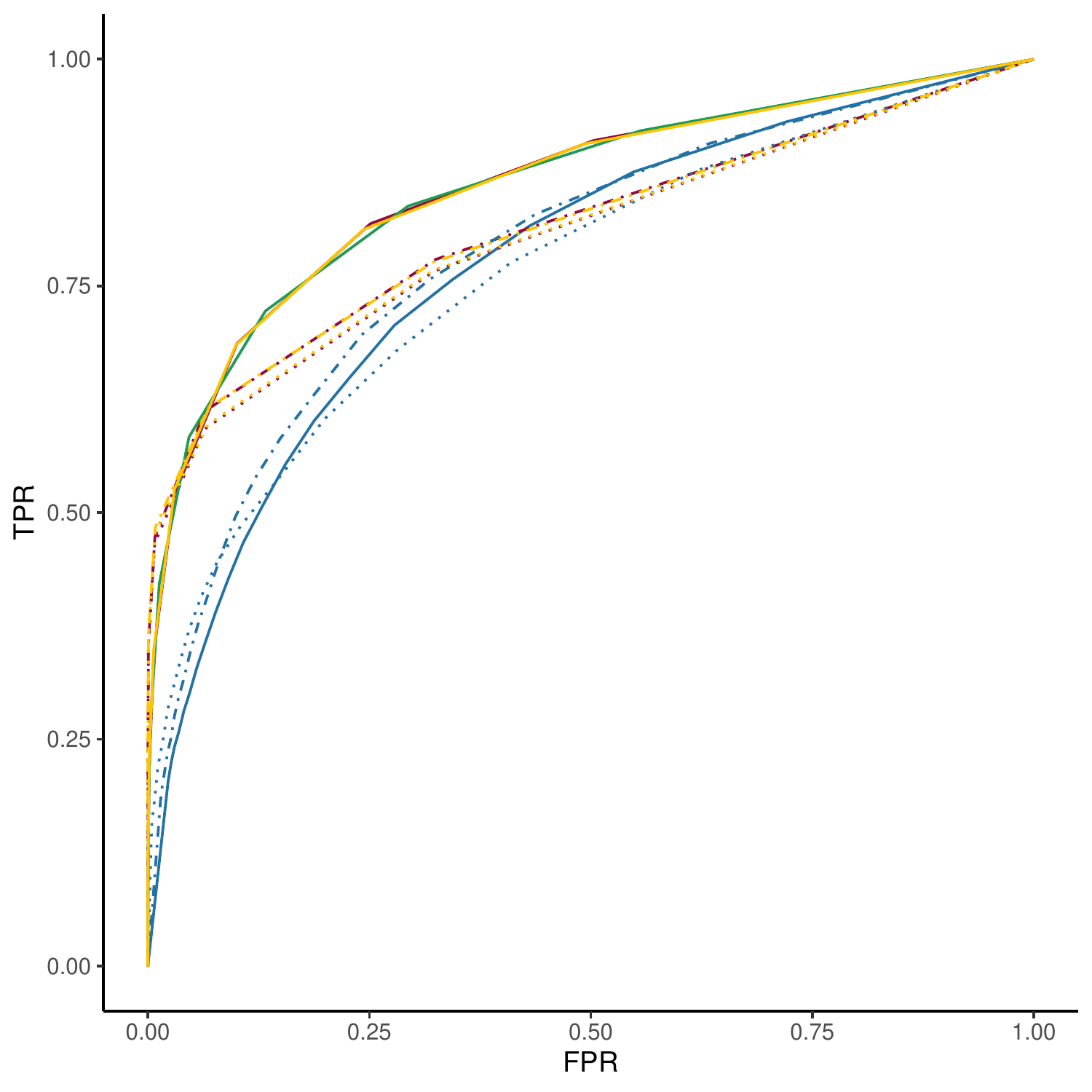}\hfill
\includegraphics[width=0.25\textwidth]{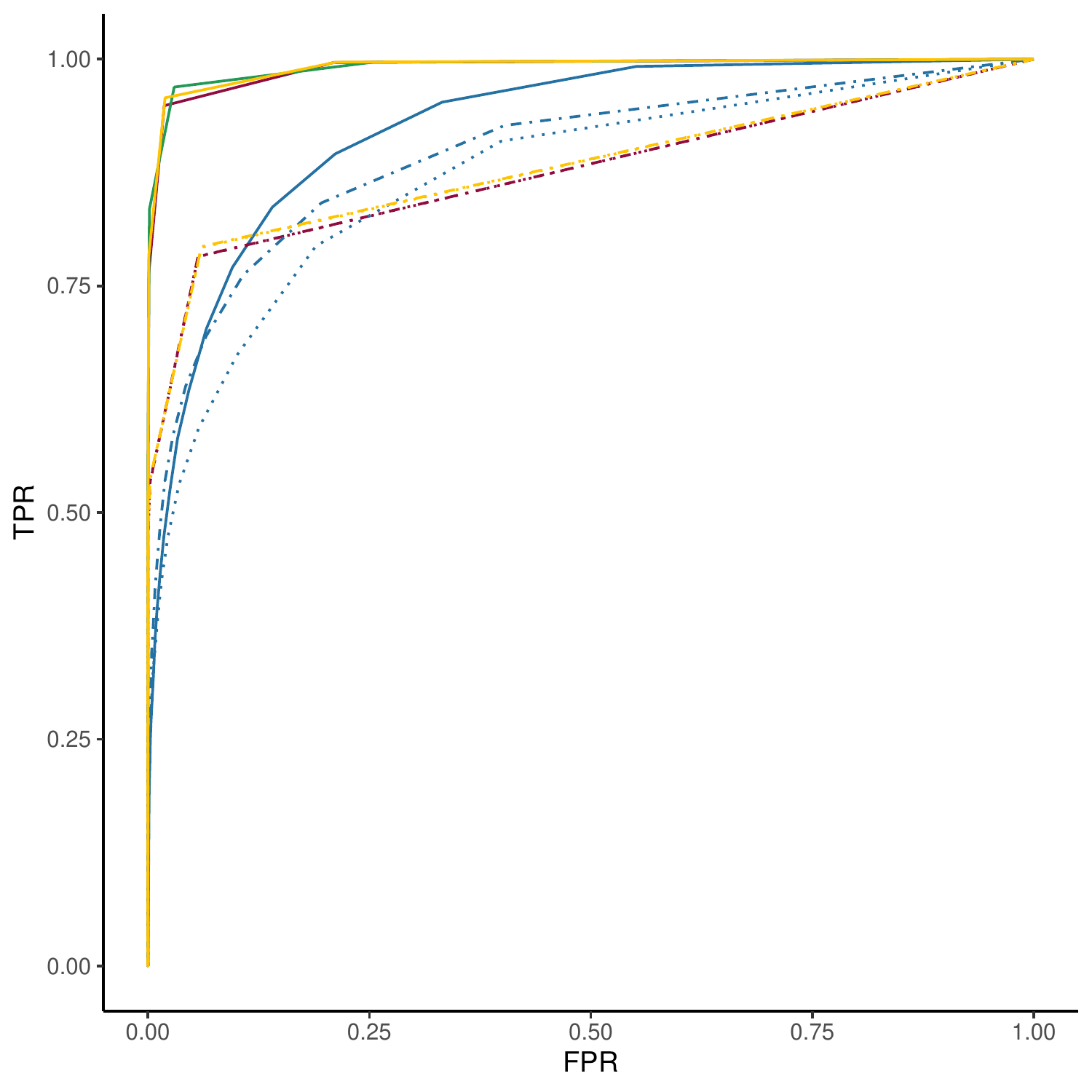}\hfill
\text{Cluster network, $p = 100$, $\rho = 1$}\\
\includegraphics[width=0.25\textwidth]{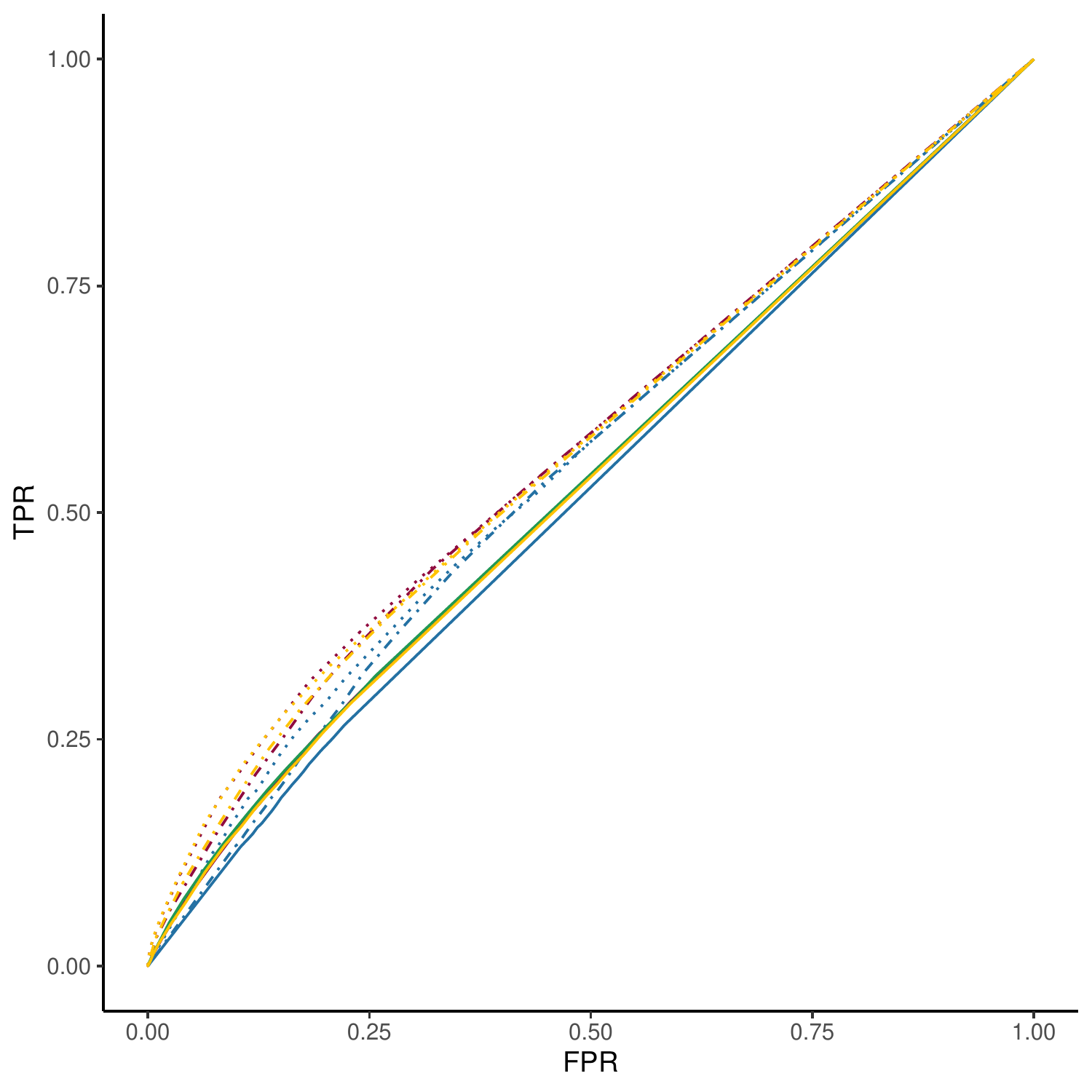}\hfill
\includegraphics[width=0.25\textwidth]{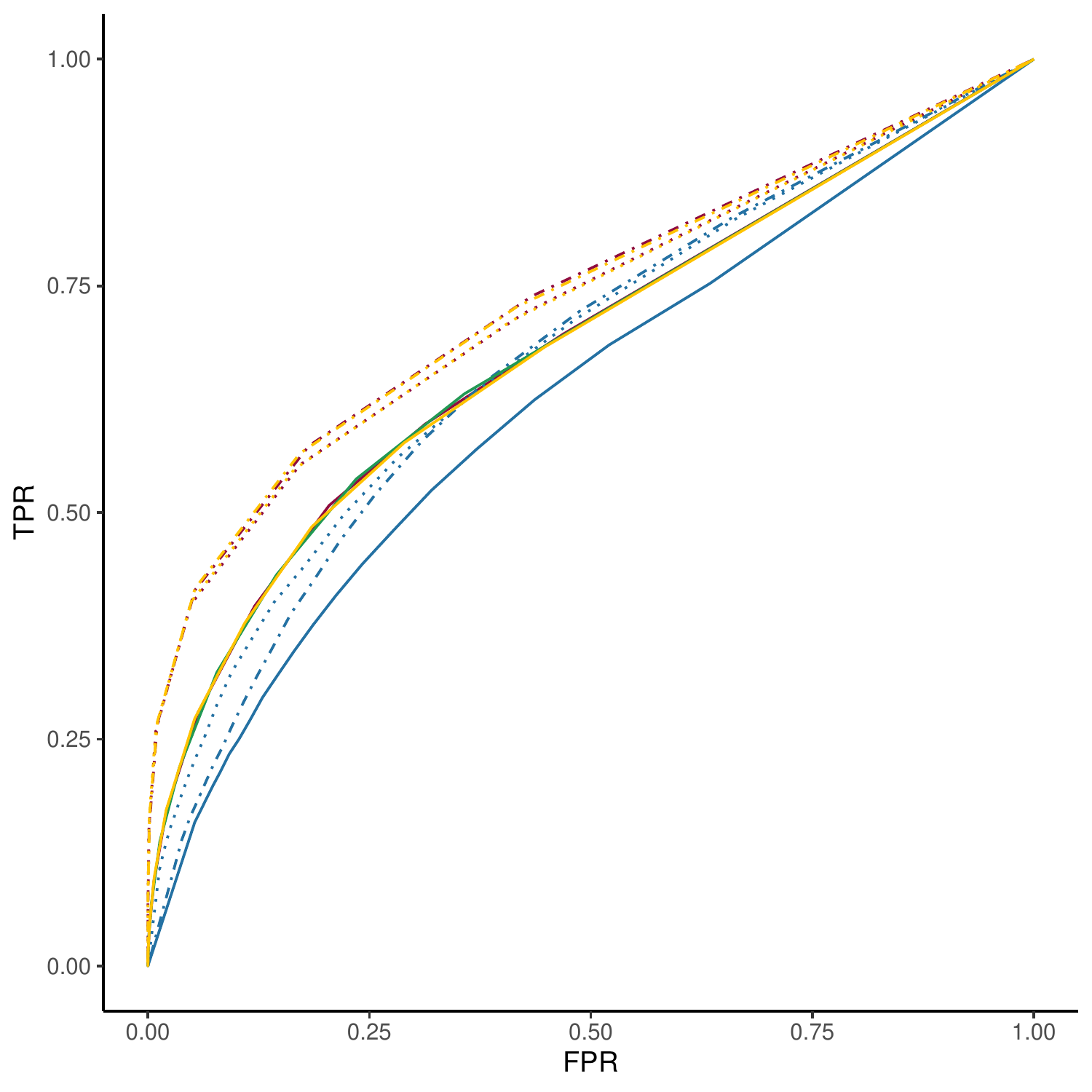}\hfill
\includegraphics[width=0.25\textwidth]{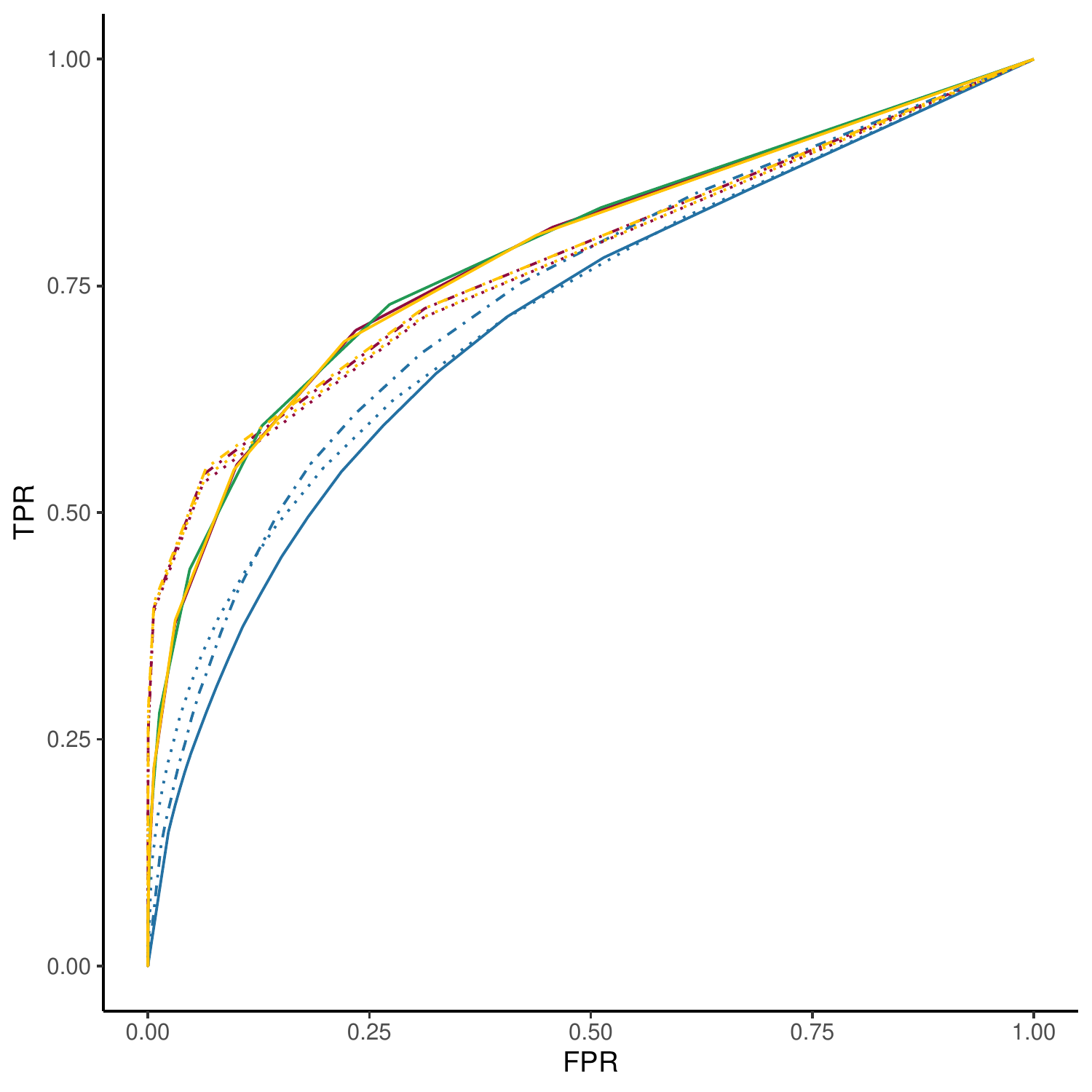}\hfill
\includegraphics[width=0.25\textwidth]{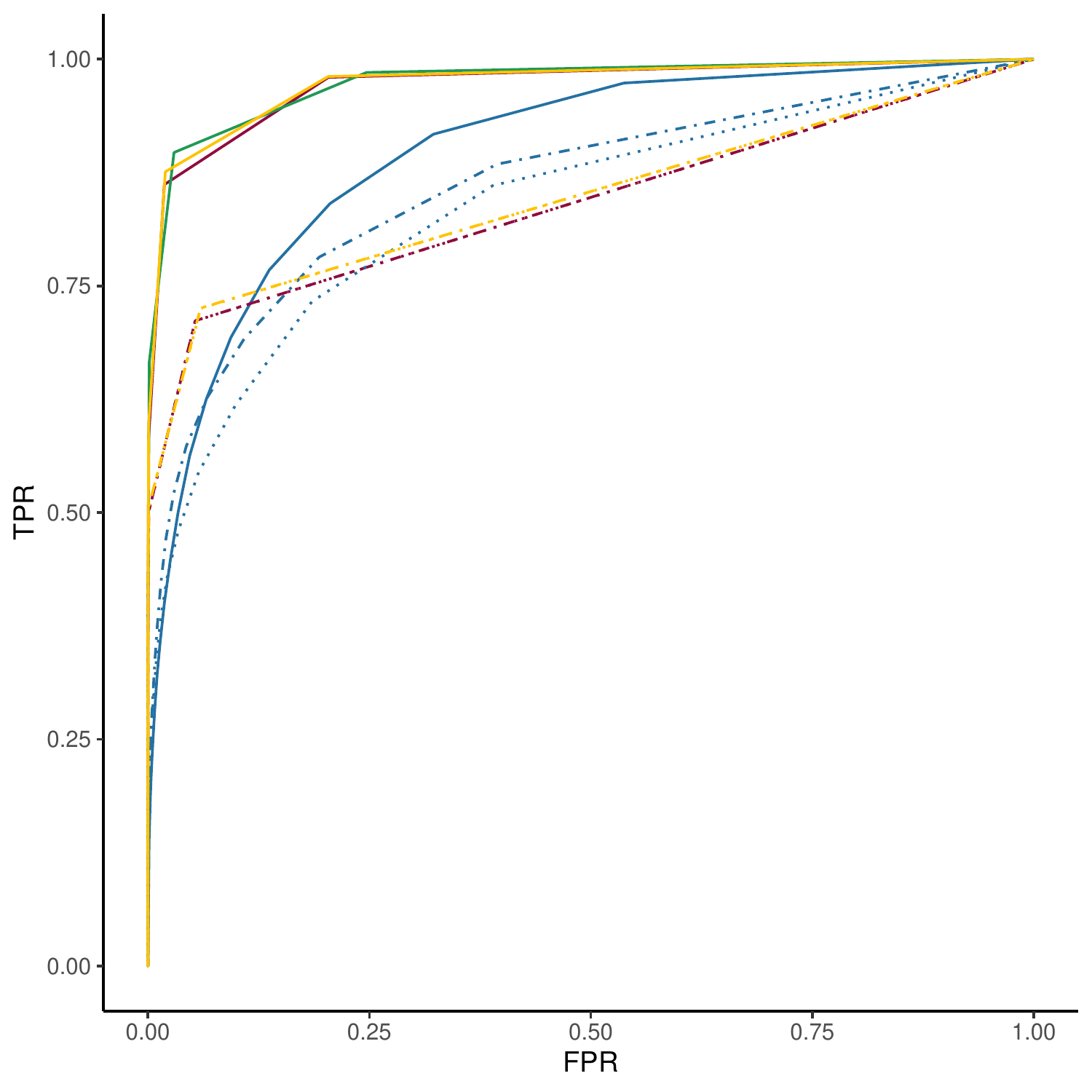}\hfill
\caption[]{ROC curves plotted over values of $\lambda_1$. The line type denotes the penalty for $\lambda_2$: $\lambda_2 = 0$ \begin{tikzpicture}\draw [thick] (0,2) -- (1,2);\end{tikzpicture}, $\lambda_2 = 0.1$  \begin{tikzpicture}\draw [dashdotted] (0,2) -- (1,2);\end{tikzpicture} and $\lambda_2 = 1$ \begin{tikzpicture}\draw [dotted] (0,2) -- (1,2);\end{tikzpicture}. The colors represent the method used: Gibbs method \begin{tikzpicture}\draw [colorGibbs, thick] (0,2) -- (1,2);\end{tikzpicture}, Approximate method \begin{tikzpicture}\draw [colorApprox, thick] (0,2) -- (1,2);\end{tikzpicture}, Fused graphical lasso \begin{tikzpicture}\draw [colorFGL, thick] (0,2) -- (1,2);\end{tikzpicture} and GLASSO \begin{tikzpicture}\draw [colorGLASSO, thick] (0,2) -- (1,2);\end{tikzpicture}. From left to right the values of $n$ are respectively 10, 50, 100 and 500.}
\label{fig:roccurves_cl}
\end{figure}

\begin{table}[H]
\centering
  \begin{threeparttable}
  \caption{Simulation results for scale-free networks, where AUC stands for area under the curve, FL stands for Frobenius loss, EL stands for entropy loss and the bc suffix stands for best choice, i.e. the best result of that respective metric (highest for AUC and lowest for EL and FL) for a particular value of $\lambda_2$. The value corresponding to the winning method is written in bold.}
  \label{tab:simscalefree}
     \begin{tabular}{lcccccc}
        \toprule
         \multicolumn{4}{r}{Gibbs method/Approximate method} 
         &
         \multicolumn{3}{c}{Fused graphical lasso/GLASSO}\\
        \midrule
         \textbf{$n, p$ $\rho$} & \textbf{AUC} &\textbf{FL} & \textbf{EL}& \textbf{AUC} & \textbf{FL} & \textbf{EL}\\ \midrule
$10, 50, 0.25$ & 0.56/\textbf{0.57} & \textbf{0.68}/0.76 & \textbf{10.25}/11.63 & 0.53/0.53 & 3.99/1.49 & 46.29/20.20\\ 
$50, 50, 0.25$ & \textbf{0.77}/0.77 & \textbf{0.13}/0.13 & \textbf{3.21}/3.22 & 0.69/0.69 & 1.08/0.28 & 33.11/7.84\\ 
$100, 50, 0.25$ & \textbf{0.85}/0.85 & \textbf{0.12}/0.12 & \textbf{2.84}/2.84 & 0.76/0.80 & 1.03/0.26 & 33.23/7.19\\ 
$500, 50, 0.25$ & 0.94/0.94 & \textbf{0.11}/0.11 & \textbf{2.70}/2.71 & 0.90/\textbf{0.97} & 1.04/0.25 & 33.45/6.95\\ 
$10, 100, 0.25$ & \textbf{0.55}/0.55 & \textbf{1.11}/1.38 & \textbf{26.45}/33.17 & 0.54/0.53 & 6.09/2.17 & 104.97/47.47\\ 
$50, 100, 0.25$ & \textbf{0.73}/0.73 & \textbf{0.14}/0.14 & \textbf{6.16}/6.30 & 0.66/0.65 & 1.21/0.32 & 65.08/16.45\\ 
$100, 100, 0.25$ & \textbf{0.82}/0.82 & \textbf{0.10}/0.10 & \textbf{4.73}/4.74 & 0.74/0.75 & 1.09/0.25 & 65.55/13.71\\ 
$500, 100, 0.25$ & 0.93/0.93 & \textbf{0.09}/0.09 & \textbf{4.19}/4.21 & 0.88/\textbf{0.95} & 1.06/0.23 & 65.88/12.69\\ 
$10, 50, 1$ & \textbf{0.55}/0.55 & \textbf{0.65}/0.73 & \textbf{10.75}/12.11 & 0.52/0.53 & 3.76/1.42 & 46.71/20.59\\ 
$50, 50, 1$ & \textbf{0.69}/0.69 & \textbf{0.14}/0.15 & \textbf{3.82}/3.83 & 0.62/0.65 & 1.04/0.30 & 33.87/8.40\\ 
$100, 50, 1$ & \textbf{0.76}/0.76 & \textbf{0.13}/0.13 & \textbf{3.45}/3.46 & 0.69/0.74 & 1.00/0.28 & 33.82/7.76\\ 
$500, 50, 1$ & 0.86/0.85 & \textbf{0.13}/0.13 & \textbf{3.31}/3.32 & 0.83/\textbf{0.94} & 0.99/0.27 & 34.33/7.54\\ 
$10, 100, 1$ & \textbf{0.53}/0.53 & \textbf{1.06}/1.33 & \textbf{27.30}/33.95 & 0.52/0.52 & 6.00/2.08 & 107.31/48.16\\ 
$50, 100, 1$ & \textbf{0.66}/0.66 & \textbf{0.15}/0.15 & \textbf{7.09}/7.23 & 0.61/0.62 & 1.17/0.33 & 65.62/17.31\\ 
$100, 100, 1$ & \textbf{0.72}/0.72 & \textbf{0.11}/0.11 & \textbf{5.69}/5.71 & 0.67/0.70 & 1.06/0.26 & 66.48/14.62\\ 
$500, 100, 1$ & 0.84/0.83 & \textbf{0.10}/0.10 & \textbf{5.15}/5.17 & 0.81/\textbf{0.91} & 1.04/0.24 & 67.07/13.62\\ 
        \midrule
         \multicolumn{4}{r}{Gibbs method/Approximate method} 
         &
         \multicolumn{3}{c}{Fused graphical lasso}\\
        \midrule
         \textbf{$n, p$ $\rho$} & \textbf{AUC bc} & \textbf{FL bc} & \textbf{EL bc} & \textbf{AUC bc} & \textbf{FL bc} & \textbf{EL bc}\\ \midrule
$10, 50, 0.25$ & \textbf{0.59}/0.59 & \textbf{0.27}/0.28 & \textbf{5.85}/6.11 & 0.55 & 1.74 & 35.41 \\ 
$50, 50, 0.25$  & \textbf{0.81}/0.81 & \textbf{0.12}/0.12 & \textbf{2.86}/2.86 & 0.74 & 1.04 & 32.59 \\ 
$100, 50, 0.25$ & 0.87/\textbf{0.88} & \textbf{0.11}/0.11 & \textbf{2.74}/2.75 & 0.80 & 1.02 & 33.11 \\ 
$500, 50, 0.25$ & \textbf{0.97}/0.97 & \textbf{0.11}/0.11 & \textbf{2.70}/2.70 & 0.91 & 1.04 & 33.42 \\ 
$10, 100, 0.25$ & \textbf{0.57}/0.57 & \textbf{0.44}/0.52 & \textbf{15.08}/17.11 & 0.55 & 2.76 & 78.08 \\
$50, 100, 0.25$ & 0.77/\textbf{0.78} & \textbf{0.10}/0.10 & \textbf{4.83}/4.84 & 0.71 & 1.10 & 62.70 \\ 
$100, 100, 0.25$ & \textbf{0.85}/0.85 & \textbf{0.09}/0.09 & \textbf{4.33}/4.34 & 0.78 & 1.07 & 65.05 \\
$500, 100, 0.25$ & \textbf{0.95}/0.94 & \textbf{0.09}/0.09 & \textbf{4.17}/4.20 & 0.89 & 1.06 & 65.79 \\
$10, 50, 1$ & \textbf{0.56}/0.56 & \textbf{0.27}/0.28 & \textbf{6.40}/6.65 & 0.54 & 1.66 & 35.95 \\ 
$50, 50, 1$ & \textbf{0.71}/0.71 & \textbf{0.13}/0.13 & \textbf{3.49}/3.49 & 0.65 & 1.01 & 33.33 \\ 
$100, 50, 1$ & \textbf{0.77}/0.77 & \textbf{0.13}/0.13 & \textbf{3.37}/3.38 & 0.70 & 0.99 & 33.70 \\ 
$500, 50, 1$ & \textbf{0.94}/0.94 & \textbf{0.13}/0.13 & \textbf{3.28}/3.29 & 0.85 & 0.99 & 34.25 \\ 
$10, 100, 1$ & \textbf{0.54}/0.54 & \textbf{0.43}/0.51 & \textbf{16.04}/18.02 & 0.53 & 2.68 & 79.36 \\
$50, 100, 1$ & \textbf{0.68}/0.68 & \textbf{0.11}/0.11 & \textbf{5.79}/5.80 & 0.64 & 1.07 & 63.27 \\ 
$100, 100, 1$ & \textbf{0.74}/0.74 & \textbf{0.11}/0.11 & \textbf{5.33}/5.34 & 0.68 & 1.04 & 65.98 \\
$500, 100, 1$ & \textbf{0.91}/0.91 & \textbf{0.10}/0.10 & \textbf{5.14}/5.16 & 0.83 & 1.04 & 66.91 \\
        \bottomrule
     \end{tabular}
  \end{threeparttable}
\end{table}

\begin{figure}[H]
\centering
\text{Scale-free network, $p = 50$, $\rho = 0.25$}\\
\includegraphics[width=0.25\textwidth]{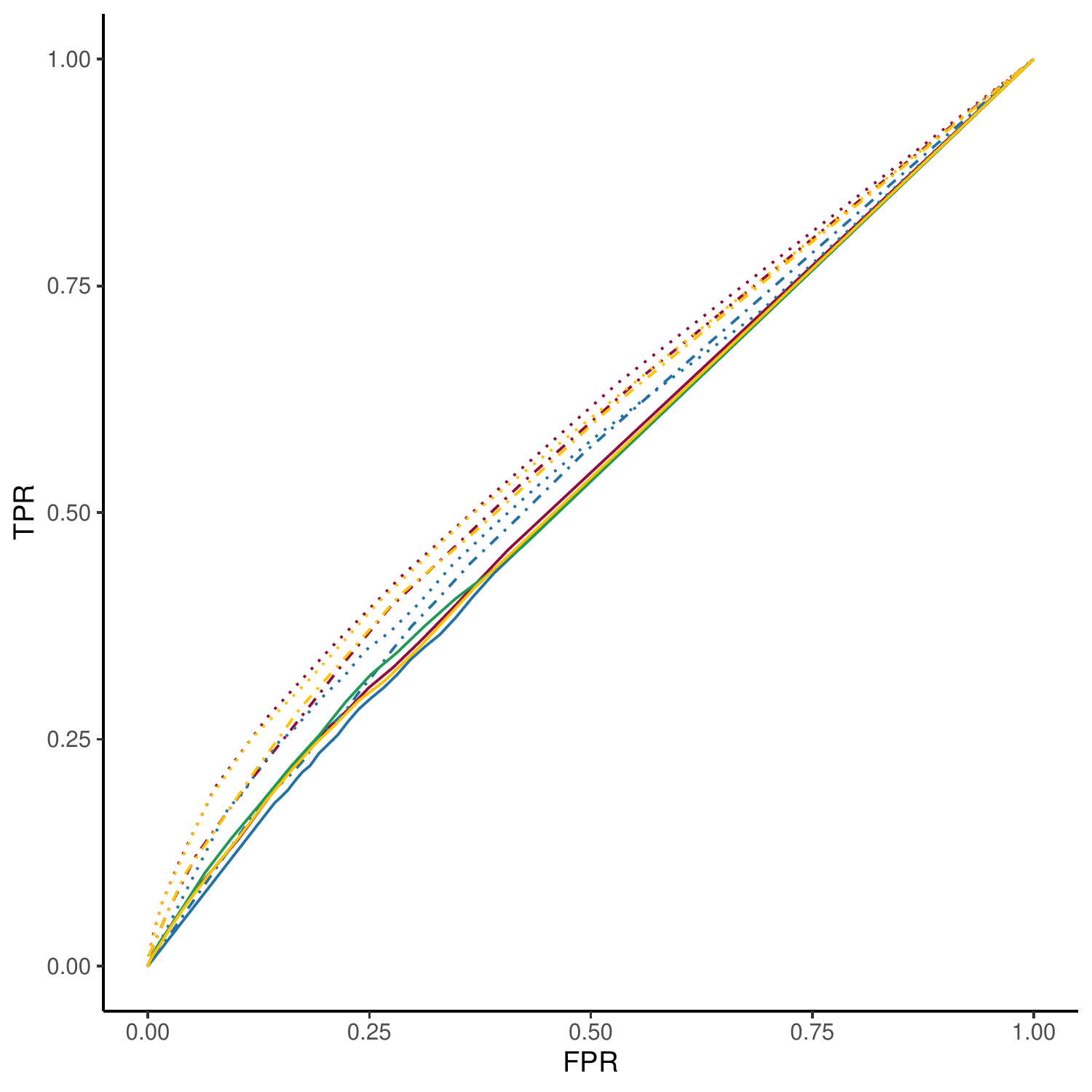}\hfill
\includegraphics[width=0.25\textwidth]{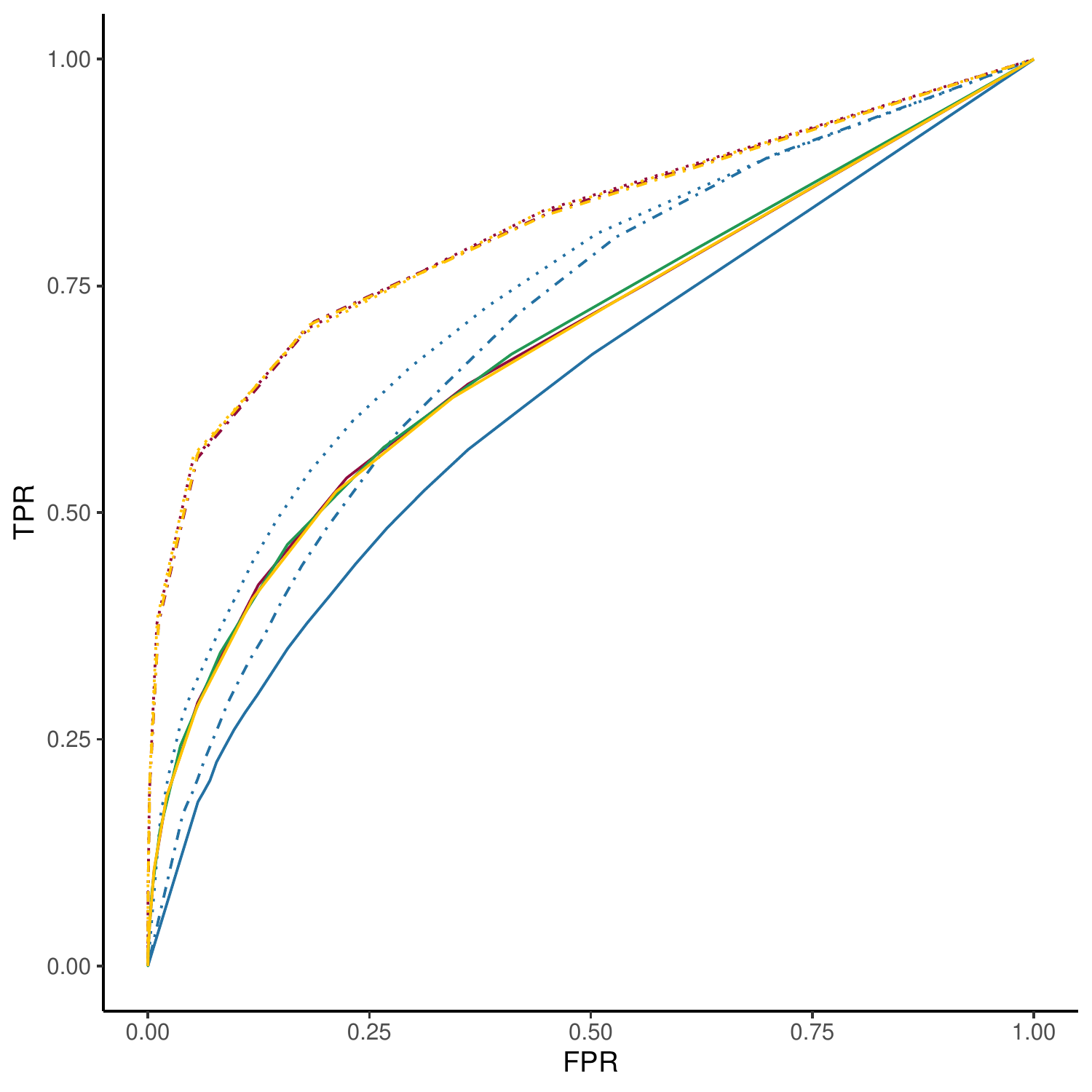}\hfill
\includegraphics[width=0.25\textwidth]{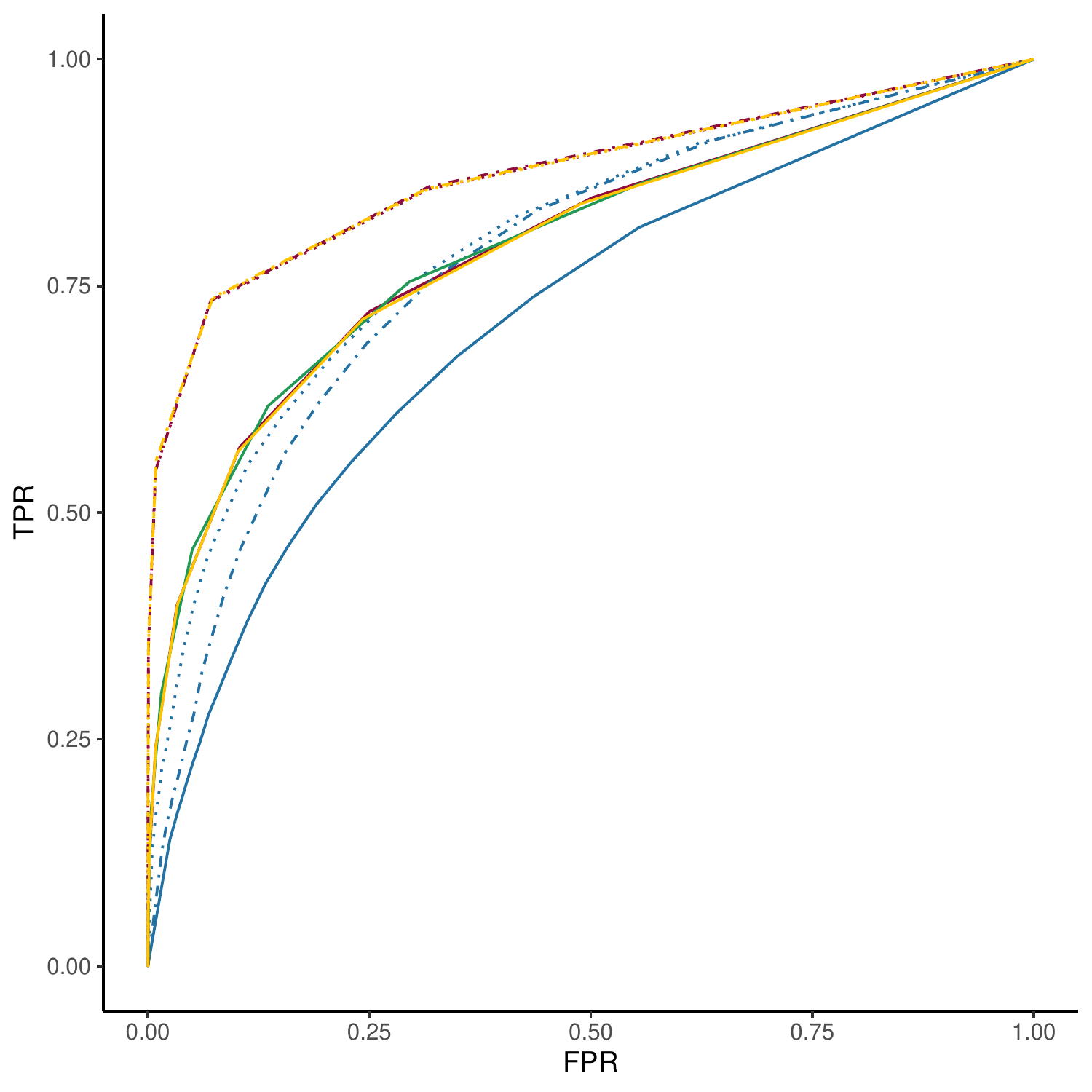}\hfill
\includegraphics[width=0.25\textwidth]{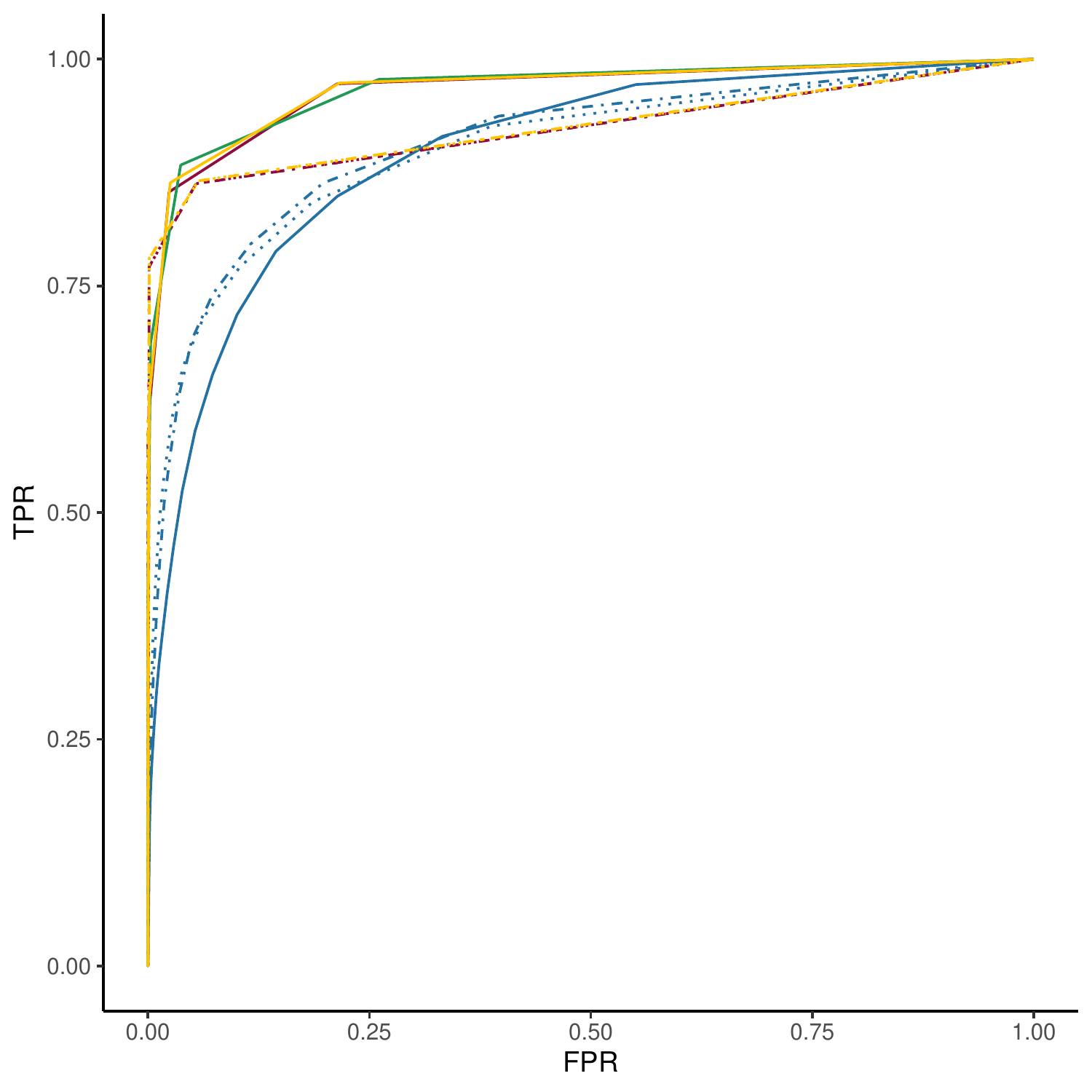}\hfill
\text{Scale-free network, $p = 100$, $\rho = 0.25$}\\
\includegraphics[width=0.25\textwidth]{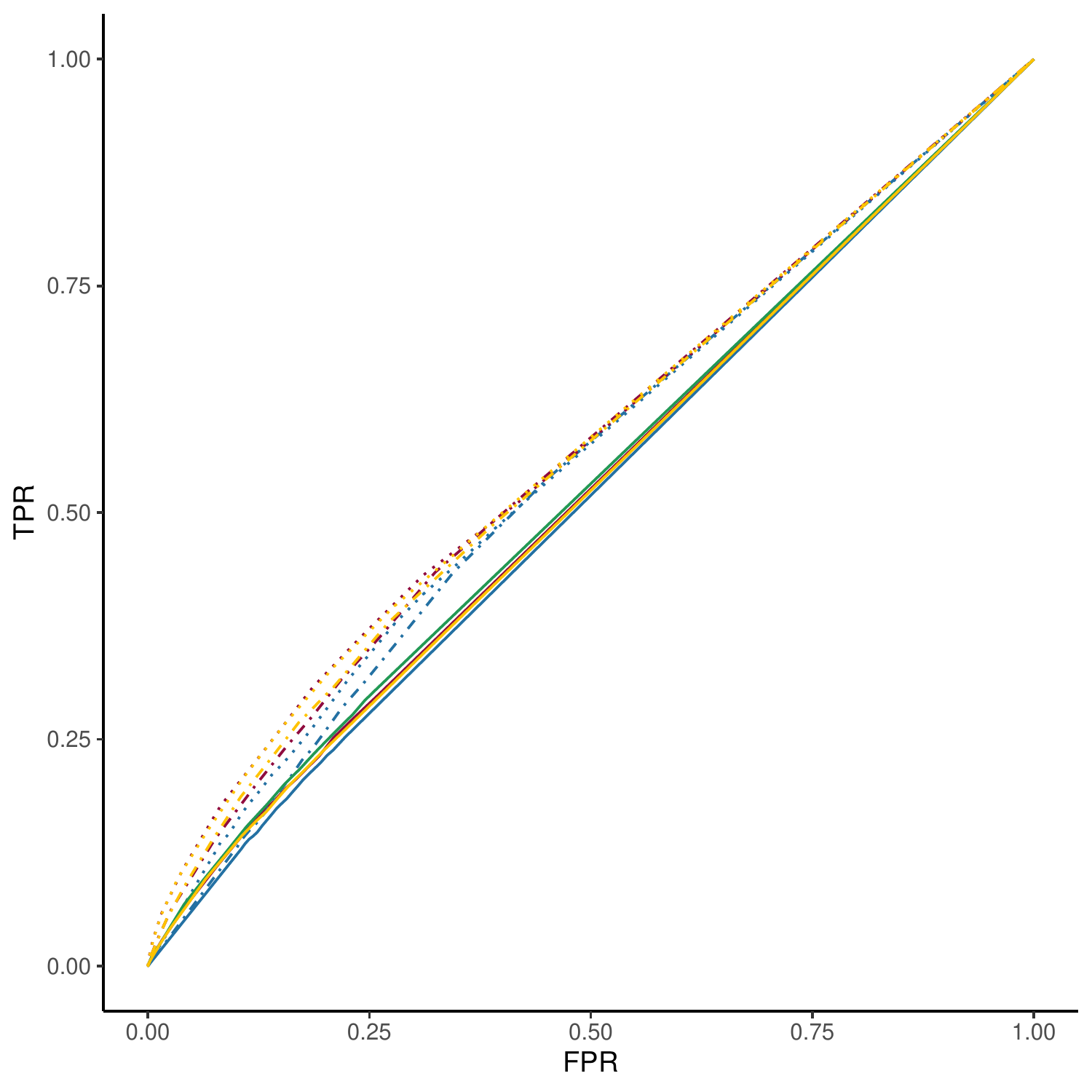}\hfill
\includegraphics[width=0.25\textwidth]{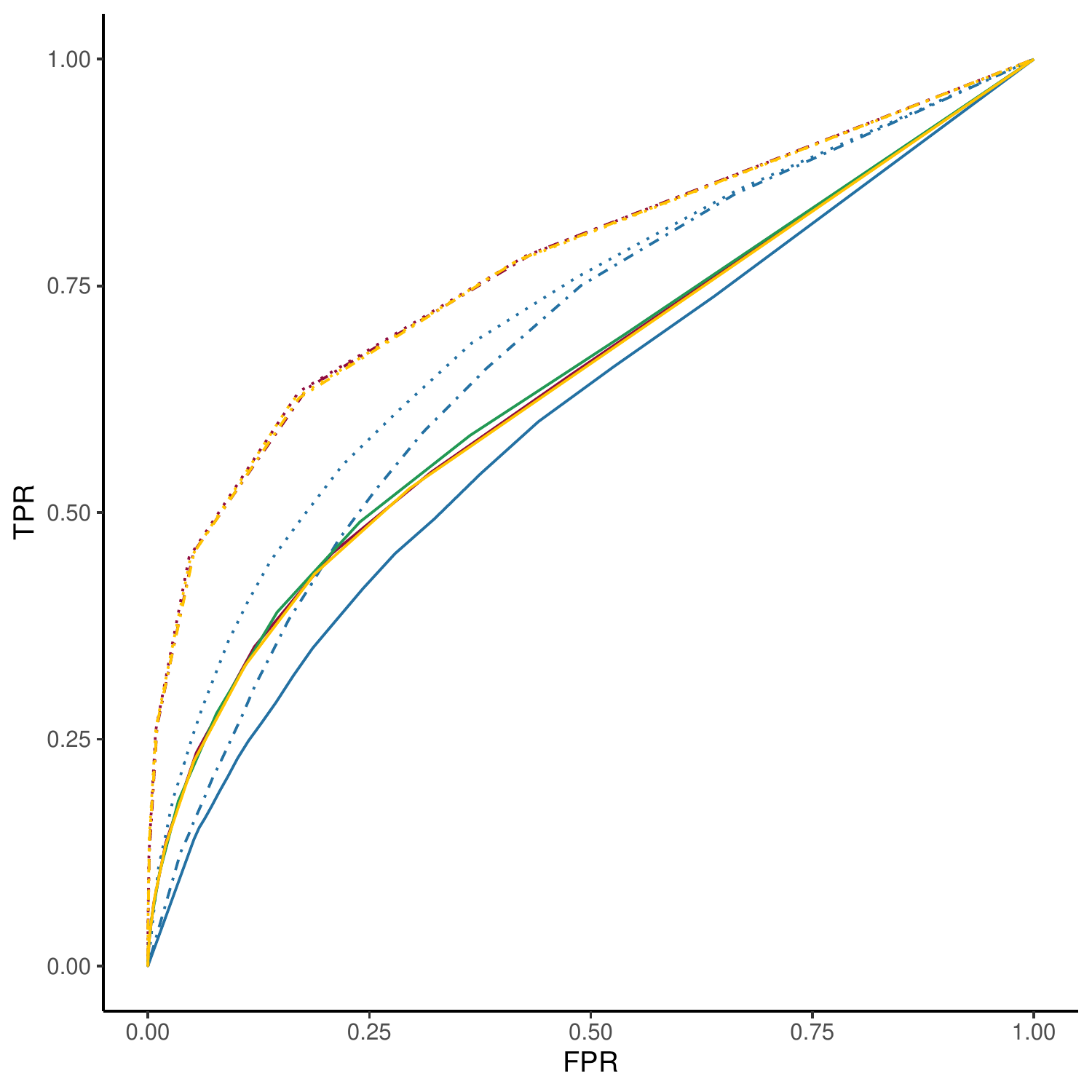}\hfill
\includegraphics[width=0.25\textwidth]{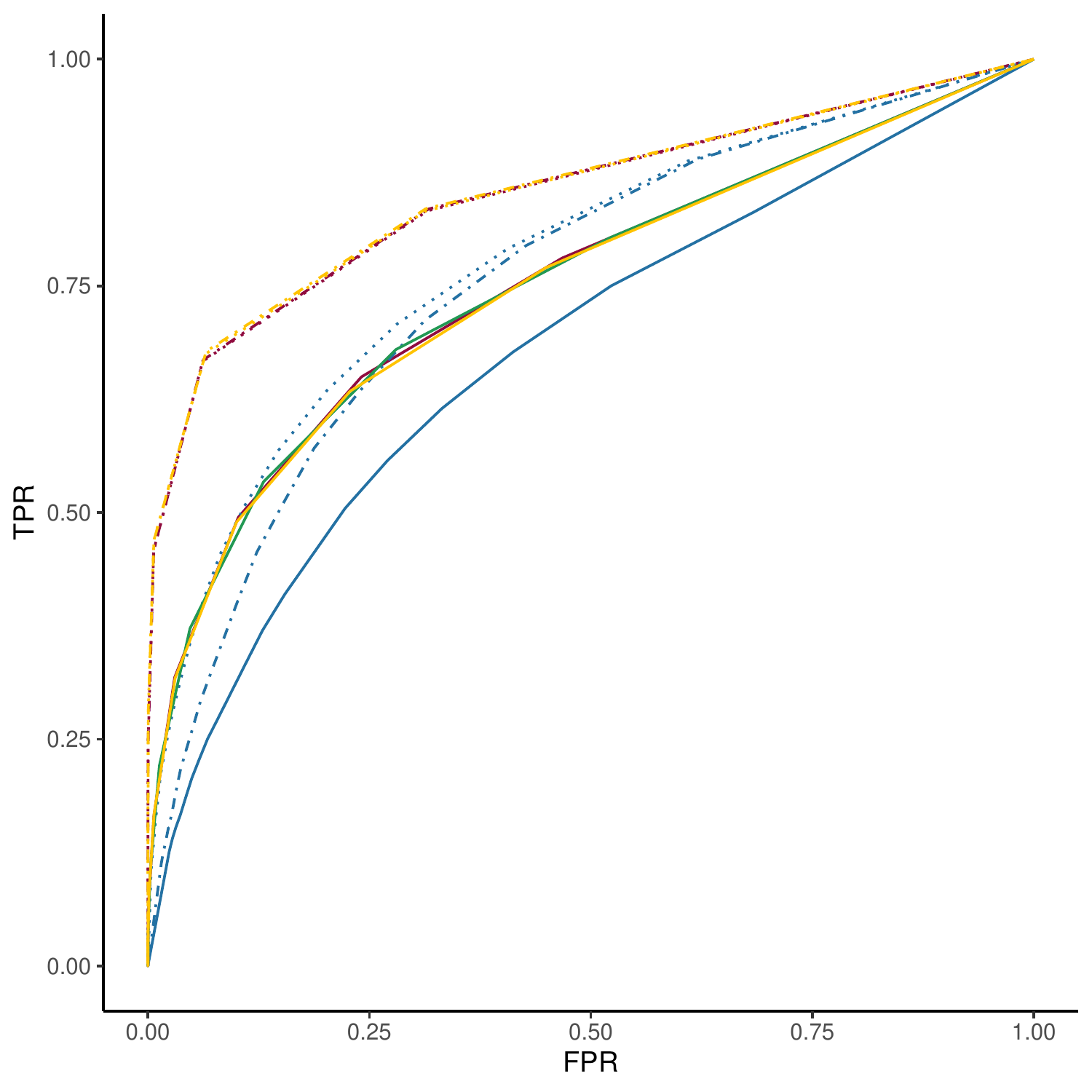}\hfill
\includegraphics[width=0.25\textwidth]{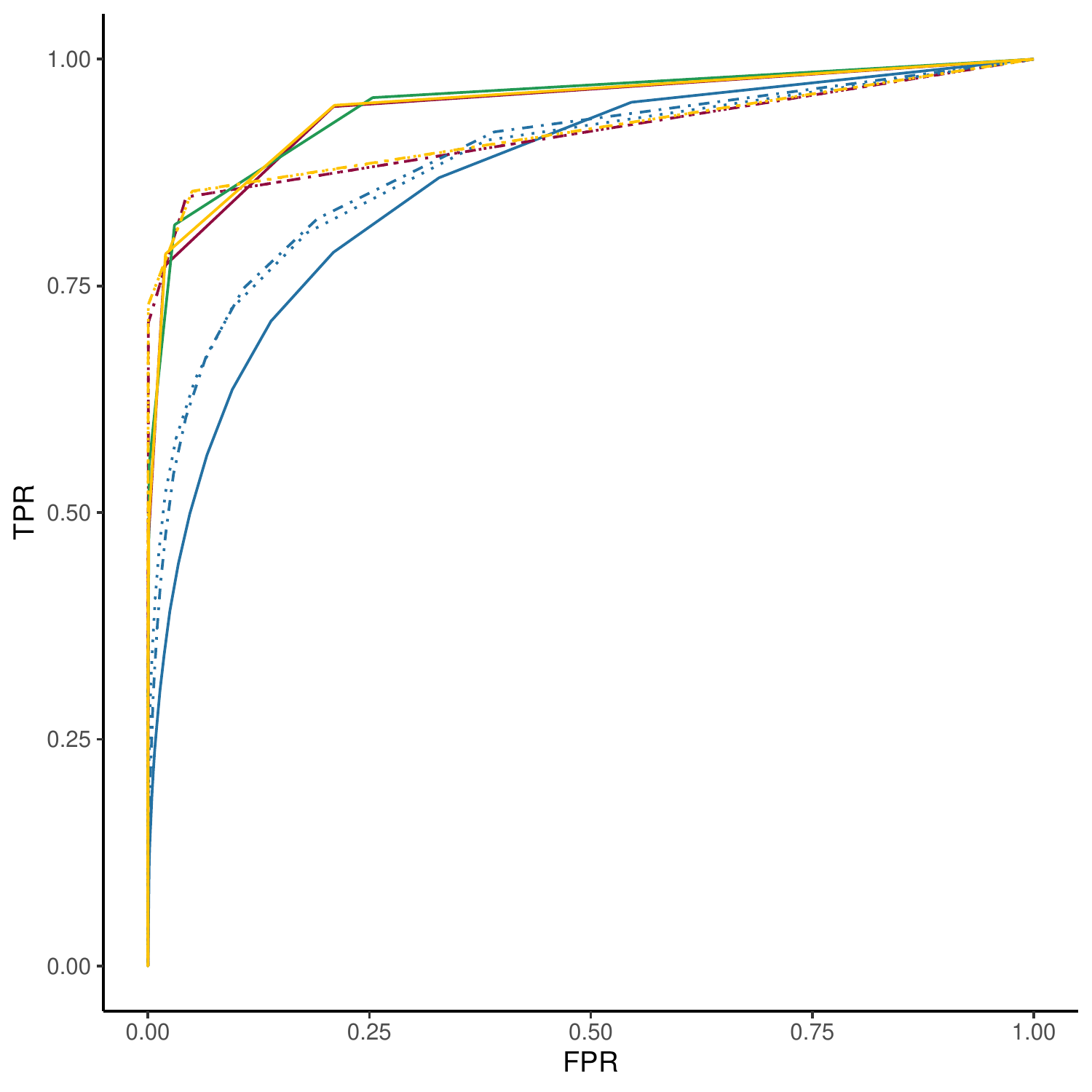}\hfill
\text{Scale-free network, $p = 50$, $\rho = 1$}\\
\includegraphics[width=0.25\textwidth]{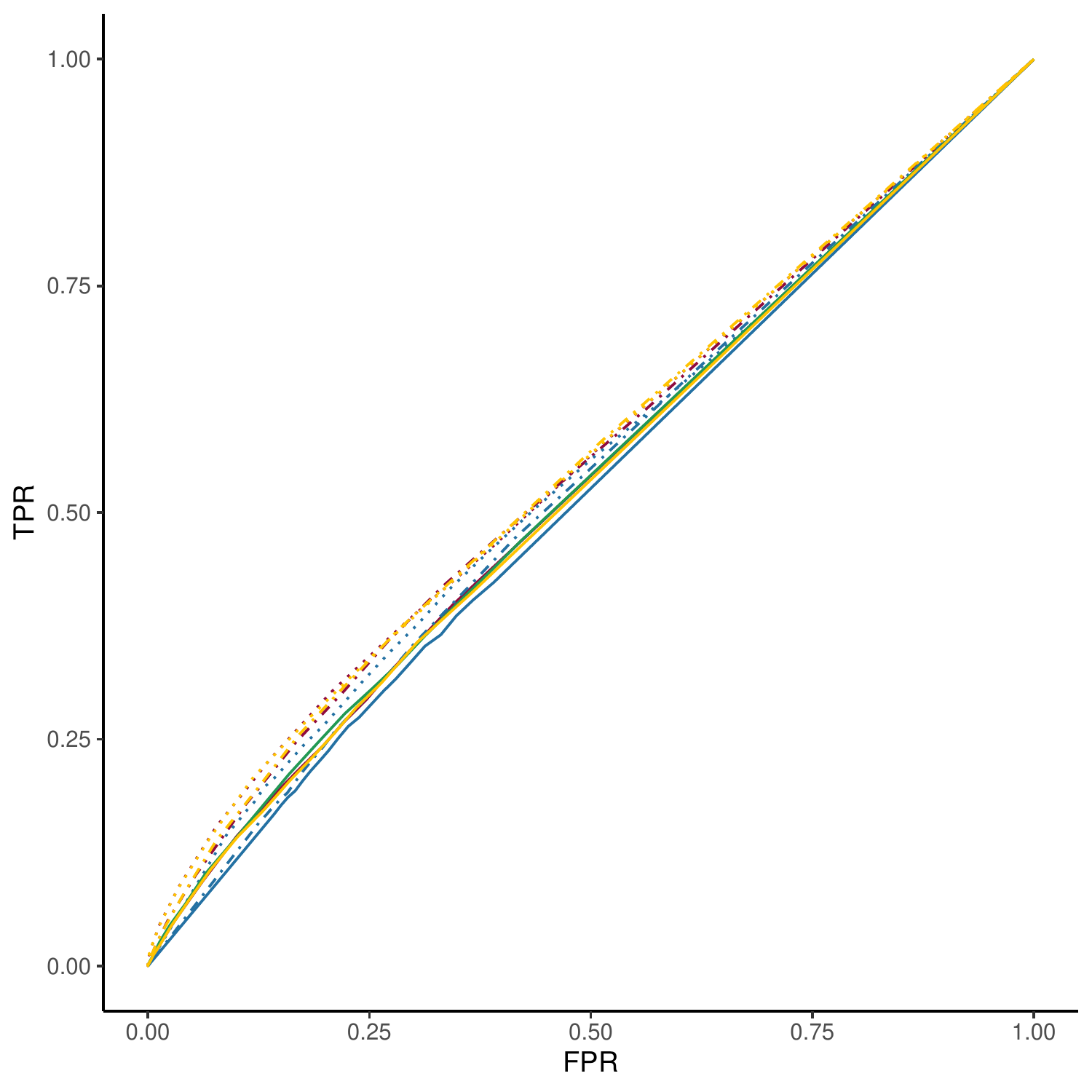}\hfill
\includegraphics[width=0.25\textwidth]{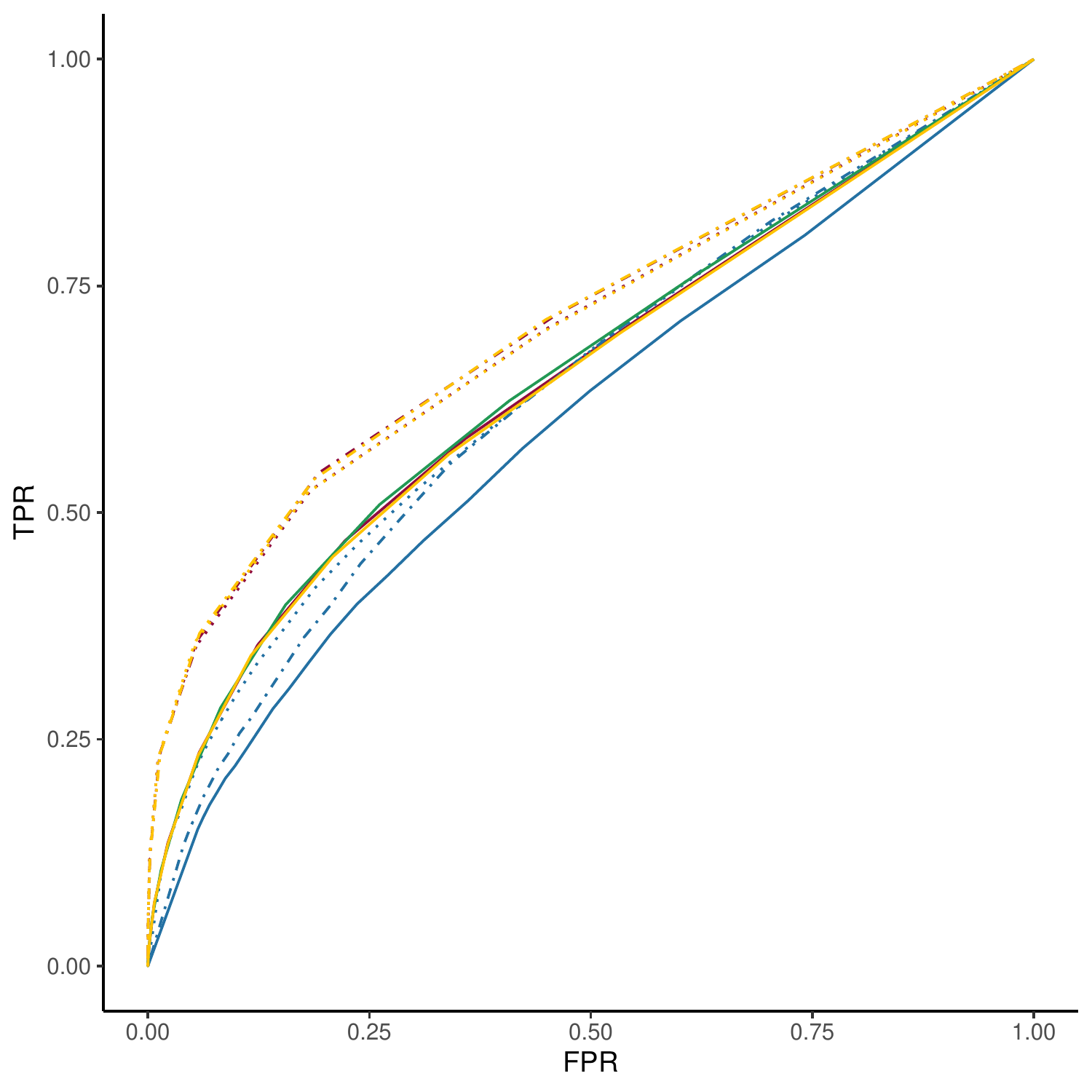}\hfill
\includegraphics[width=0.25\textwidth]{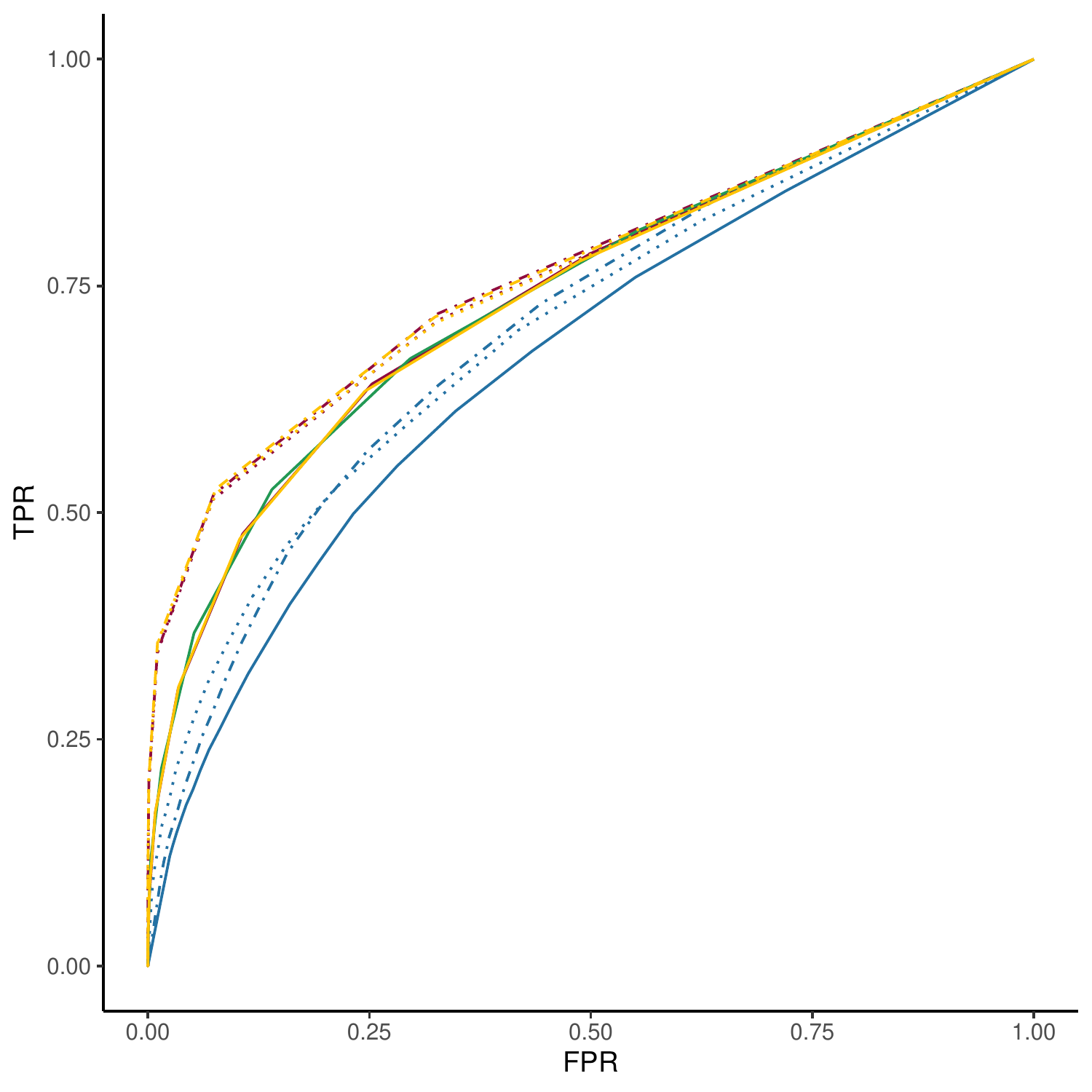}\hfill
\includegraphics[width=0.25\textwidth]{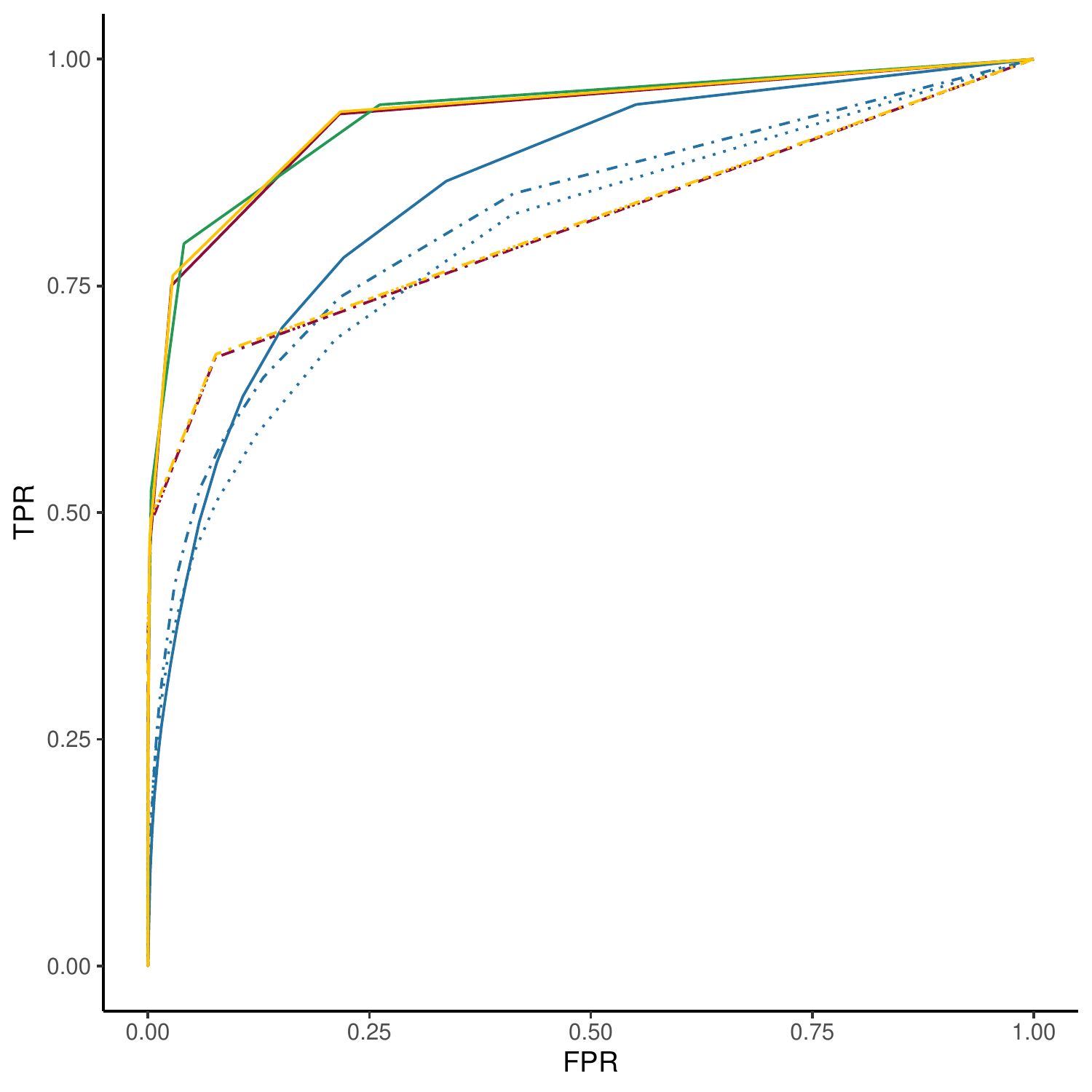}\hfill
\text{Scale-free network, $p = 100$, $\rho = 1$}\\
\includegraphics[width=0.25\textwidth]{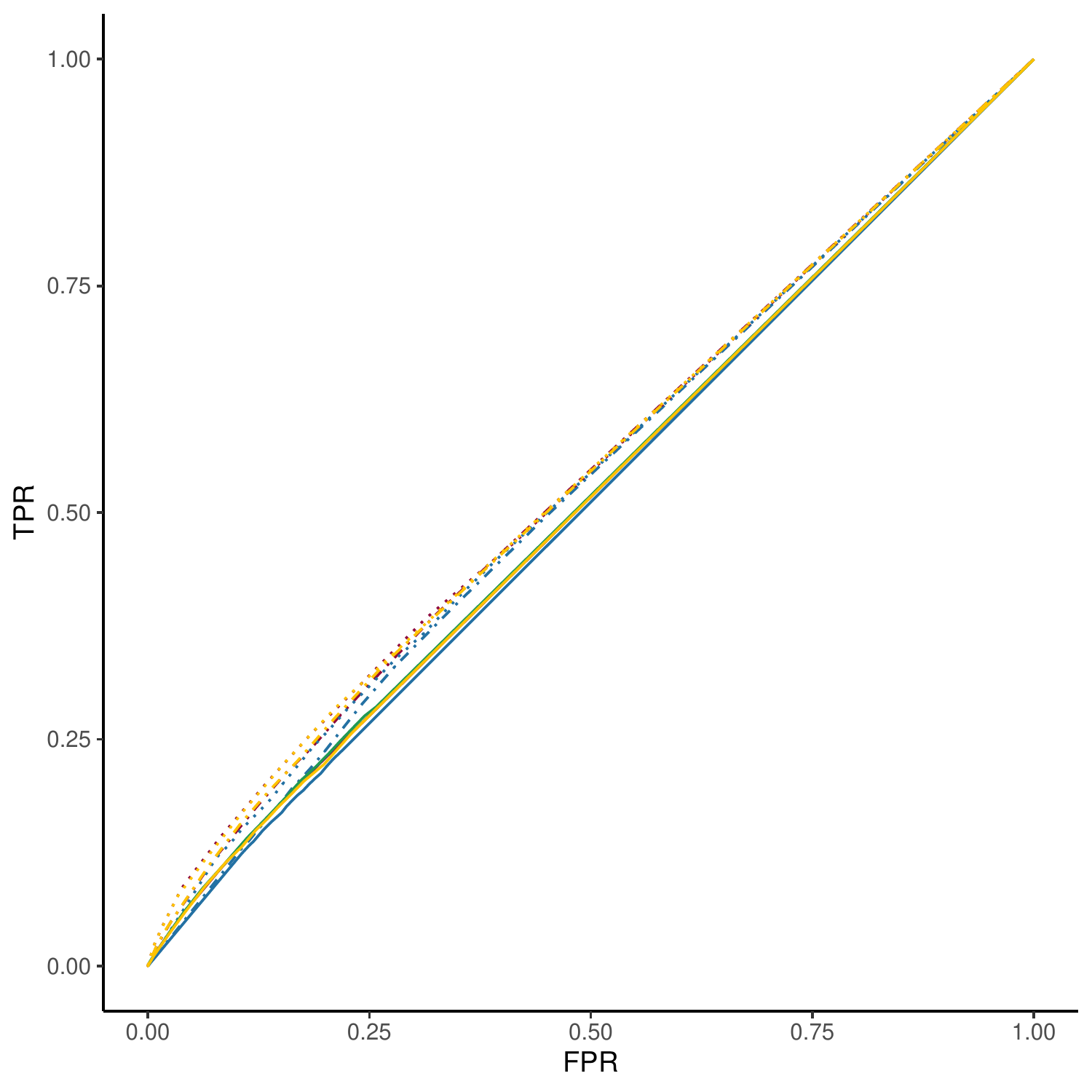}\hfill
\includegraphics[width=0.25\textwidth]{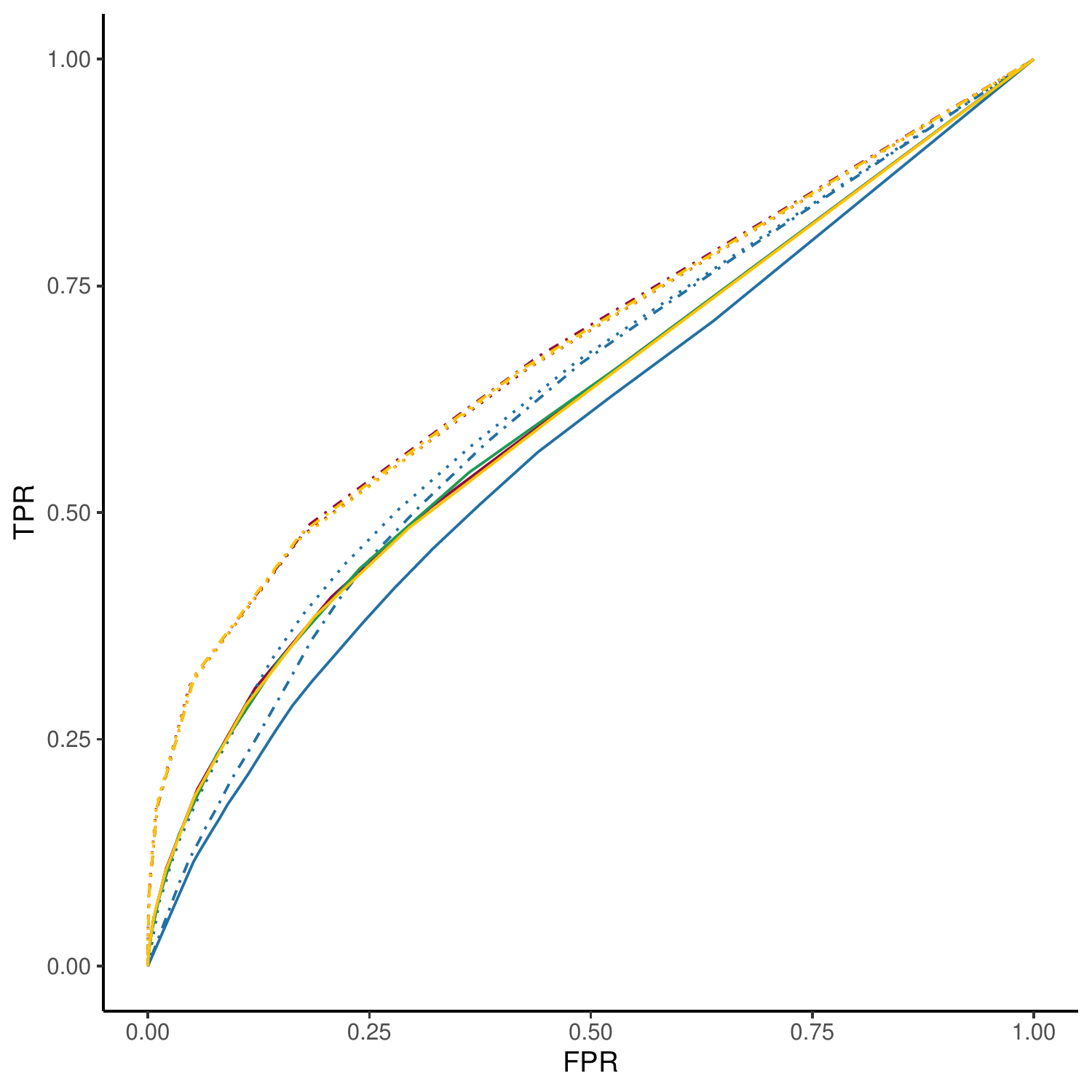}\hfill
\includegraphics[width=0.25\textwidth]{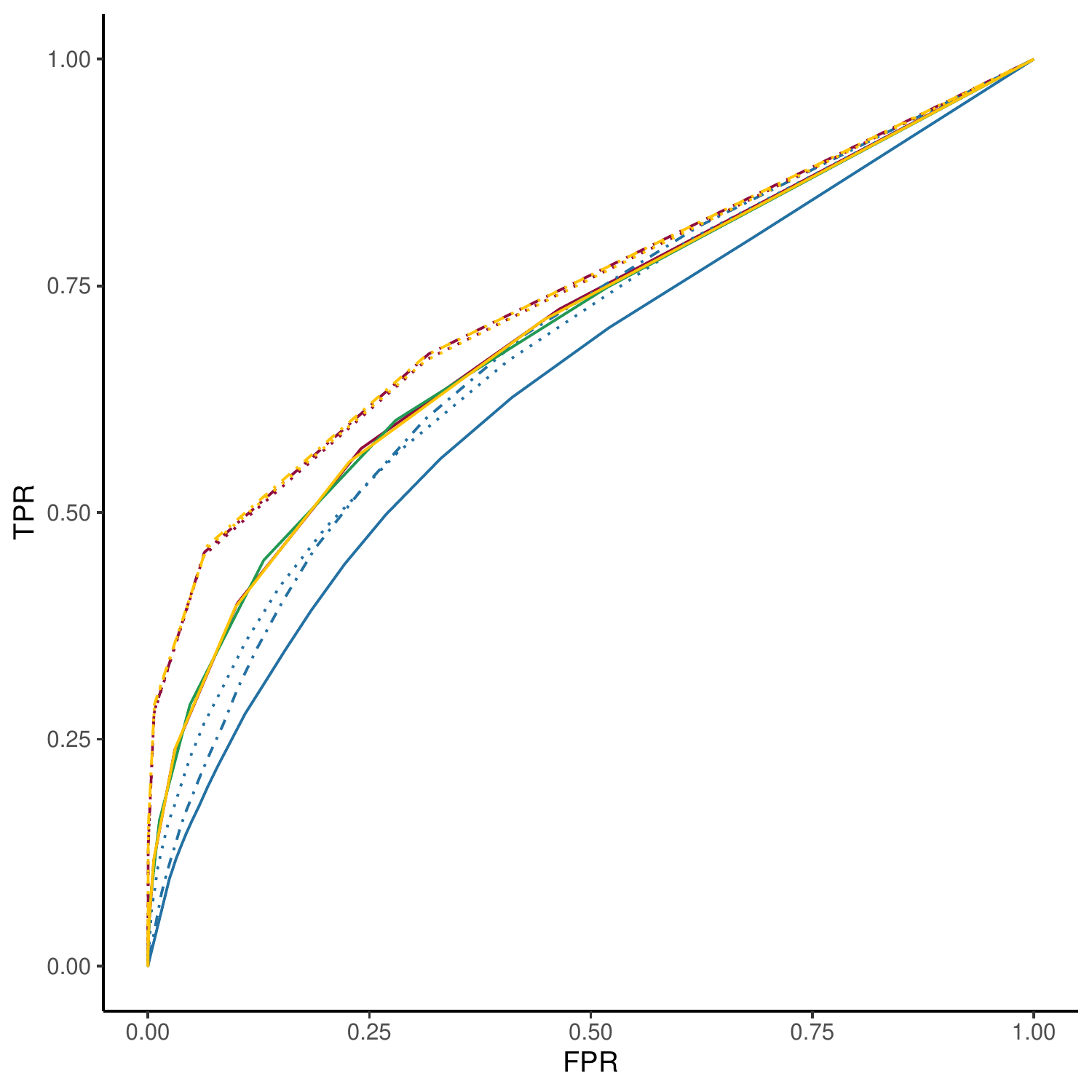}\hfill
\includegraphics[width=0.25\textwidth]{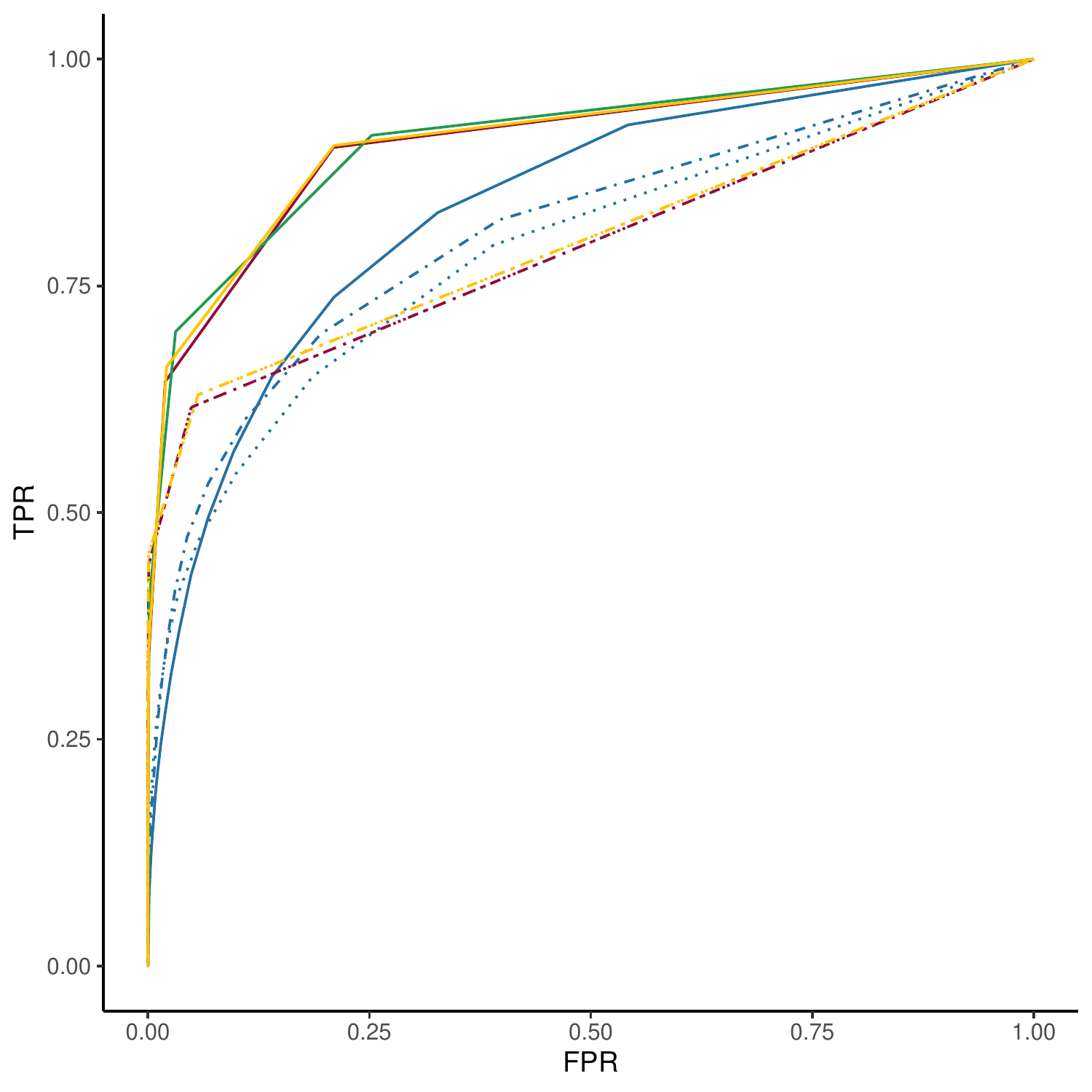}\hfill
\caption[]{ROC curves plotted over values of $\lambda_1$. The line type denotes the penalty for $\lambda_2$: $\lambda_2 = 0$ \begin{tikzpicture}\draw [thick] (0,2) -- (1,2);\end{tikzpicture}, $\lambda_2 = 0.1$  \begin{tikzpicture}\draw [dashdotted] (0,2) -- (1,2);\end{tikzpicture} and $\lambda_2 = 1$ \begin{tikzpicture}\draw [dotted] (0,2) -- (1,2);\end{tikzpicture}. The colors represent the method used: Gibbs method \begin{tikzpicture}\draw [colorGibbs, thick] (0,2) -- (1,2);\end{tikzpicture}, Approximate method \begin{tikzpicture}\draw [colorApprox, thick] (0,2) -- (1,2);\end{tikzpicture}, Fused graphical lasso \begin{tikzpicture}\draw [colorFGL, thick] (0,2) -- (1,2);\end{tikzpicture} and GLASSO \begin{tikzpicture}\draw [colorGLASSO, thick] (0,2) -- (1,2);\end{tikzpicture}. From left to right the values of $n$ are respectively 10, 50, 100 and 500.}
\label{fig:roccurves_sf}
\end{figure}

\end{appendices}
\end{document}